\documentclass[twocolumn]{aastex631} 
\usepackage{enumitem}
\usepackage{xcolor}
\usepackage{amsmath}
\usepackage{xcolor, soul}
\sethlcolor{yellow}
\usepackage{subfigure}
\usepackage{tablefootnote}
\usepackage{bm}
\hypersetup{backref=true,pagebackref=true,  hyperindex=true,colorlinks=blue,breaklinks=true,  urlcolor=violet,linkcolor= red, bookmarks=true, 	bookmarksopen=true,  filecolor=cyan, citecolor=blue, linkbordercolor=blue}
\shorttitle{Microlensing due to Isolated Black holes towards the MCs}
\shortauthors{Sajadian, Makler}

\begin{document}
\title{Detection and Characterization of Microlensing Events due to Isolated Black Holes towards the Magellanic Clouds in the Rubin Observations}

\author[0000-0002-0167-3595]{Sedighe Sajadian}
\affiliation{Department of Physics, Isfahan University of Technology, Isfahan 84156-83111, Iran, \url{s.sajadian@iut.ac.ir}}
\affiliation{Department of Astrophysics, Cosmology and Fundamental Interactions, Brazilian Center for Research in Physics, Centro Brasileiro de Pesquisas F\'{i}sicas, Brazil}

\author[0000-0002-0167-3595]{Mart\'{i}n Makler}
\affiliation{Department of Astrophysics, Cosmology and Fundamental Interactions, Brazilian Center for Research in Physics, Centro Brasileiro de Pesquisas F\'{i}sicas, Brazil}

\begin{abstract}
Gravitational microlensing surveys potentially discover isolated black holes (IBHs) at large distances through their gravitational effects on apparent brightness and motion of collinear stars. Here, we study detecting and characterizing IBHs within the mass range $[3,~5000]M_{\odot}$ with either stellar or dark matter origins in the upcoming observations by Vera~C.~Rubin Observatory towards the Large and Small Magellanic Clouds (LMC and SMC). We consider four lens mass functions (MFs:~$dN/dM\propto M^{-\beta}$ for $\beta=0,~0.5,~1,~2$), and generate long-duration microlensing events due to IBHs detectable by Rubin. By assuming the fraction of IBHs's mass in total Galactic mass as $\mathcal{F}$, Rubin potentially detects $\sim0.3,~\rm{and}~2$ microlensing events towards SMC and LMC due to IBHs if $\beta=1,~\rm{and}~\mathcal{F}=5\times10^{-3}$. These IBHs are inside our galaxy with the probability $\gtrsim70\%$. These events have on average $\theta_{\rm{E}}\sim30-50~\rm{mas}$, so that the probabilities of resolving their lensing-induced images through the Rubin astrometric observations are $\sim1.4,~\rm{and}~16.3\%$ towards LMC and SMC. In $\sim1,~5.2\%$ of these events their astrometric deflections are realizable. The probabilities of discerning their parallax are $\sim40,~25\%$. We evaluate the relative errors with simulating synthetic data points and by assuming the true models as the best-fitted ones. We conclude for a log-uniform MF for $\lesssim0.1,~\rm{and}~0.3\%$ of photometrically detectable microlensing events towards in LMC and SMC the relative errors in the lens mass, distance and velocity are $\lesssim3\%$. We calculate the number of detectable IBHs versus their mass by assuming their complete contribution of compact objects in halos' dark matter, and conclude Rubin specifies $95\%$~C.L. upper-limit on IBHs exclusion with masses $\lesssim208,~8M_{\odot}$ for $\mathcal{F}\simeq4\times10^{-3}$ by observing LMC and SMC.
\end{abstract}
\keywords{Gravitational Lensing:  Micro --  Black Halos --  Magellanic Clouds -- Numerical method}

\section{Introduction}\label{sec1}
Any compact object with radius equal to or smaller than Schwarzschild radius, i.e. $R_{\rm{sch}}=2 G M/c^{2}$, is the so-called black hole (BH). Here $M$ is the object mass, $G$ is the gravitational constant, and $c$ is the light speed. Such a compact object has an escape velocity equal to or higher than the light velocity, which means that nothing (even light) can escape from its surface. Hence, they are dark and can be realized through their gravitational effects, interacting with nearby bright companions or their accretion disks \citep{2006bookCaroll}.  

BHs are generated through three main channels which are listed here. (i) Massive stars with masses higher than $\simeq20M_{\odot}$ are converted to BHs at the end of their life when their internal thermonuclear fuels are finished. Through this channel, which is the so-called core collapse of massive stars, stellar-mass black holes are formed with masses in the range $\sim[3,~100]M_{\odot}$ \citep{SMBHs2017}. The smallest BH with the mass $\simeq3.3M_{\odot}$ was discovered in a non-interacting binary system with a giant star \citep{smallestMBHS}. (ii) Super massive black holes (SMBHs) can be generated through either direct collapse of massive gas clouds which mostly reside at the center of massive galaxies such as our Milky Way or merging less-massive black holes \citep{2003ApJLoeb,2014YueSMBHs}. They usually have masses $\gtrsim10^{5}M_{\odot}$. BHs with masses in the range $[100,~10^{5}]$ solar mass (between two introduced categories) are the so-called massive black holes (MBHs) and can be found in dense stellar clusters. Several collisions and mergers of stars in these dense regions potentially create MBHs \citep{2020IMBHSGreene}. (iii) Hypothetical primordial black holes could be generated in very early universe because of density fluctuations \citep{2024PBHs,2024PBHSArbey}.

Detection and characterization of BHs lie in the forefront of astrophysics, because they have important roles in (I) testing general relativity \citep{2016CQGraTESSGR}, and quantum mechanic, (II) study Galaxies' formation, structure, and evolution \citep{2009NaturBHGalaxy}, (III) observations of gravitational waves \citep{2016PhRvLGWdetection}, Hawking radiation \citep{1974NaturHawking}, and dark matter \citep{2021FrASSBHDarkmatter}, etc. There are several methods for detecting BHs depending on their masses and surrounding mater, which are listed in follows.  

\begin{itemize}[leftmargin=2.0mm]
\item Massive and super massive BHs usually have accretion disks which falling and hot gas and dust in these disks radiate throughout the electromagnetic spectrum from $X$-ray to the radio band. For instance, actively accretion disks around distant and early SMBHs are extremely luminous, the so-called quasars. Therefore, detection of $X$-ray emissions from the center of galaxies is a sign for SMBHs \citep[e.g., see, ][]{1967ApJXRayBHS,2017ApJZhang}.  

\item BHs which live in binary or multiple systems with stellar companions disturb orbits of their companions rotating them. If such systems are close with a significant falling matter from donors to compact objects (or accretors), they emit $X$-ray radiations. Therefore, BHs potentially are realizable through detecting $X$-ray emissions which are usually in conjunction with distorting motion of their bright companions \citep{2006csxs.bookTauris,2007AAXrayBinary}. Through time-varying spectroscopic observations by the Gaia telescope \citep{2016bgaia}, several non-interacting binary systems including black holes have been reported \citep[e.g., see, ][]{2023AJgaiachakrabarti,2023MNRASElbadri,2023MNRASElbardi2,2024AAGaiaPanuzzo}.

\item Merging compact objects in binary systems including white dwarfs, neutron stars and black holes produces gravitational waves detectable by precise interferometers such as the Laser Interferometer Gravitational-wave Observatory (LIGO), and Virgo \citep{2016PhRvLGWdetection,2016PhRvLGWdetection2}. 

\item Detecting isolated stellar-mass black holes (ISMBHs) is possible through their lensing effects on the apparent brightness of background stars while they have the same lines of sight with BHs' ones. It is worth noting that isolated black holes are uniquely detectable through gravitational microlensing surveys. In the lensing formalism, by considering a point-like lens and a point-like source star, the lens equation (a known relation between the images' and source' positions projected on the sky plane with respect to the lens location) is converted to a simple second-order equation which means two images are generated with an angular separation in order of some ten micro arc seconds for a common lens object \citep{Einstein1936,1964MNRASrefsdal,Liebes1964}.  
\end{itemize}

The first unambiguous detection of an ISMBH was reported in 2022 \citep{2022ApJSahu,2022ApJCasey}. For this event, simultaneous observations by ground-based telescopes and the Hubble Space Telescope (HST) telescope revealed the parallax amplitude and the astrometric deflection in its source trajectory. Accordingly, the mass and distance of its dark lens object were uniquely indicated as $7.1~M_{\odot}$ and $1.58~$kpc, respectively. In addition, some more ISMBHs were discovered through microlensing surveys, although in most of them the physical parameters of BHs were estimated through Bayesian analysis owing to the degeneracy problem \citep[see, e.g., ][]{2025AAHowil,2002ApJBennett,2002MNRASMao}. Degeneracy in microlensing observations means that we are not able to determine the physical parameters of lens and source stars uniquely. Resolving this degeneracy is possible by measuring both lensing-induced astrometric deflection in the source trajectory and the parallax amplitude \citep{2000ApJDominik,2011Sahuproposal}. Instead of astrometric deflection in the source trajectory, resolving two images in high-magnification microlensing events leads to break the microlensing degeneracy \citep{2019ApJDong}.

In the regard of detecting BHs through microlensing surveys, performing realistic Monte-Carlo simulations from these events helps to predict, test models, adjust future observations to improve the detection efficiencies, and find applicable methods for resolving degeneracy. For instance, \citet{2023AJSajadianSahu} simulated microlensing events due to ISMBHs detectable in the Roman observations and proposed some sparse observations during its long gap with a $\sim10$-day cadence to measure astrometric deflections with a higher efficiency. Also, \citet{2023AJSajadianArya} investigated possibility of measuring the parallax amplitudes through follow-up astrometric observations, e.g., by the Extremely Large Telescope (ELT) telescope, in long-duration microlensing events due to ISMBHs.      

In this work, we study detection and characterization of isolated stellar-mass and massive black holes (throughout the paper we use IBHs for them) numerically. We consider a wide mass range for IBHs and while simulating microlensing events we apply survey observations by Vera C. Rubin observatory which was previously known as the Large Synoptic Survey Telescope \citep[LSST, ][]{lsstbook}, towards the Magellanic Clouds (MCs). This telescope will survey all southern-sky during $\sim3$ nights and its mission will last $10$ years. According to its observing strategy, we simulate microlensing light curves due to IBHs and their astrometric source trajectories. We aim to answer some main questions from our numerical calculations: 
(i)What is the Rubin efficiency for detecting and characterizing these massive objects in the Galactic halo during its mission?
(ii)What are the probabilities of detecting parallax effects, and lensing-induced astrometric deflections in the source trajectories, and resolving two images? 
(iii)What is the number of detectable and characterizable IBHs during the Rubin survey observations? 
(iv)What are $95\%$ confidence level (C.L.) upper-limits on the exclusion of IBHs in the Galactic halo dark-matter?

The outline of the paper is as follows. 
In Section \ref{sec2}, we review the formalism for simulating microlensing light curves and astrometrc source trajectories by considering the parallax effect. 
In Section \ref{sec3}, we explain the details of Monte-Carlo simulations from microlensing events due to IBHs toward MCs, and mention our detectability criteria.  
In section \ref{sec4}, we offer the results from simulations, investigate the lensing, physical and statistical properties of detectable events, and evaluate the relative errors of the extracted parameters from observations numerically.  
In Section \ref{sec5}, we re-perform simulations (by considering discrete values for the lens mass in the logarithmic scale and assuming dark-matter halos are completely generated from IBHs) to determine high-confidence upper limits on IBHs' mass exclusion in dark matter halos.  
In Section \ref{sec6}, we review results and conclude.   

\section{Formalism}\label{sec2}
Several valuable review papers and books about the microlensing formalism and its applications have been published up to now \citep[see, e.g.,][]{1992grle.bookschneider,2006glswambsganss,2012gaudireview,2012RAAmao}. Based on them, we briefly review the known microlensing formalism here, by explaining the parallax effect and lensing-induced astrometric deflections in source trajectories.

\noindent In the lensing formalism, there is a known equation which gives the source position as a function of the images' position projected on the sky plane with respect to the lens location which is the so-called lens equation. If angular separations of lens, source and the observer are very small and the lens and source star are point-like objects, the lens equation is converted to a simple second-order equation. Therefore, two images are formed whose angular distances from the lens position are given by:  
\begin{eqnarray}
\boldsymbol{\theta}_{\pm}= \frac{\theta_{\rm E}\hat{u}}{2}\Big(u\pm \sqrt{u^{2}+4}\Big), 
\label{theta}
\end{eqnarray}
\noindent here, the bold font indicates a vector, and $\hat{u}$ is a unit vector showing the direction of the relative lens-source separation projected on the sky plane. $\theta_{\rm E}$ is the length scale in the lensing formalism which is the so-called angular Einstein radius and is given by:  
\begin{eqnarray}
\theta_{\rm{E}}=\sqrt{\kappa~M_{\rm l}~\pi_{\rm{rel}}},~~ \pi_{\rm{rel}}(\rm{mas})=\frac{1}{D_{\rm l}(\rm{kpc})}-\frac{1}{D_{\star}(\rm{kpc})},
\label{tetE}
\end{eqnarray}
where, $\kappa=8.13$ mas is a constant, $M_{\rm l}$ is the lens mass, and $\pi_{\rm{rel}}$ is the amplitude of the relative parallax effect. In Equation \ref{tetE}, by inserting the lens mass in the unit of the solar mass, and $\pi_{\rm{rel}}$ in the unit of mas, the resulting $\theta_{\rm E}$ will be in the unit of mas. $D_{\rm l}$ and $D_{\star}$ are the lens and source distances from the observer. However, the length scale in the lensing formalism is additionally specified by $R_{\rm E}=D_{\rm l}\theta_{\rm E}$, i.e., the Einstein radius which is the radius of the images' ring projected on the lens plane when the lens, source and observer are completely aligned. 

\noindent In Equation \ref{theta}, $u=\sqrt{u_{0}^{2}+(t-t_{0})^{2}\big/t_{\rm E}^{2}}$ is the relative lens-source separation projected on the lens plane and normalized to the Einstein radius which is a function of time $t$. Here, $u_{0}$ is the so-called lens impact parameter which is the closest lens-source separation normalized to the Einstein radius and projected on the lens plane, $t_{0}$ is the time of the closest approach, and $t_{\rm E}=\theta_{\rm E}/\mu_{\rm{rel}}$ is the known time scale in the lensing formalism, i.e., the time duration of crossing the angular Einstein radius by the angular lens-source relative velocity ($\mu_{\rm{rel}}$) which is given by:
\begin{eqnarray}
\boldsymbol{\mu}_{\rm{rel}}= \boldsymbol{\mu}_{\star}-\boldsymbol{\mu}_{\rm l}=\frac{\boldsymbol{v}_{\star}-\boldsymbol{v}_{\sun}}{D_{\star}}-\frac{\boldsymbol{v}_{\rm l}-\boldsymbol{v}_{\sun}}{D_{\rm l}},
\end{eqnarray}  
where, $v_{\sun}$ refers to the Sun velocity projected on the sky plane, and $v_{\star}$ and $v_{\rm l}$ are the source and lens velocities projected on the sky plane, respectively. 

According to Equation \ref{theta}, the images' separation reaches $\Delta \theta_{0}=\theta_{\rm E}\sqrt{u_{0}^{2}+4}$ at the time of the closest approach. In common microlensing events with $u_{0}\lesssim1$, this separation is $\sim2\theta_{\rm{E}}$. For lensing observations towards the Galactic bulge due to stellar-mass lens objects, $\theta_{\rm E}\sim0.3$ mas, which is unresolvable in ground-based observations. In such events, we receive the total flux due to both images which is magnified with respect to the source flux at the baseline by the factor $A$ (the so-called magnification factor) as: 
\begin{eqnarray}
A=A_{+}+A_{-}=\frac{u^{2}+2}{u\sqrt{u^{2}+4}},~A_{\pm}=\frac{u^{2}+2}{2u\sqrt{u^{2}+4}}\pm\frac{1}{2},
\end{eqnarray}   
which is a function of time. Here, $A_{\pm}$ are the magnification factors of the two images formed at the positions $\theta_{\pm}$.

In this work, we study microlensing events due to IBHs towards MCs. For these events $D_{\star}\simeq50-60$ kpc. The lens objects could be inside either our galaxy with $D_{\rm l}\lesssim8$ kpc (the so-called halo-lensing events) or MCs with $D_{\rm l}\gtrsim47$ kpc (the so-called self-lensing events). For these events $\pi_{\rm{rel}}\sim0.1-10,~10^{-3}$ mas for halo- and self-lensing events, respectively. By assuming IBHs with masses in the range $M_{\rm l}\in100-1000M_{\sun}$, halo- and self-lensing events have $\theta_{\rm E}\sim10-100,~0.1-2$ mas, respectively. 

\noindent The Rubin astrometric accuracy to resolve two close stars is $\sim100-1000$ mas \citep{lsstbook}, Hence, the microlensing events due to IBHs inside our galaxy (halo-lensing events) towards MCs barely have resolvable images. In addition the separated images may be too faint to be discerned, or the blending effect prevents us to detect them. The magnification factor of the secondary image ($A_{-}$) which is located at $\theta_{-}$ (with the negative parity) rapidly reaches zero while the source is moving away from the lens object, so resolving images and directly measuring $\theta_{\rm{E}}$ are hardly possible in high-magnification events and at the time of the maximum magnification. For high-magnification microlensing events, both images are magnified at the magnification peak.

Equation \ref{theta} manifests that images' positions are not over the source location (which is $\boldsymbol{u}~\theta_{\rm{E}}$) and there is a time-varying deflection between true position of the source star and the images' brightness center which is: 
\begin{eqnarray}
\boldsymbol{\delta\theta}_{\rm c}=\frac{\boldsymbol{\theta}_{+}A_{+}+\boldsymbol{\theta}_{-}A_{-}}{A_{+}+A_{-}}-\boldsymbol{u} \theta_{\rm E}= \frac{\boldsymbol{u}~\theta_{\rm E}}{u^{2}+2},
\end{eqnarray}
where, its size is proportional to $\theta_{\rm{E}}$. Discerning this deflection in the source trajectory causes measuring $\theta_{\rm E}$ which is a function of the lens mass and its distance from the observer. For microlensing events due to IBHs, the astrometric shift is large since its amplitude is scaled by $\theta_{\rm E}\propto \sqrt{M_{\rm l}}$.  

Parallax effect refers to the Earth's motion around the Sun, because the observer is rotating the Sun with the Earth. The Earth motion makes a periodic deviation in the lens-source projected separation normalized to the Einstein radius $\boldsymbol{\Delta u}$ (and as a result the magnification factor) with the amplitude $\pi_{\rm E}$ as given by:  
\begin{eqnarray}
\boldsymbol{\Delta u}=\pi_{\rm E}\int_{t_{0}}^{t}\boldsymbol{\mu}_{\oplus, \rm n} dt;~~\pi_{\rm E}=\frac{\pi_{\rm{rel}}}{\theta_{\rm E}}=\sqrt{\frac{\pi_{\rm{rel}}}{\kappa~M_{\rm l}}},
\label{parallax}
\end{eqnarray}\\
where, $\boldsymbol{\mu}_{\oplus, \rm{n}}$ is the Earth's angular velocity with respect to the Sun versus time, projected on the sky plane and normal to the observer's line of sight. The parallax effect makes similar periodic deviations in the astrometric deflection as well, but with the amplitude $\pi_{\rm{rel}}$ \citep{2023AJSajadianArya}. 
  
By measuring three parameters $t_{\rm E}$, $\theta_{\rm E}$, and $\pi_{\rm E}$ one can uniquely specify $M_{\rm l}$, $D_{\rm l}$, and $\mu_{\rm{rel}}$. To derive the lens angular velocity, the source angular velocity $\boldsymbol{\mu}_{\star}$ should be extracted from the source trajectory projected on the sky plane which is given by: 
\begin{eqnarray}
\boldsymbol{\theta}_{\star}=\boldsymbol{\mu}_{\star}(t-t_{0})+ \boldsymbol{u_{0}} \theta_{\rm E}+\boldsymbol{\delta\theta}_{\rm{c}}+ \pi_{\star}\boldsymbol{\Delta u}, 
\label{tets}
\end{eqnarray}
where, $\boldsymbol{u_{0}}=u_{0}\big(-\sin \xi,~\cos \xi\big)$, and $\xi$ is the angle between the source trajectory and the horizontal axis which in our formalism is the intersection axis of the Galactic plane and sky plane \citep[see, Appendix of ][]{2023AJSajadianSahu}. $\pi_{\star}(\rm{mas})=1/D_{\star}(\rm{kpc})$ is the so-called source parallax amplitude.  
In the next section, based on the introduced formalism we simulate microlensing events toward MCs due to IBHs which potentially are detected in the Rubin survey observations.

\section{Microlensing due to IBHs towards Magellanic Clouds}\label{sec3}
We simulate microlensing events towards MCs by limiting observing directions to two square-shape fields with the right ascension and declination ranges $\alpha\in[75,~90]$ degree, $\delta\in[-75, -60]$ degree for Large Magellanic Cloud (LMC), $\alpha\in[5.7,~20.7]$ degree, $\delta\in[-78.7,~-63.7]$ degree for Small Magellanic Cloud (SMC). LMC has three structures including disk, bulge and halo, while SMC has two structures halo and disk. In the two following paragraphs, we explain how to generate source and lens populations, respectively.

The source distance is specified based on the overall stellar mass versus the radial distance from the observer ($D_{\star}$) for any given line of sight, i.e., $dM(D_{\star}, \alpha, \delta)\big/dD_{\star}=\rho^{\rm{ST}}_{\rm{tot}}(D_{\star}, \alpha, \delta)\times D_{\star}^{2}~d\Omega$, where $\rho_{\rm{tot}}^{\rm{ST}}=\rho_{\rm b}+\rho_{\rm{d}}^{\rm{Thin}}+\rho_{\rm{d}}^{\rm{Thick}}+\rho_{\rm{h}, \rm{ST}}+\rho^{\rm{MC}}_{\rm{d}}+\rho^{\rm{MC}}_{\rm{b}}+\rho^{\rm{MC}}_{\rm{h}, \rm{ST}}$ is the overall stellar spatial distribution in a given line of sight ($\alpha, \delta$) and distance. Here, the indexes b, d, and h for densities refer to the bulge, disk and halo contributions in density. Also, ST for the densities refers to their stellar contribution. We use the spatial distributions in our Milky Way galaxy \citep{Robin2003,Robin2012} and MCs \citep[see, e.g., ][]{2000ApJGyuk,Mancini2004,2013MNRAS.435.1582C} which are reviewed in Appendix \ref{app1}. After specifying the source distances, we determine their physical and photometric parameters based on the Besan\c{c}on Galaxy model \citep{Robin2003,Robin2012}. In this model, the stars in Galactic thin and thick disks, Galactic bulge and its halo have different ages which are $\rm{Age}\in[0-10],~11,~10$ and $14$ Gyr, respectively. We assume physical and photometric properties of stars inside the disk, bulge and halo of MCs are similar to ones within the thin disk, bulge, and halo of our galaxy. The blending effect towards MCs especially for the deep Rubin observations is significant. So, we estimate the blending effect by accounting stars whose lines of sight are within the source PSF (Point Spread Function) \citep[see, Eq. (2) of ][]{sajadian2019}. We note that the stellar PSF area depends on the applied filter.

The lens population from IBHs within the mass range $[3,~5000] M_{\odot}$ either have a stellar-origin or is a part of dark matter in form of compact massive halo objects (we specify them here as dark-matter IBHs). For both stellar-origin (SO) and dark-matter (DM) IBHs, we determine their distances from the observer ($D_{\rm l}$) using the lensing event rate function which is proportional to $\Gamma\propto R_{\rm E}~\rho_{\rm{tot}}(D_{\rm l}, \alpha, \delta)$, where $R_{\rm{E}}\propto \sqrt{D_{\rm l}(D_{\star}-D_{\rm l})/D_{\star}}$.  Also, $\rho_{\rm{tot}}=\rho_{\rm{tot}}^{\rm ST}+f \times \rho_{\rm{h}, \rm{DM}}+f\times \rho_{\rm{h}, \rm{DM}}^{\rm MC}$, which is the overall (stellar and dark-matter) spatial densities for each given direction and distance. The spatial distribution of the Galactic and MC dark-matter halos ($\rho_{\rm{h}, \rm{DM}}$ and $\rho_{\rm{h}, \rm{DM}}^{\rm MC}$) are reported in Appendix \ref{app1}. We assume $f=5\%$ of the dark-matter Galactic halo and MCs' halos are made of massive compact objects. 

\noindent The numbers of detected and isolated SO or DM IBHs (compact and massive objects in the form of MACHOs, primordial BHs, etc.) are low and there is no known mass function for them. Although, the mass function for sub-halos in dark-matter has been investigated in several references and proposed as a power-law function with the power-index of $\in[-2, -1]$ \citep{subhalo2,subhalo1,Subhalomass2}, here we simulate microlensing events due to compact objects in halo dark-matter. Hence, to determine the lens mass for both SO and DM IBHs we consider four mass functions as given by:
\begin{eqnarray}
\rm{MF}_{1}&:&~~~~~\frac{dN}{dM_{\rm l}}= \rm{const},\nonumber\\
\rm{MF}_{2}&:&~~~~~\frac{dN}{dM_{\rm l}}\propto M_{\rm l}^{-0.5},\nonumber\\
\rm{MF}_{3}&:&~~~~~\frac{dN}{dM_{\rm l}}\propto M_{\rm l}^{-1},\nonumber\\
\rm{MF}_{4}&:&~~~~~\frac{dN}{dM_{\rm l}}\propto M_{\rm l}^{-2}.
\label{mfs}
\end{eqnarray}
During $10$-year Rubin mission long-duration microlensing events due to such massive IBHs (with $M_{\rm l}\sim 1000M_{\odot}$) are likely realizable, so we include such massive BHs in simulations as well. The lens impact parameter is chosen uniformly from the range $u_{0}\in[0,~u_{0,\rm{m}}]$, and we choose $u_{0,\rm{m}}=3$. Although, the efficiency for detecting microlensing events with large impact parameters is small, but it is not zero.   

We determine the lens and source (global and dispersion) velocities in our galaxy and MCs based on their known distributions which are reported in Appendix \ref{app2}. We calculate the parallax effect numerically and by integrating the projected velocity vector of the Earth over time (given by Equation \ref{parallax}). For simulating microlensing events we ignore finite-source sizes \citep{1994wittmoa}, because of large angular Einstein radii for these long-duration events.

\begin{figure}
    \centering
    \includegraphics[width=0.49\textwidth]{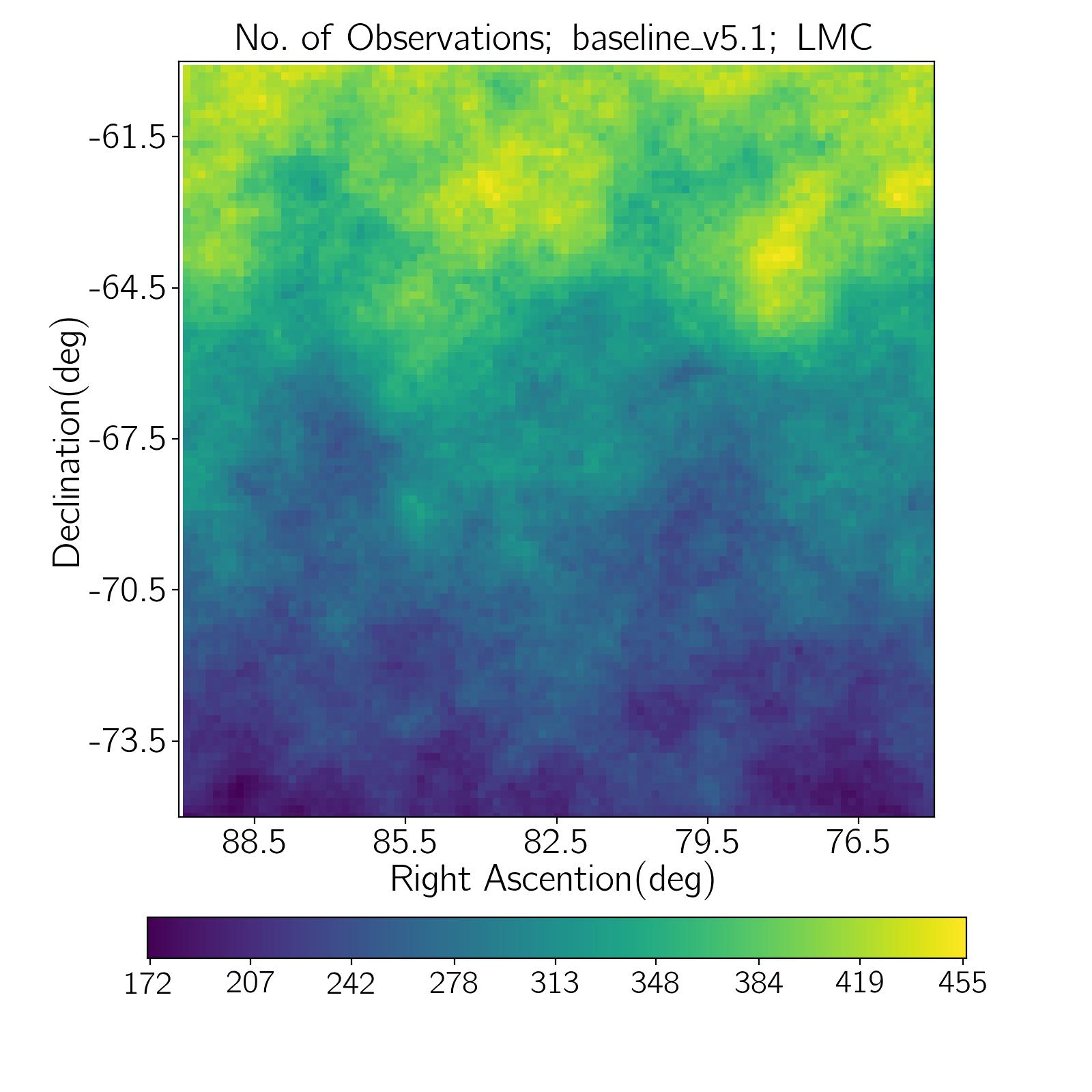}
    \includegraphics[width=0.49\textwidth]{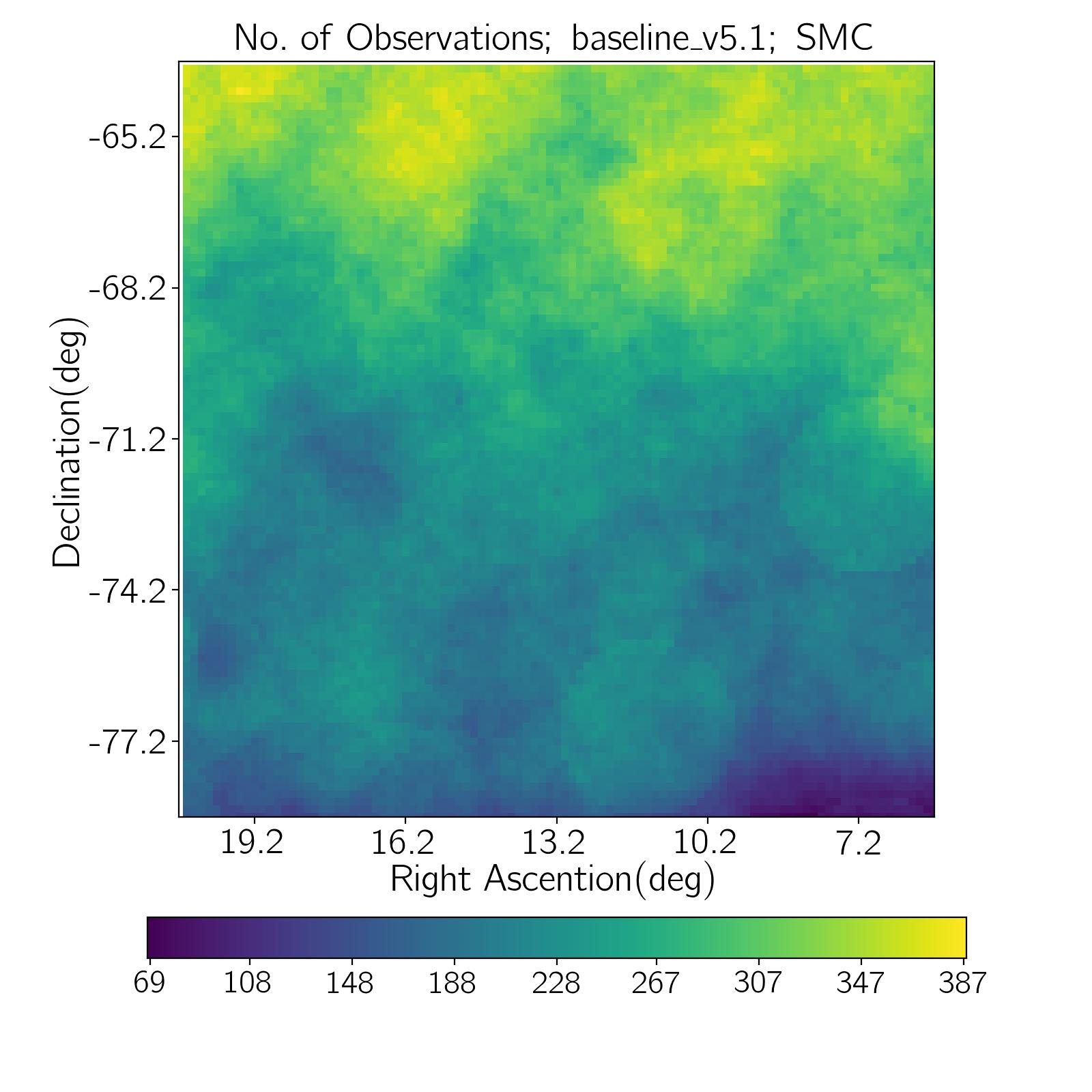}
    \caption{Maps of number of observations $N_{\rm{data}}$ will be recorded by Rubin during its 10-year mission towards LMC (top panel) and SMC (bottom panel) according to the \texttt{baseline}$\_$\texttt{v5.1} strategy developed in the OpSim simulator.}\label{basev5}
\end{figure}

After generating microlensing light curves and astrometric source trajectories, we produce synthetic Rubin data for them. In this regard, we make sequences of observing cadences, applied filters, and the $5\sigma$ depth (which specifies the Rubin photometric accuracies) using the Rubin Operational Simulations (OpSim) and its \texttt{baseline}$\_$\texttt{v5.1} version for any observing directions towards LMC and SMC. In these simulations, the number of data points (the visit number) in any given direction is different. In Figure \ref{basev5}, we show two maps of number of the Rubin observations ($N_{\rm{data}}$) towards LMC and SMC which is offered by \texttt{baseline}$\_$\texttt{v5.1} strategy. Hence, Rubin is planned to observe LMC and SMC during its $10$-year mission around $\sim300$-$200$ times. The photometric accuracies in the Rubin observations ($\sigma_{\rm p}$) depend on source apparent magnitudes in different filters and the $5\sigma$ depth as given by Equation (6) of \citet{lsstbook}. In simulations, we determine the Rubin astromtric accuracy ($\sigma_{\rm a}$) as a function of stellar $r$-band apparent magnitudes based on Table (3) of \citet{lsstbook}.

We then extract detectable source stars. For detectability of stars we apply three criteria as following. The first criterion (i) is that the source star should be detectable at the magnification peak at least in two Rubin filters, i.e., the stellar apparent magnitude in the $k$th observing filter at the lensing peak which is given by
\begin{equation} m_{\star,~k}=m_{\rm{base},~k}-2.5\log_{10}\big[f_{\rm{b},~k}A_{\rm{m}}+1-f_{\rm{b},~k}\big]
\label{appm}
\end{equation}
should be less than the Rubin detection threshold in that filter. These thresholds are the Rubin single visit depths, i.e., $23.4$, $24.6$, $24.3$, $23.6$, $22.9$, $21.7$ mag related to the Rubin filters $ugrizy$ \citep[reported in Table (1) of ][]{lsstbook}. Here, $A_{\rm{m}}=A(u_{0})$ is the maximum magnification factor, and $f_{\rm{b},~k}=F_{\star,~k}/F_{\rm{base},~k}$ is the blending parameter in the $k$th observing filter which is the ratio of the source flux to the baseline flux at the source position, and $m_{\rm{base},~k}=-2.5\log_{10}[F_{\rm{base}, k}]$ is the apparent magnitude at the baseline.

\noindent As the second criterion (ii) we consider the blending parameter in $r$-band (the Rubin middle filter, i.e., $f_{\rm b, r}$) as a weight function while choosing detectable stars. Because, in the real Rubin observations all blending stars with separations smaller than the size of a typical PSF are assumed as one star, so the probability of realizing each of them is $1/N_{\rm{b}}$, where $N_{\rm{b}}$ is the number of blending stars. In the third criterion (iii) we label simulated microlensing events which have low number of synthetic data points (less than three) as the ones whose source stars are not detectable. Although, the number of such short-duration microlensing events or the ones mostly take place during the Rubin season gaps (which have less than three data points) in simulations is low, this number increases with the power index of the IBHs mass function (i.e. $\beta$ in $dN\big/dM_{\rm{l}}\propto M_{\rm{l}}^{-\beta}$). Because, by increasing the power index $\beta$ the simulated events have on average shorter durations. In our simulations, we evaluate the Rubin efficiency for detecting stars (which pass through these mentioned criteria) for each given observing direction as $\epsilon_{\rm{D}}$ (see the forth column of Table \ref{tab2}).

\begin{figure*}
    \centering
    \includegraphics[width=0.48\textwidth, height=0.25\textheight]{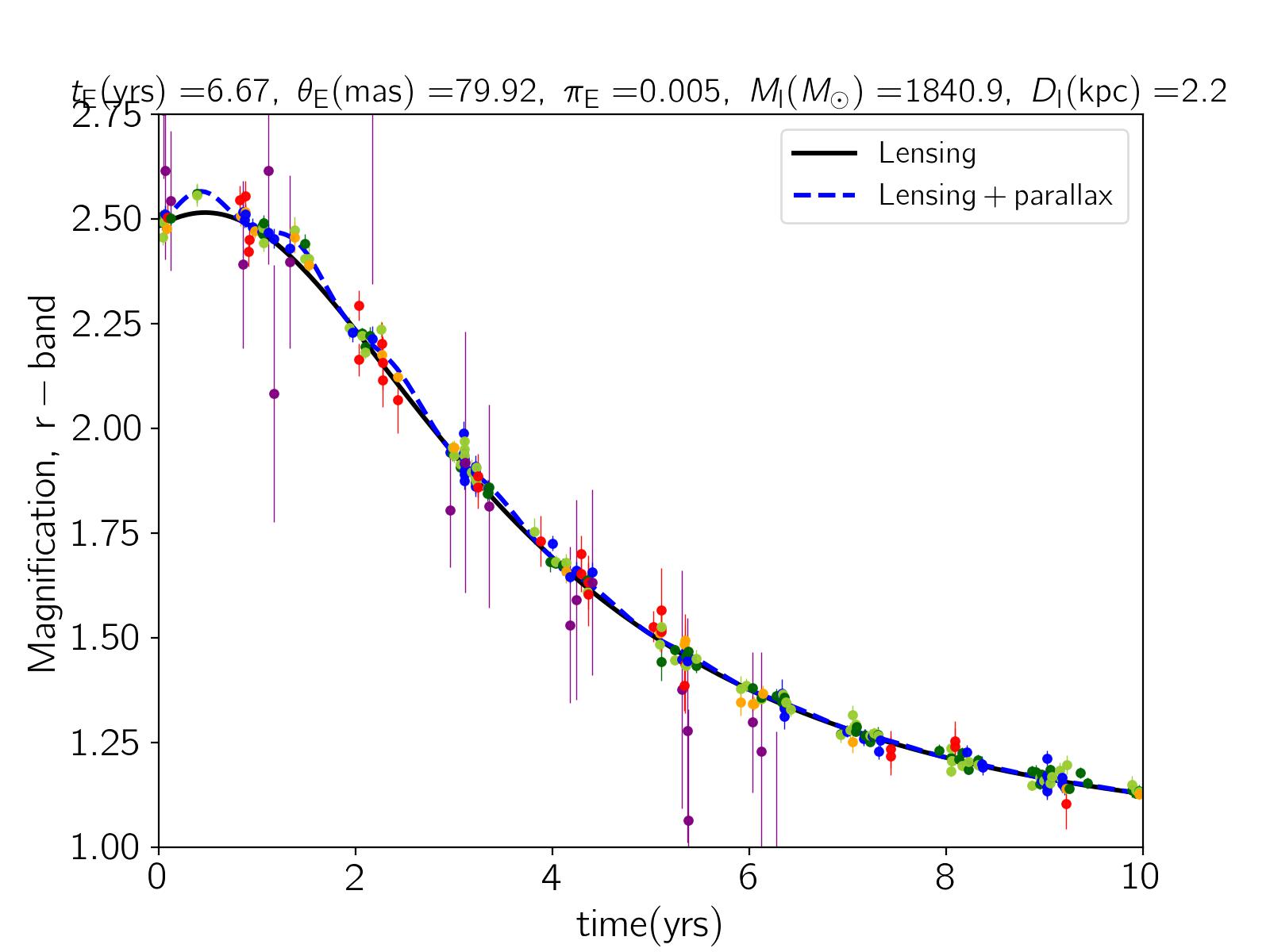}
    \includegraphics[width=0.48\textwidth, height=0.25\textheight]{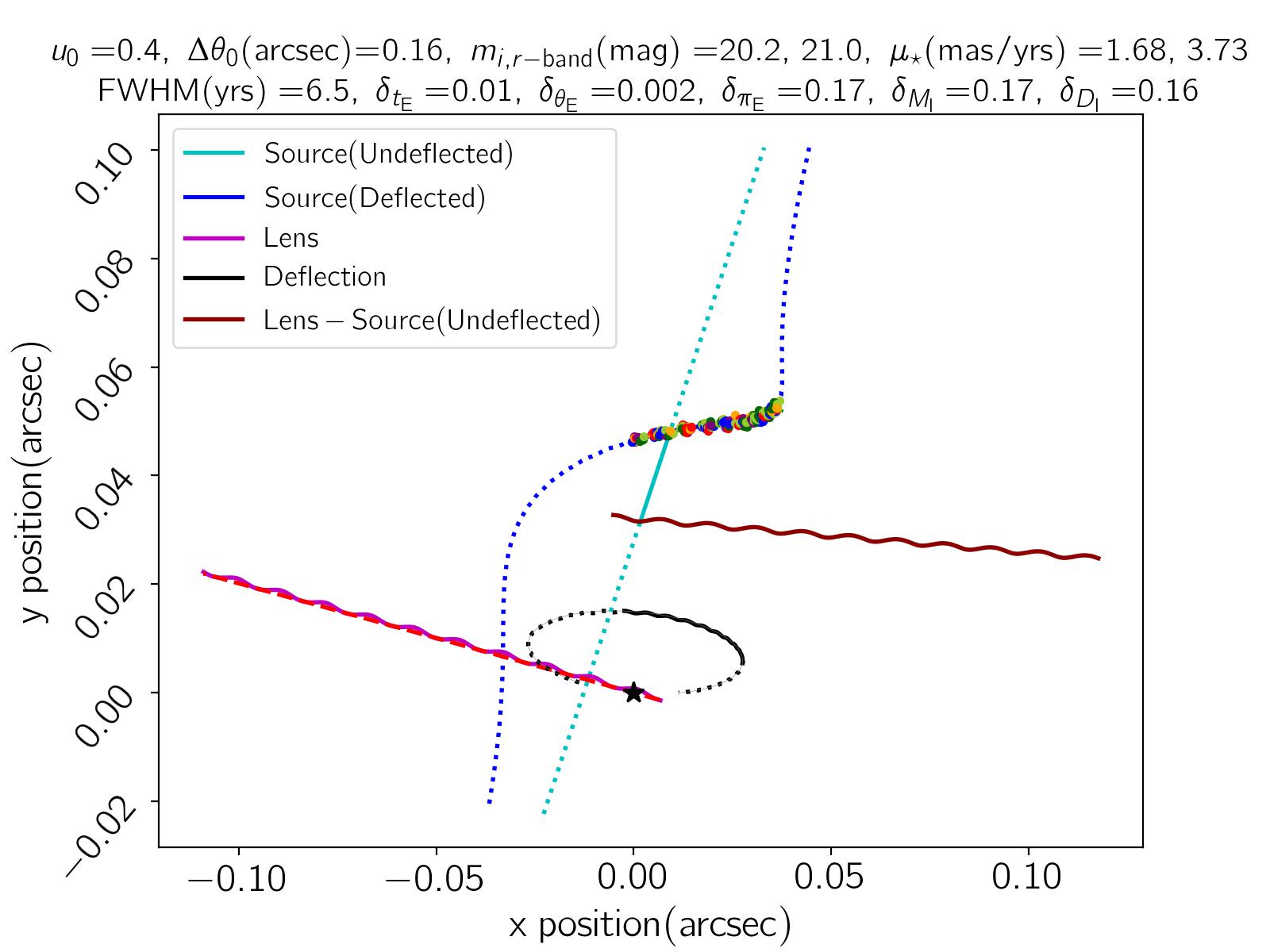}
    \includegraphics[width=0.48\textwidth, height=0.25\textheight]{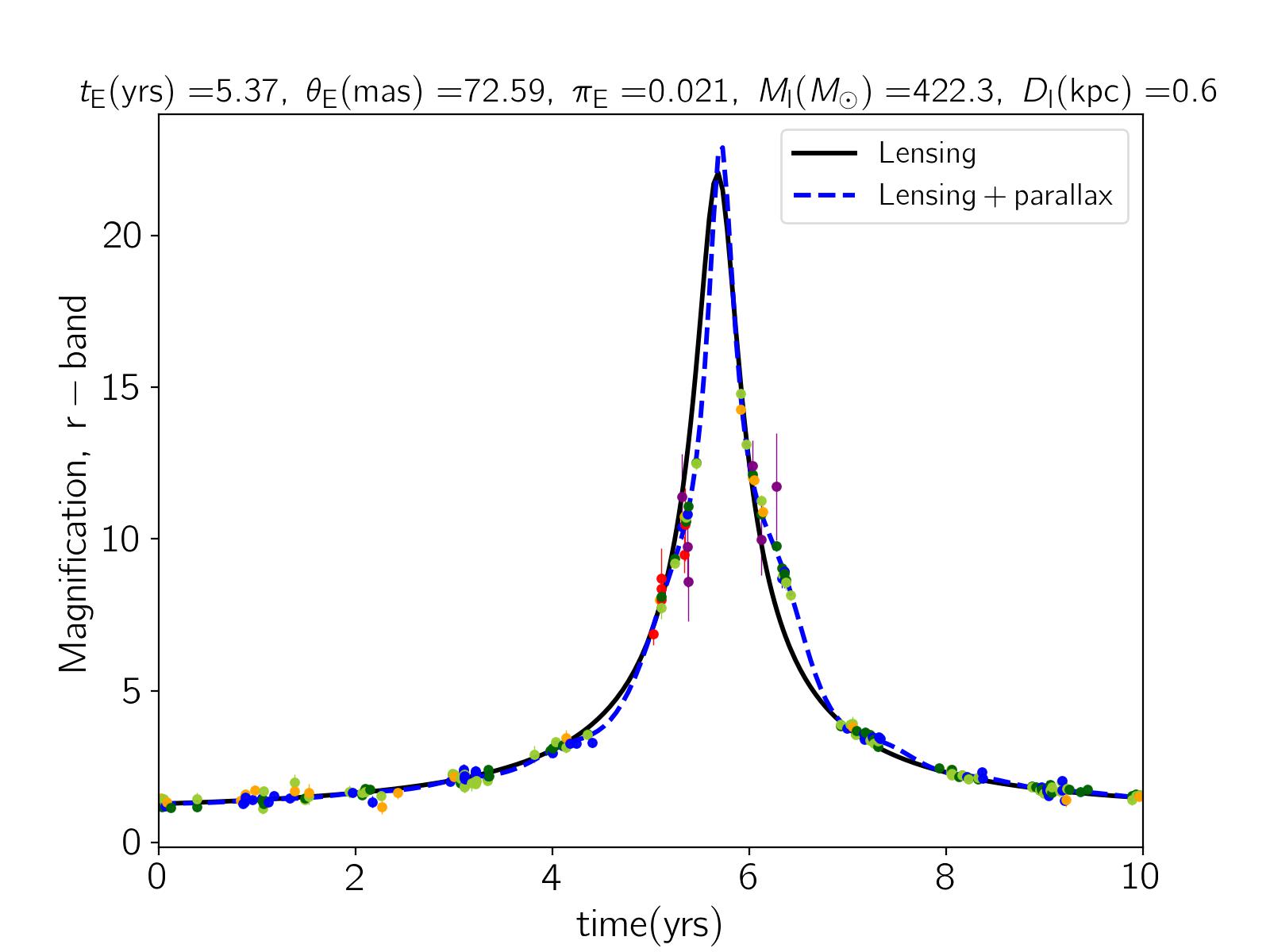}
    \includegraphics[width=0.48\textwidth, height=0.25\textheight]{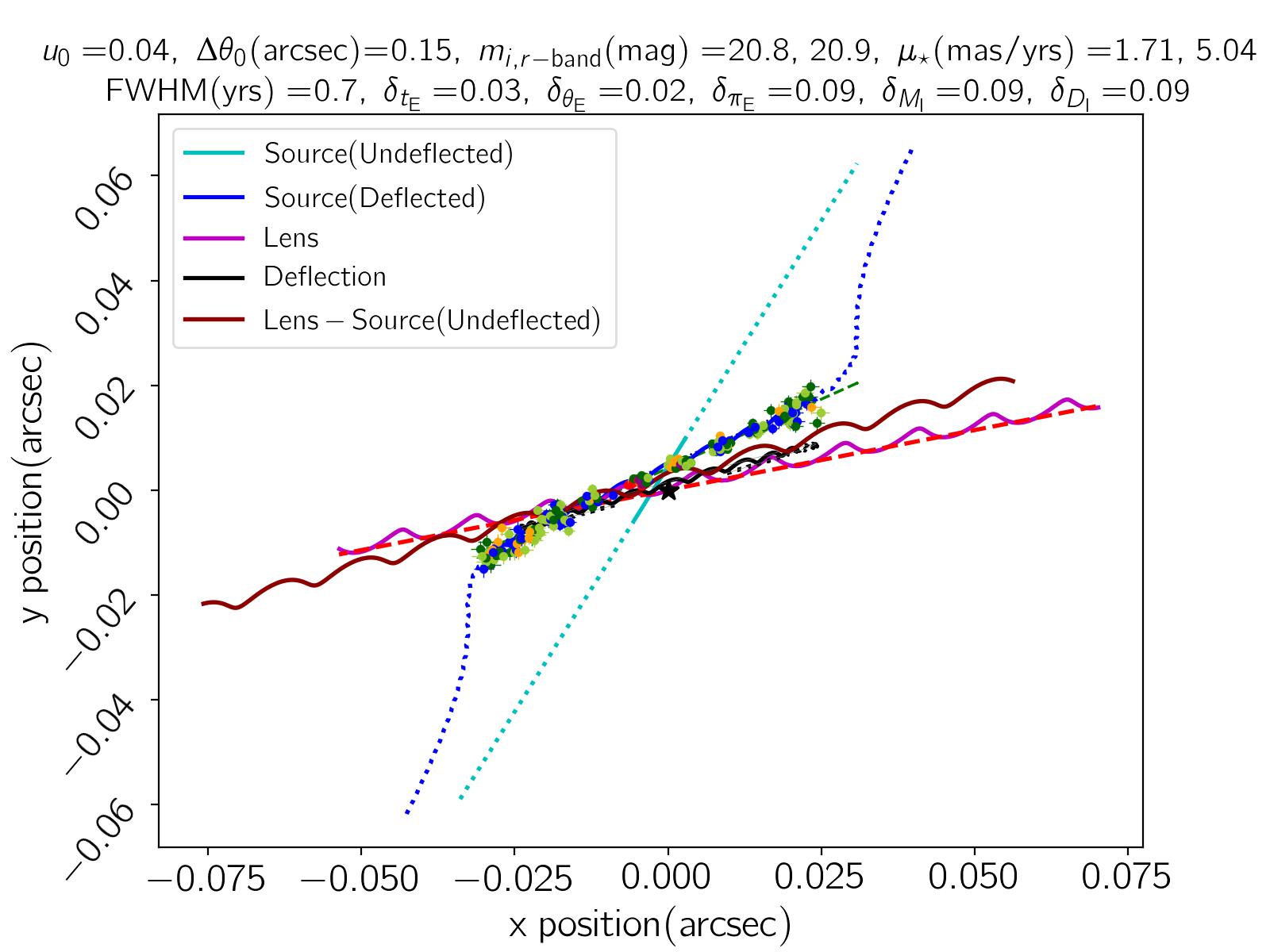}
    \includegraphics[width=0.48\textwidth, height=0.25\textheight]{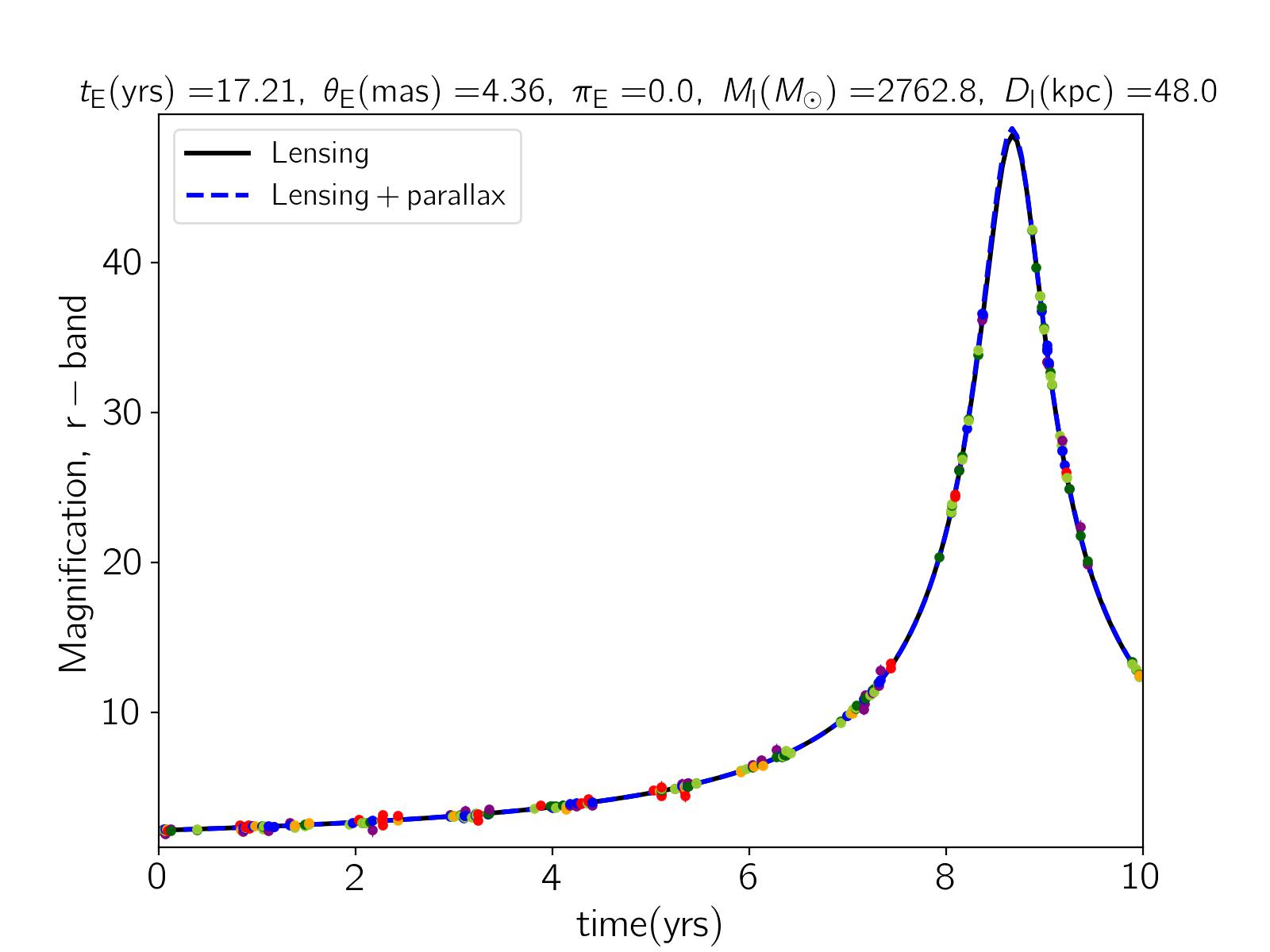}
    \includegraphics[width=0.48\textwidth, height=0.25\textheight]{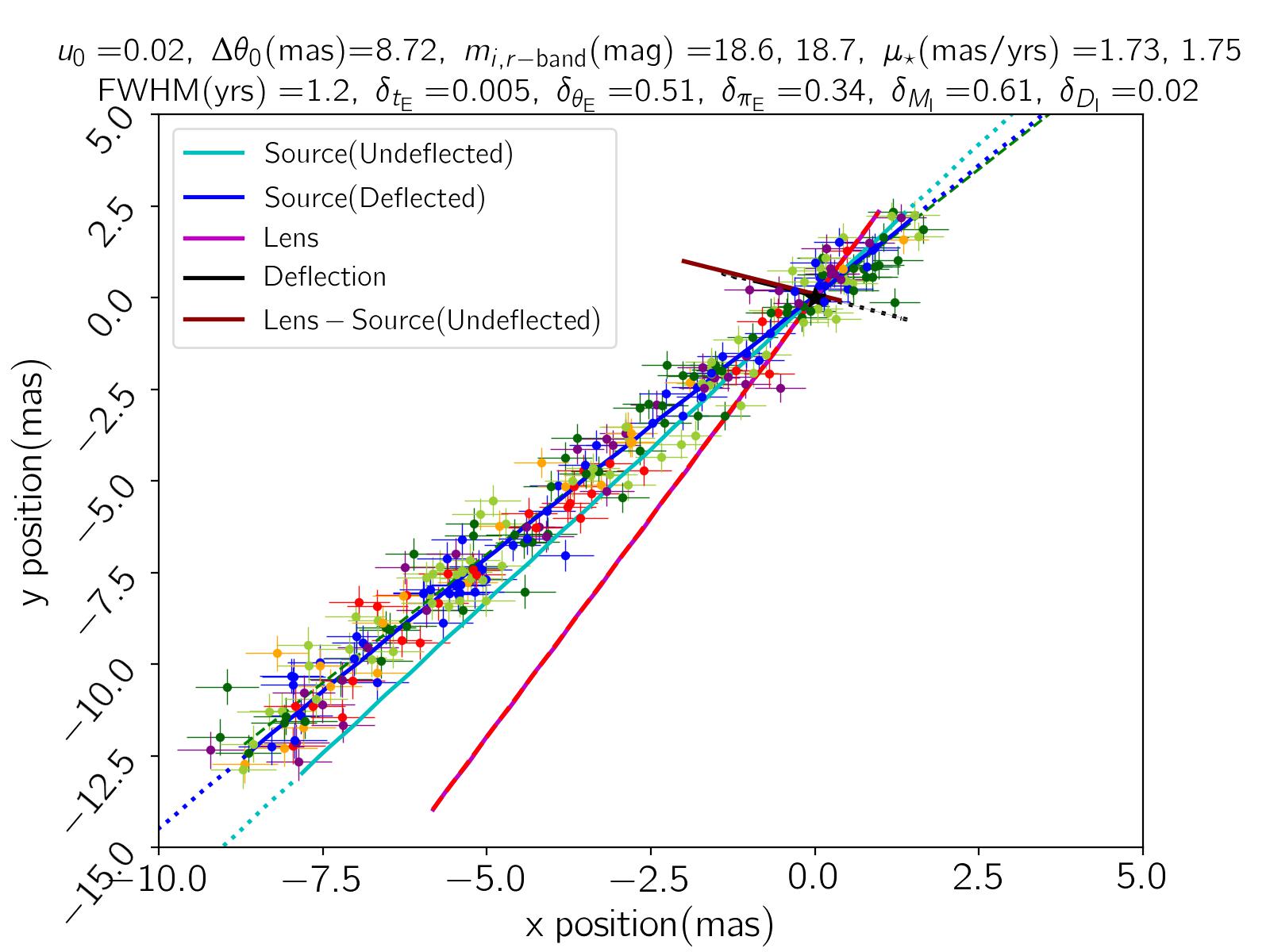}
    \caption{Three examples of microlensing light curves due to IBHs towards LMC without (solid black curves) and with (dashed blue curves) parallax effects are shown in left panels. Their corresponding astrometric trajectories are represented in their right-side panels. In right panels, the lens trajectory with (solid magenta curves) and without (dashed red curves) parallax effects, the deflected (blue) and undeflected (cyan curves) source trajectories, the relative lens-source trajectory (solid dark-red curves), and the astrometric deflection (black curves) are represented. Source trajectories are extended from the $10$-year Rubin observing window as specified by the dotted style. Purple, blue, dark-green, light-green, orange, and red data points are taken by the Rubin telescope in the $ugrizy$ filters, respectively.}\label{fig1}
\end{figure*}

In the next step, for detectable source stars we evaluate detectability of their lensing signals by applying three following criteria. They are the photometric detection criteria and the corresponding astrometric detection criteria will be specified in Subsection \ref{astrop}.
\begin{itemize}[leftmargin=2.0mm]
\item(I)$\Delta\chi^{2}\ge2N_{\rm{data}}$, where $N_{\rm{data}}$ is the number of data points taken by Rubin for each event in a given direction.
\item(II)At least three data points should be above the baseline by at least $3\sigma_{\rm p}$.
\item(III)$\rm{FWHM}\leq T_{\rm{obs}}$. 
\end{itemize} 
In the first criterion, $\Delta\chi^{2}$ is the difference between $\chi^{2}$ values from fitting real and baseline (without lensing effects) models. According to Figure \ref{basev5}, based on the \texttt{baseline}$\_$\texttt{v5.1} strategy and towards LMC and SMC the numbers of data points that Rubin will take during its $10$-year mission are $N_{\rm{data}}\sim170-450$, and $70-390$, respectively. The threshold twice the number of data points (which is almost equal to the number of degree of freedom) is high enough to realize lensing effects.

\noindent In the second criterion (II) we demand that at least three data points should be above the baseline by at least $3\sigma_{\rm{p}}$ which guarantees that the enhancement in the stellar flux to be realizable. Since, most of these microlensing events due to IBHs have long durations we apply the third criterion (III) which is the Full Width of Half Maximum (FWHM) of light curves should be shorter than the Rubin observing window, i.e., $T_{\rm{obs}}=10$ year. We determine FWHM values of light curves using Equation (12) in \citet{2022sajadianAA}. We show the Rubin efficiency for discerning lensing effects of detectable stars by $\epsilon_{\rm{L}}$, which depends on some lensing parameters (e.g., see Figure \ref{plotef}). 

We show three examples of simulated light curves without (solid black curves) and with (dashed blue curves) parallax effects (left panels) and their corresponding astrometric trajectories (right-hand panels) in Figure \ref{fig1}. In each right panel, the lens trajectory with (solid magenta curves) and without (dashed red curves) parallax effects, the deflected (blue) and undeflected (cyan curves) source trajectories, the relative lens-source trajectory (solid dark-red curves), and the lensing-induced astrometric deflection (black curves) are represented. Source trajectories and astrometric deflections are extended from the $10$-year Rubin observing window as specified by dotted style. The synthetic data points are taken in the Rubin $ugrizy$ filters as specified by purple, blue, dark-green, light-green, orange, and red colors, respectively.

\noindent In Figure \ref{fig1}, the physical parameters are mentioned at the top of plots. For the angular source velocity two values are given which are its velocity at the baseline ($\mu_{\star}$) and its velocity by considering lensing-induced deflection, respectively. Also, $m_{i, r-\rm{band}}$ values are the apparent magnitudes of two lensing-induced images at the magnification peak in the $r$-band. $\delta_{t_{\rm{E}}}$, $\delta_{\theta_{\rm{E}}}$, $\delta_{\pi_{\rm{E}}}$, $\delta_{M_{\rm{l}}}$, and $\delta_{D_{\rm{l}}}$ are the relative errors in lensing and physical parameters based on the Fisher and Covariance matrix calculations which will be explained in Subsection \ref{error}. All of three microlensing events are realizable as they pass three mentioned criteria (i.e., I, II, and III).


Two first light curves in Figure \ref{fig1} are halo-lensing events while the last one is a self-lensing event. Accordingly, for halo-lensing events  $\theta_{\rm{E}}$ and $\pi_{\rm{E}}$ values are higher than those for self-lensing events by one or two orders of magnitude on average. Hence, measuring parallax and astrometric deflection in the source trajectory or images' angular separation for self-lensing events is a challenge. Follow-up observations even with a space-based telescope will not lead to measure parallax amplitudes for self-lensing events because the observers' distance from the Sun (which is $\sim\rm{AU}$) is considerably less than the LMC and SMC distances ($\sim 50,~60$ kpc). But follow-up observations, e.g., with the Extremely Large Telescope, will lead to measuring accurately $\theta_{\rm{E}}$ because of its great astrometric accuracy. 

After simulating many detectable microlensing events owing to IBHs towards MCs, in the next section, we study their lensing, physical, and statistical properties.

\section{Results}\label{sec4}
We divide the map of each Magellanic Cloud (with the angular area $15\times15~\rm{deg}^{2}$) into 10,000 grids, each with the area $0.023~\rm{deg}^{2}$. For each grid we make at least 150 detectable microlensing events. For detectable events we calculate Fisher and Covariance matrices, and numerically estimate the errors in physical parameters of their microlenses as will be extracted from data points. 

\noindent We first explain the lensing and statistical properties of photometrically detectable microlensing events in Subsection \ref{prop}. By performing simulations for four suggested IBHs mass functions, we study the effect of IBHs' MF on the results in Subsection \ref{MFs}. The Rubin efficiency for detecting long-duration microlensing events due to IBHs and its dependence are investigated in Subsection \ref{effi}. In Subsection \ref{astrop}, we evaluate the probabilities of resolving images, lensing-induced astrometric deflections, and parallax effects in photometrically detectable microlensing events. In Subsection \ref{error}, we study the relative errors in physical and lensing parameters of lens objects and their dependences on lensing parameters.

\subsection{Properties of Detectable Microlensing Events}\label{prop}
\begin{figure*}
\centering
\includegraphics[width=0.32\textwidth]{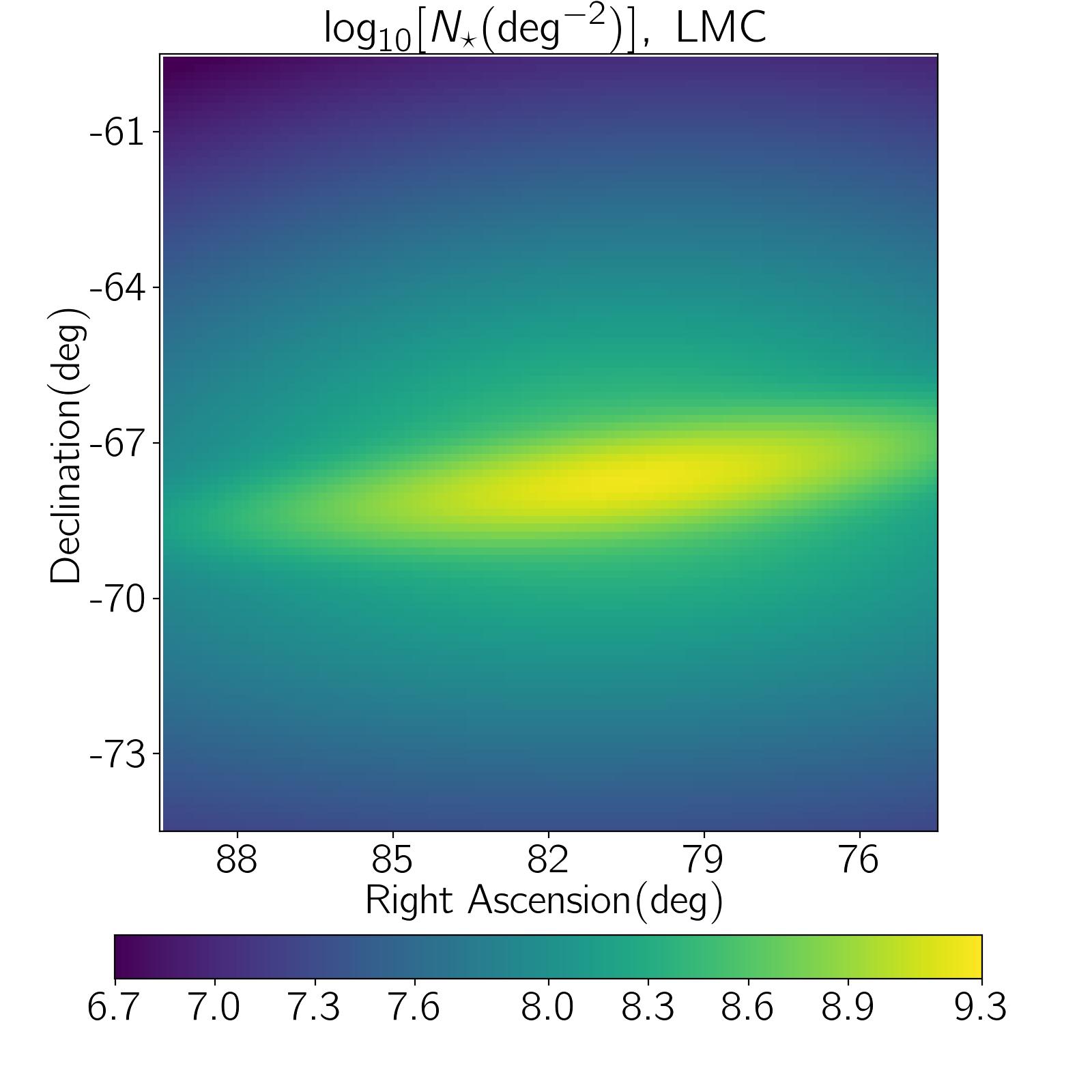}
\includegraphics[width=0.32\textwidth]{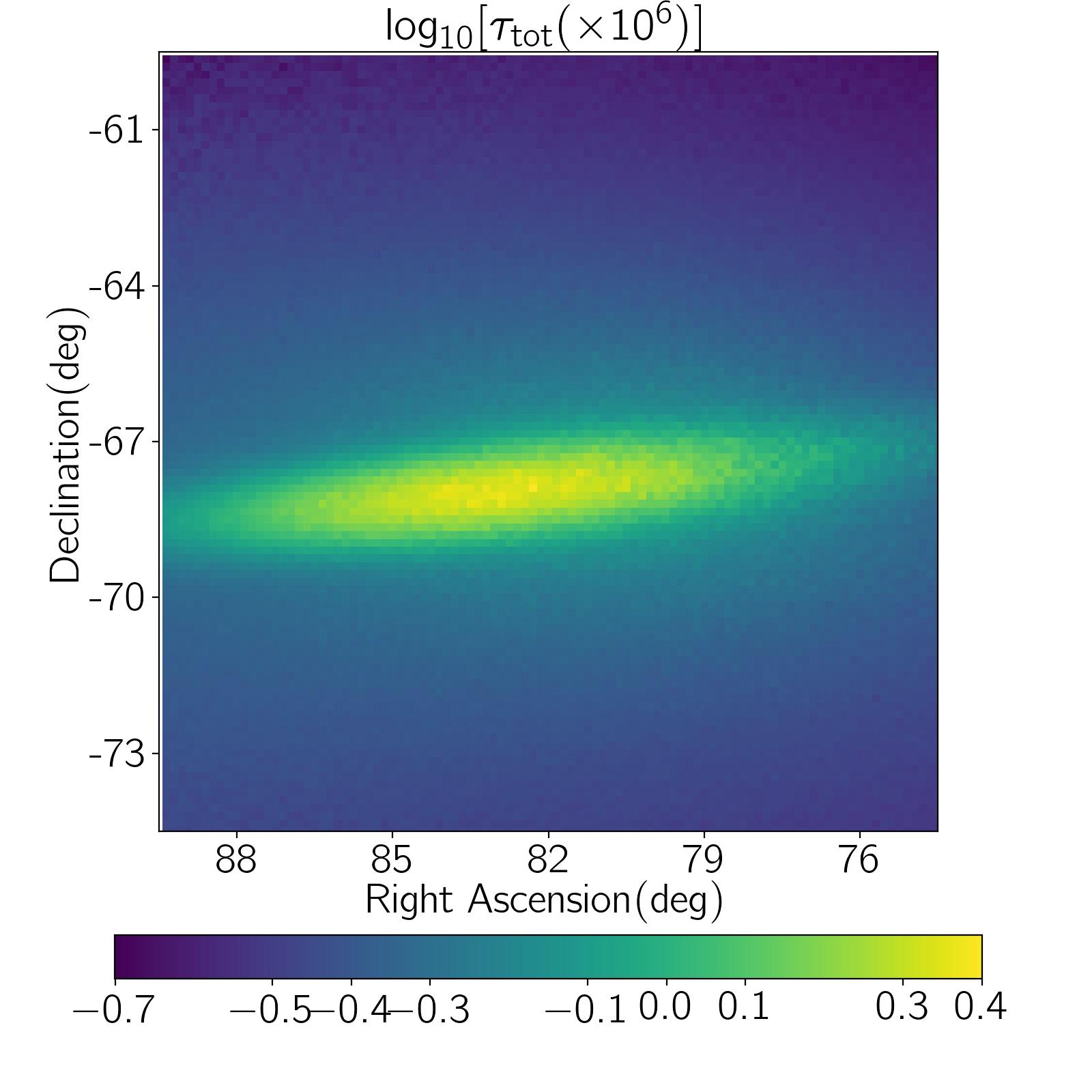}
\includegraphics[width=0.32\textwidth]{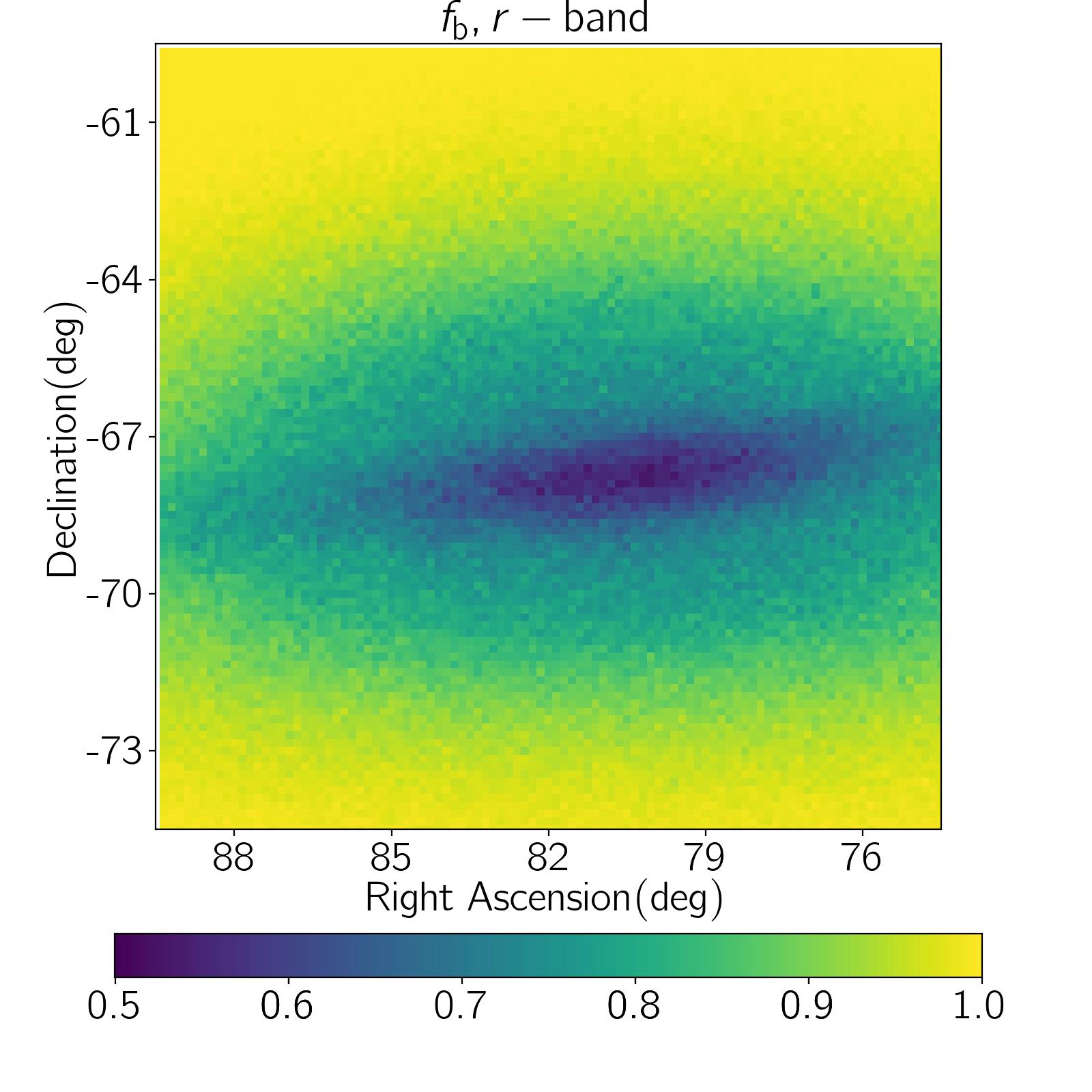}
\includegraphics[width=0.32\textwidth]{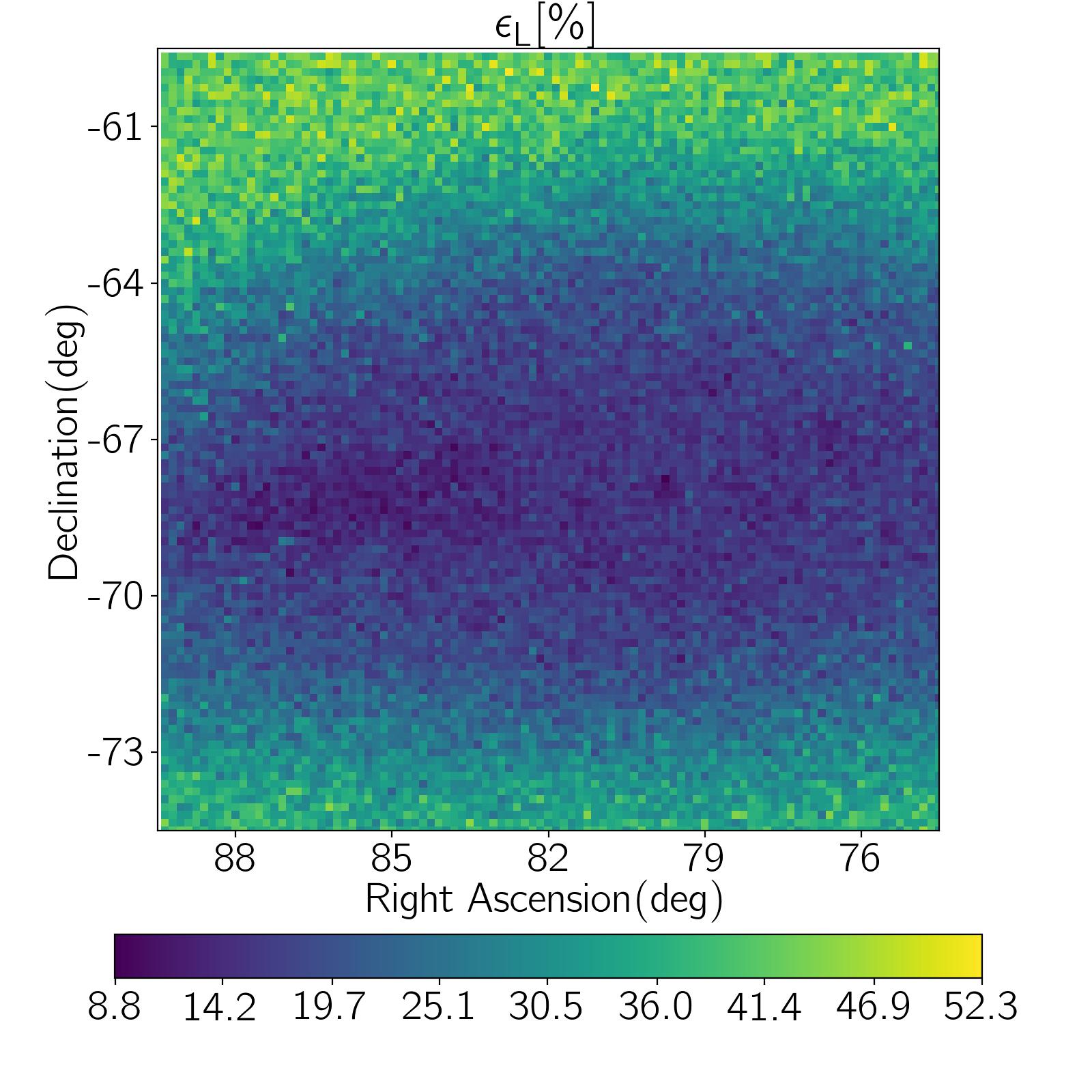}
\includegraphics[width=0.32\textwidth]{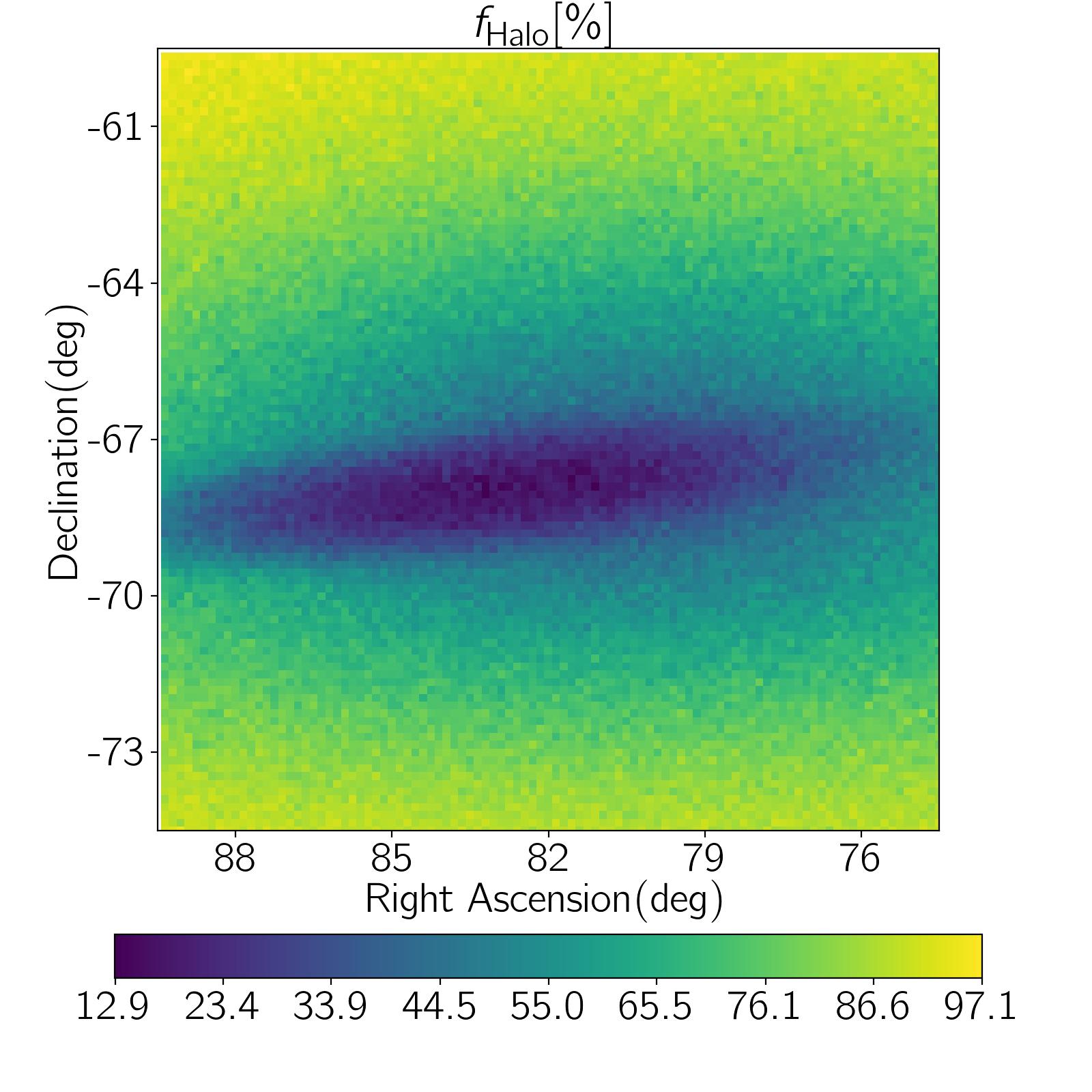}
\includegraphics[width=0.32\textwidth]{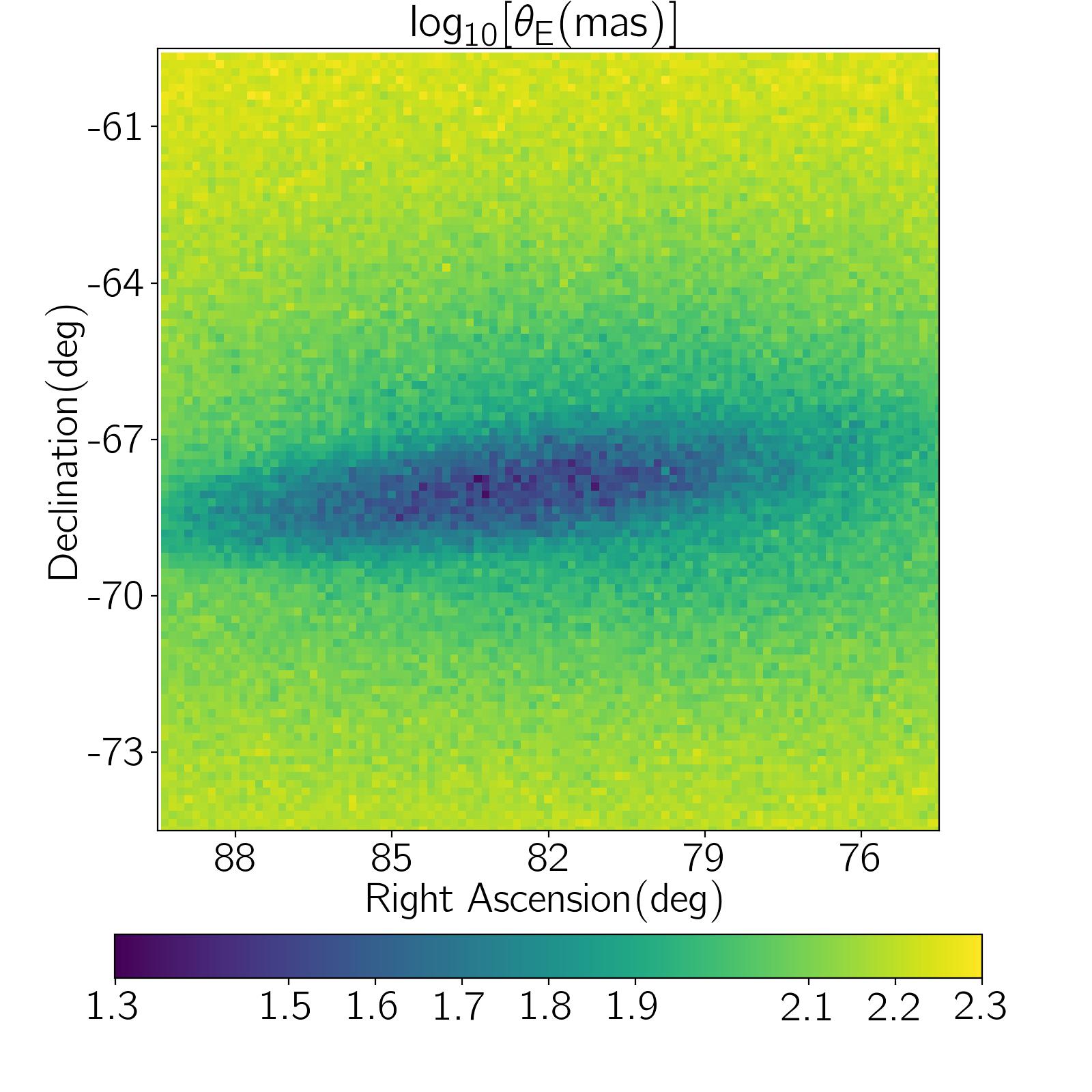}
\includegraphics[width=0.32\textwidth]{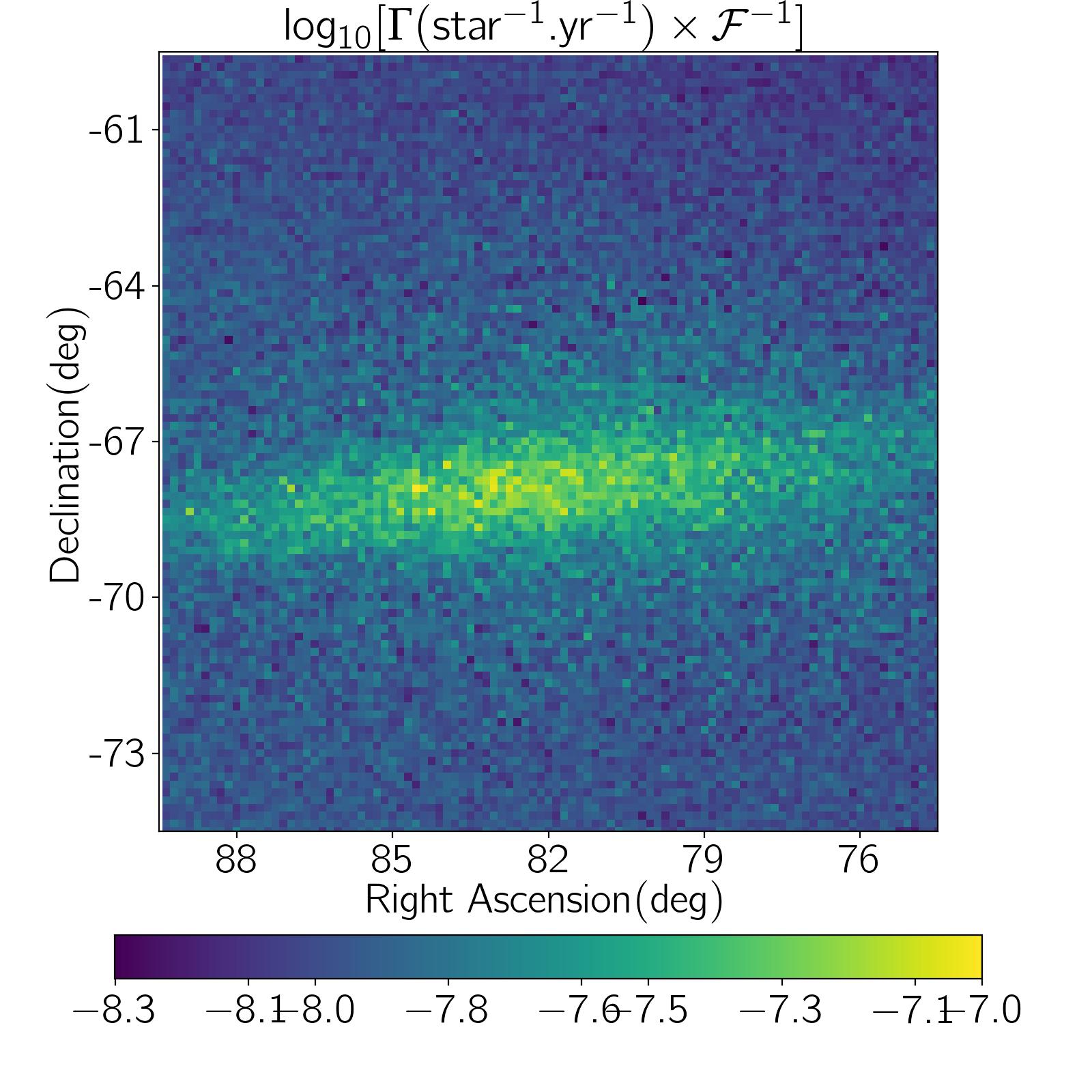}
\includegraphics[width=0.32\textwidth]{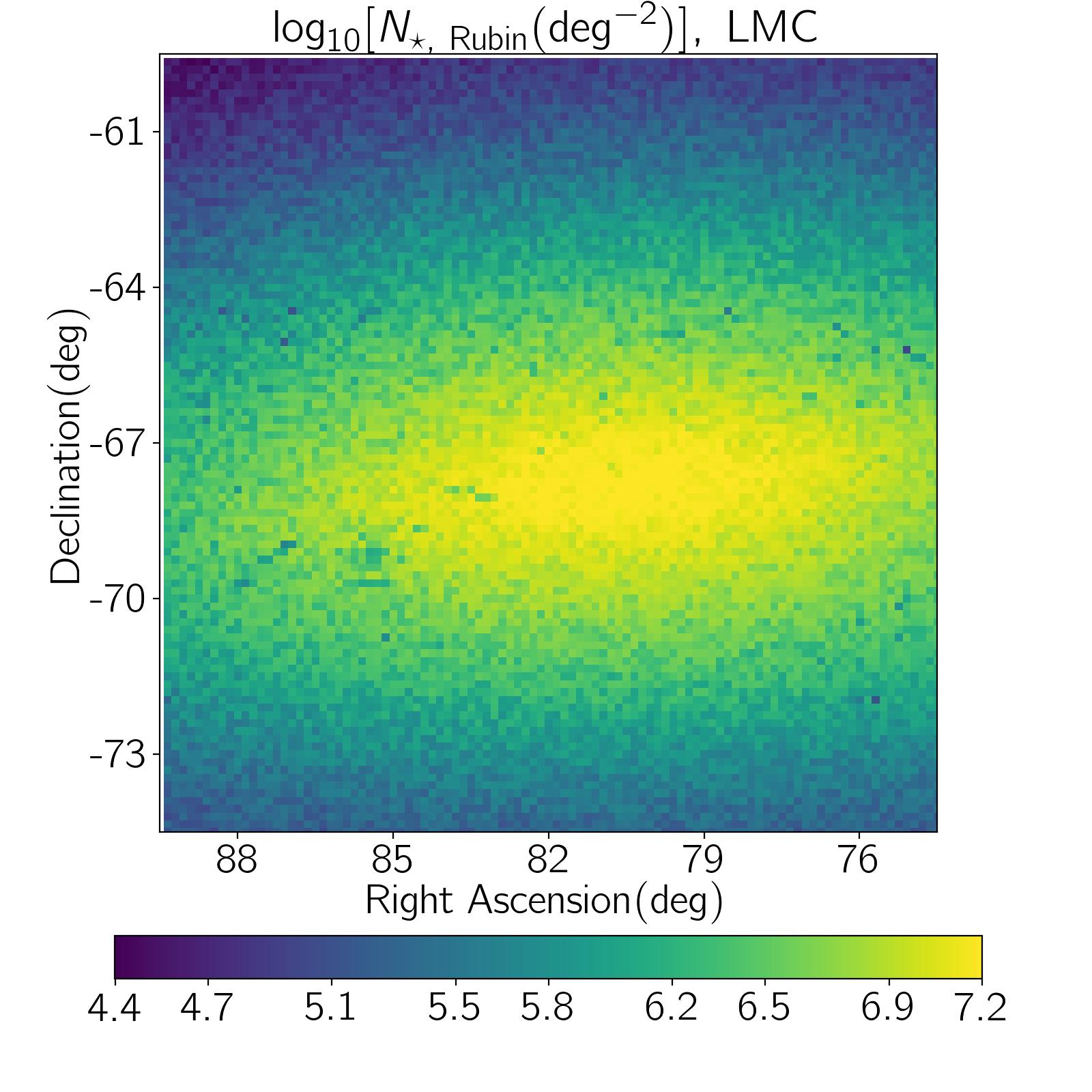}
\includegraphics[width=0.32\textwidth]{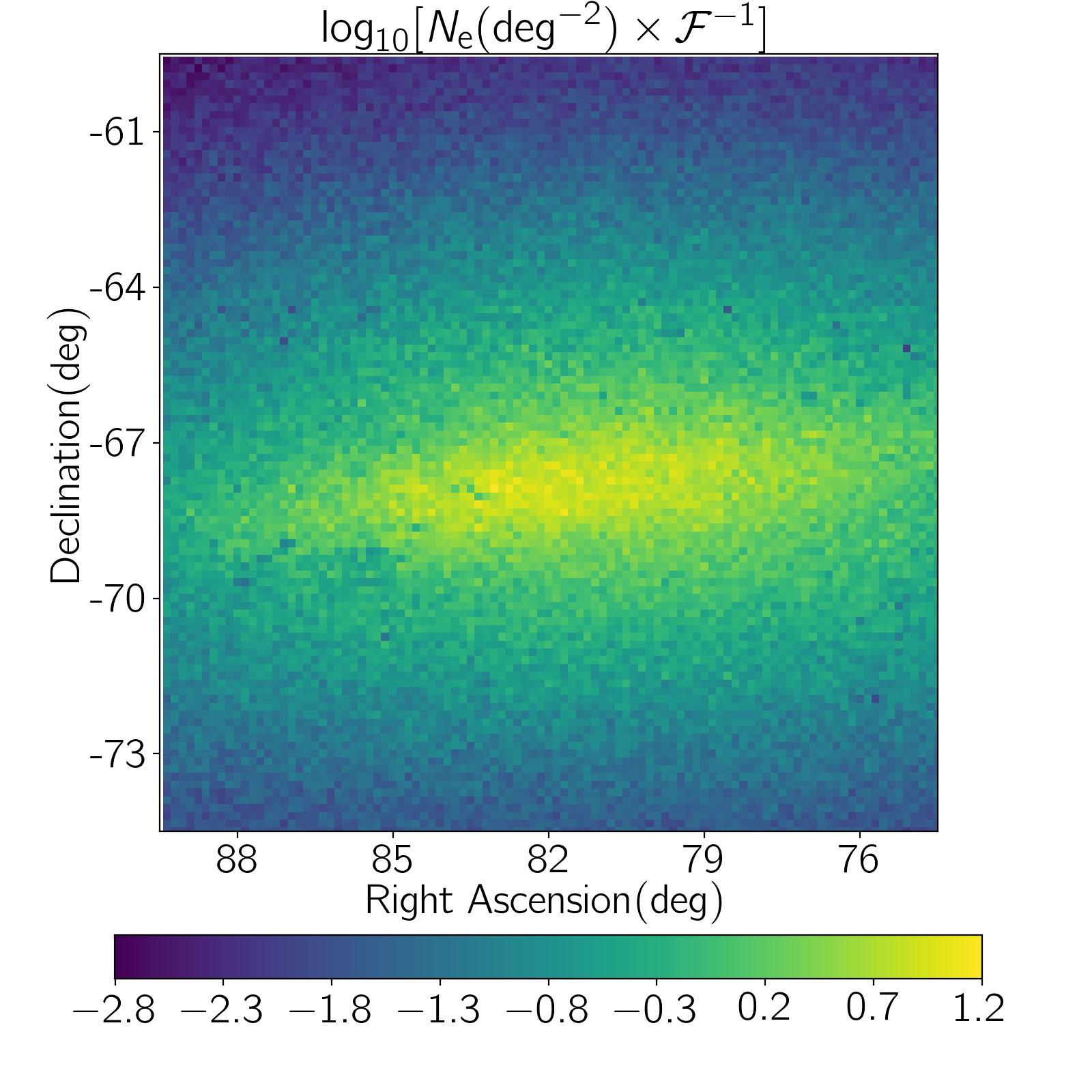}
\caption{Maps of nine statistical and lensing parameters (which are mentioned at the top of maps) by averaging over detectable microlensing events due to IBHs over the 2D space representing the sky plane towards LMC.}\label{map1}
\end{figure*}

\begin{figure*}
    \centering
    \includegraphics[width=0.32\textwidth]{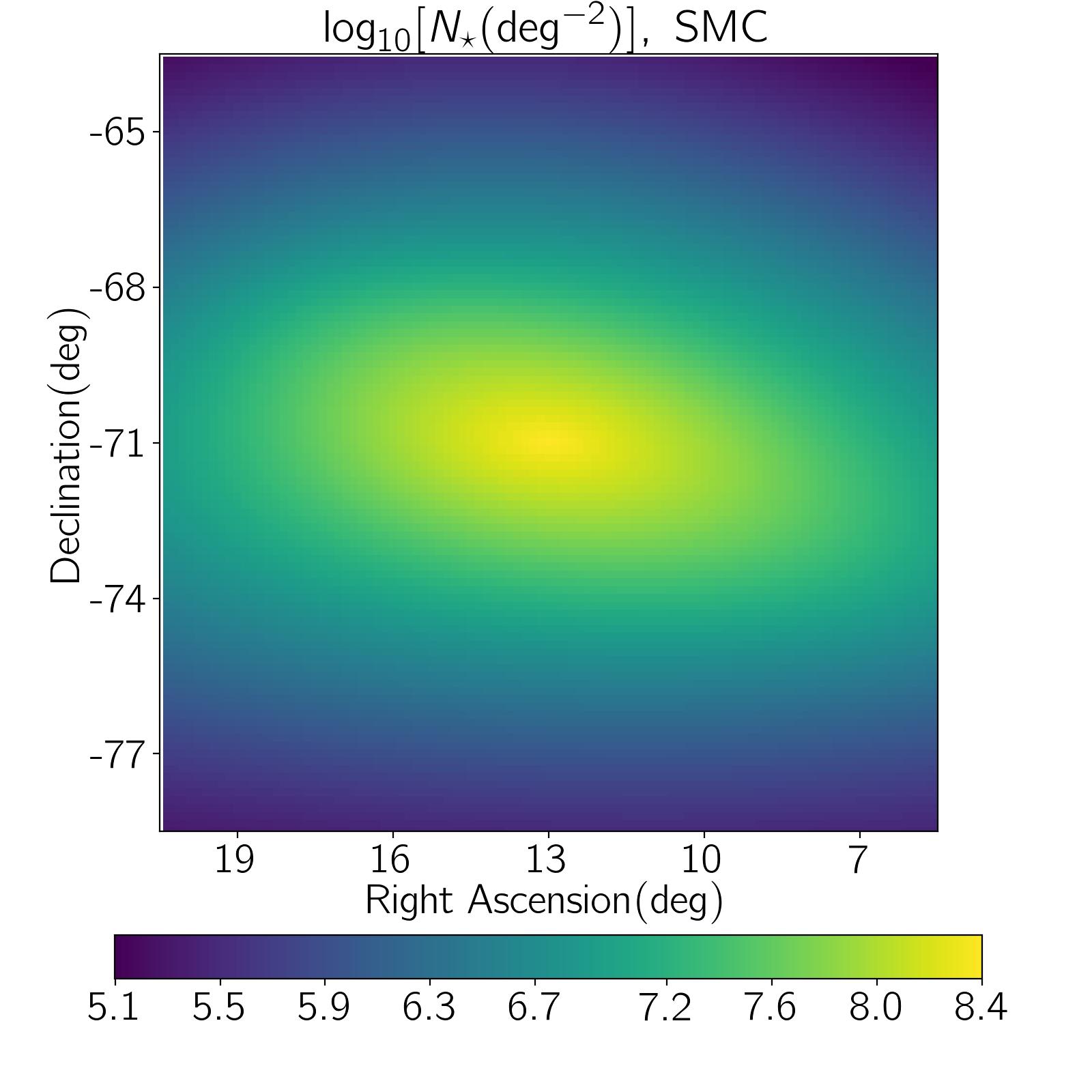}
    \includegraphics[width=0.32\textwidth]{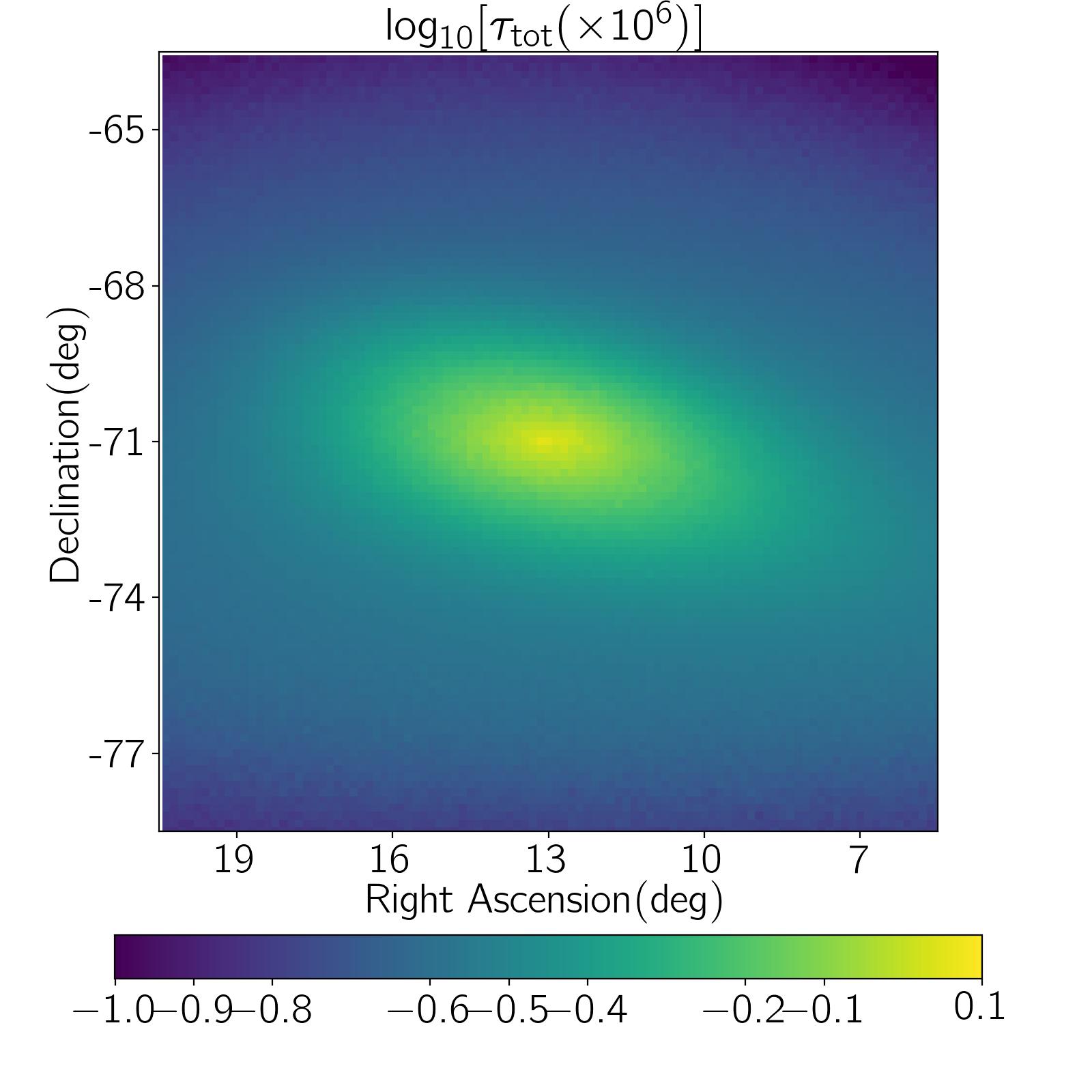}
    \includegraphics[width=0.32\textwidth]{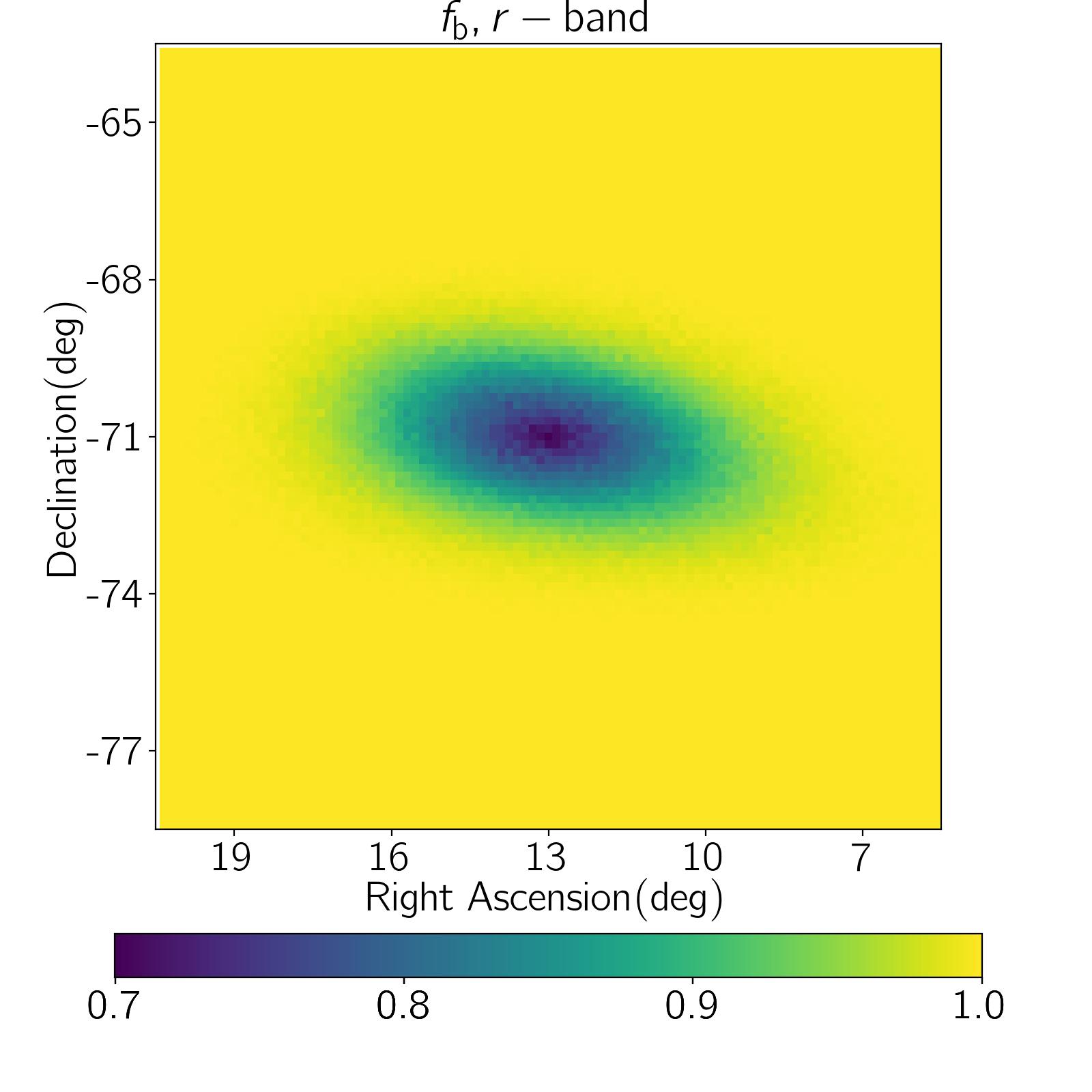}
    \includegraphics[width=0.32\textwidth]{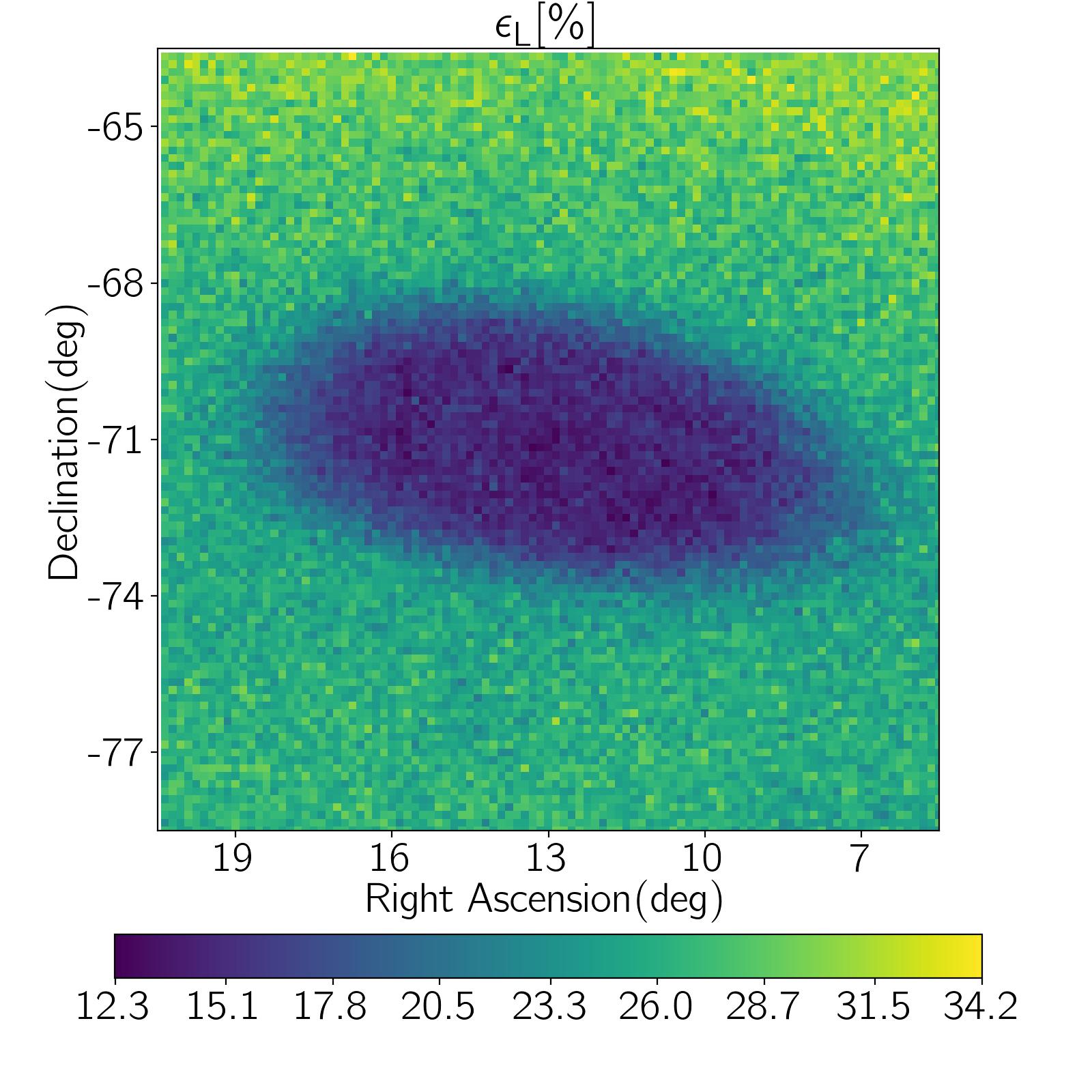}
    \includegraphics[width=0.32\textwidth]{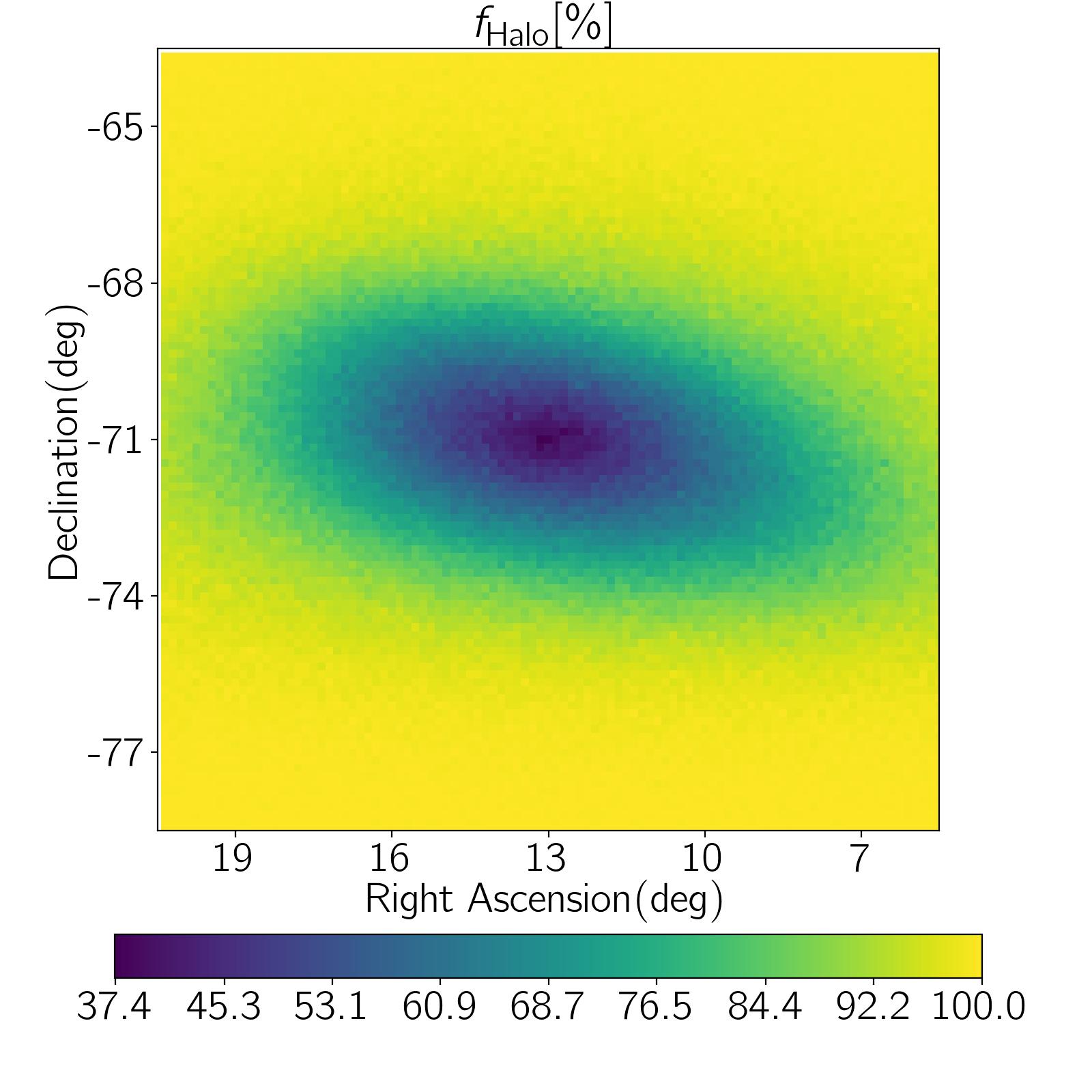}
    \includegraphics[width=0.32\textwidth]{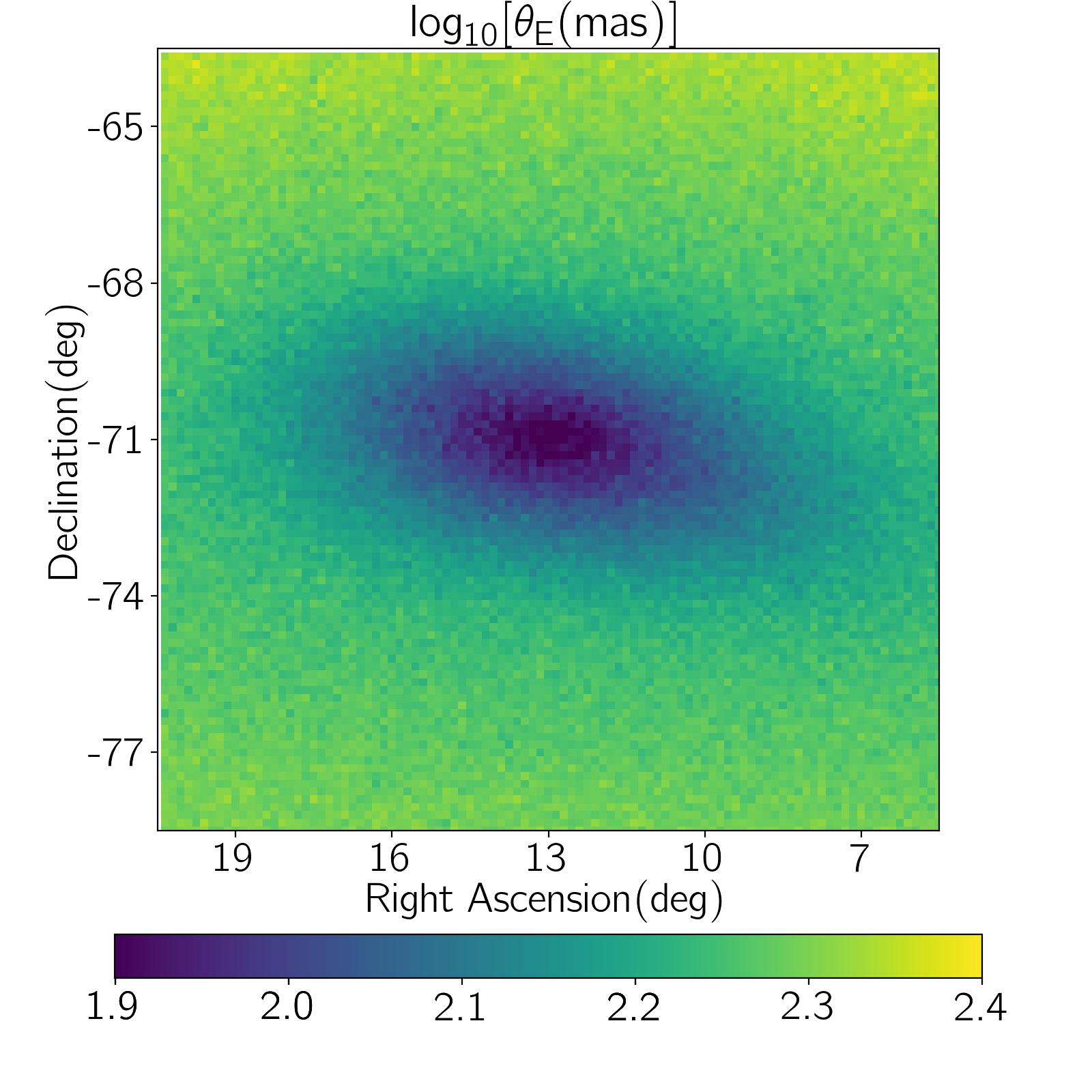}
    \includegraphics[width=0.32\textwidth]{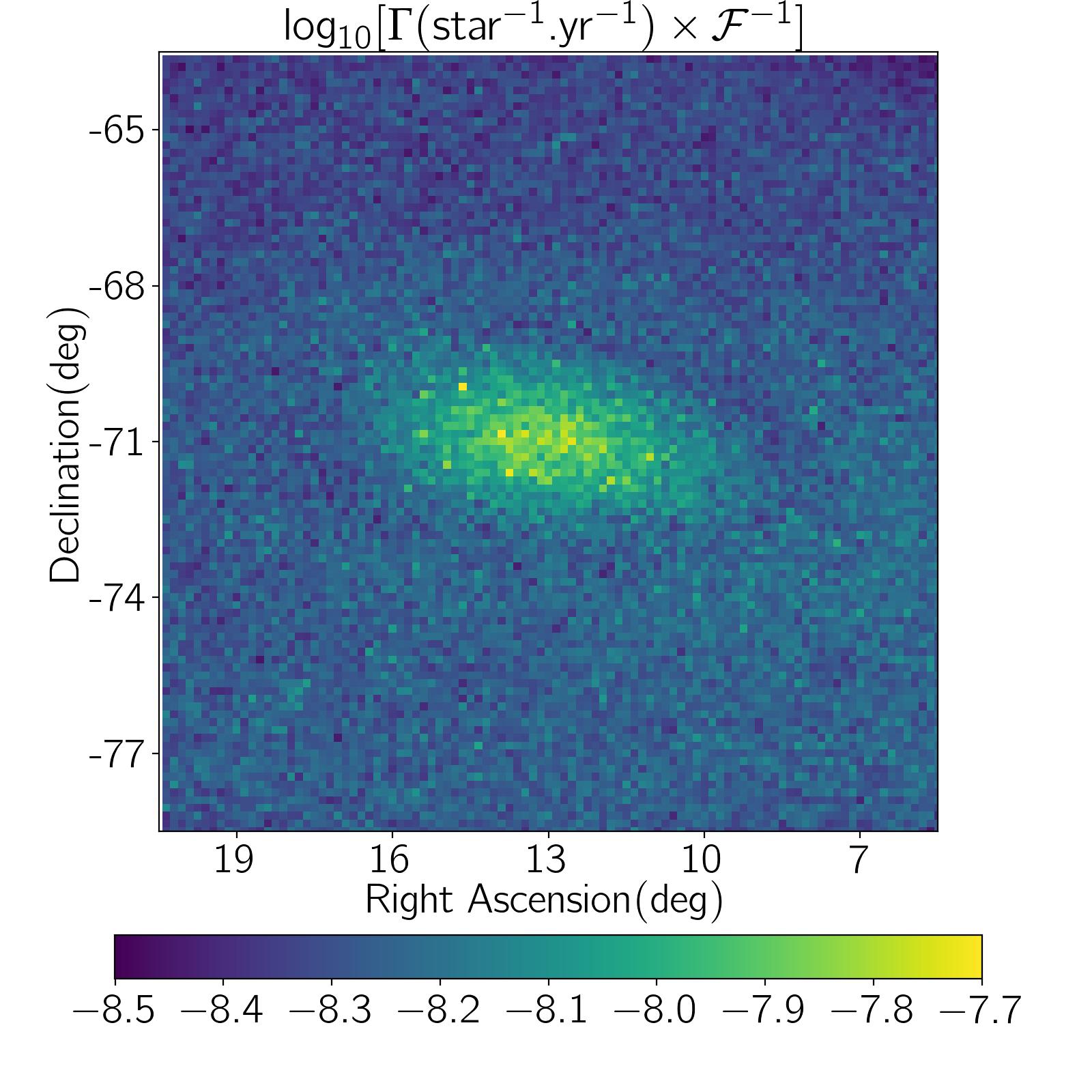}
    \includegraphics[width=0.32\textwidth]{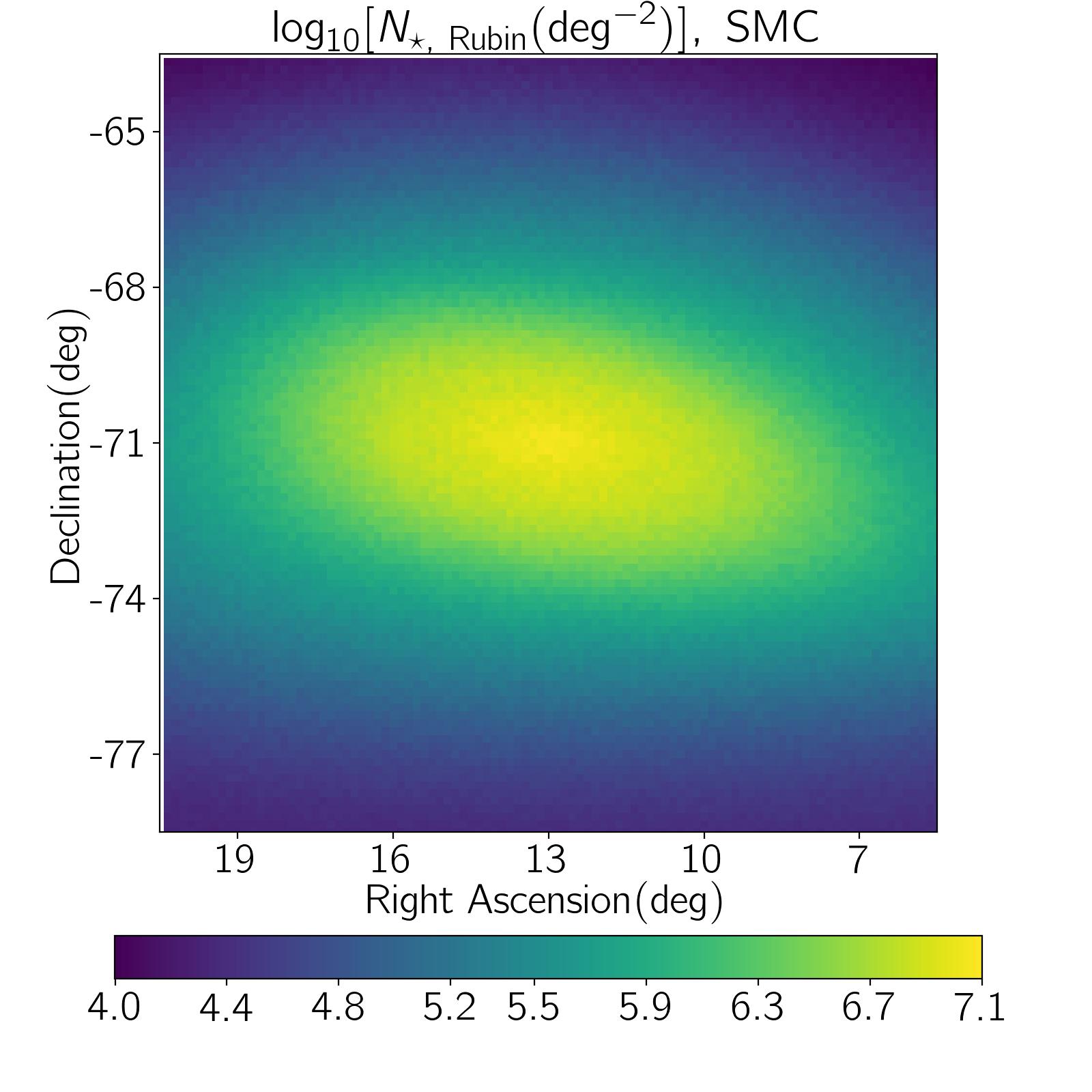}
    \includegraphics[width=0.32\textwidth]{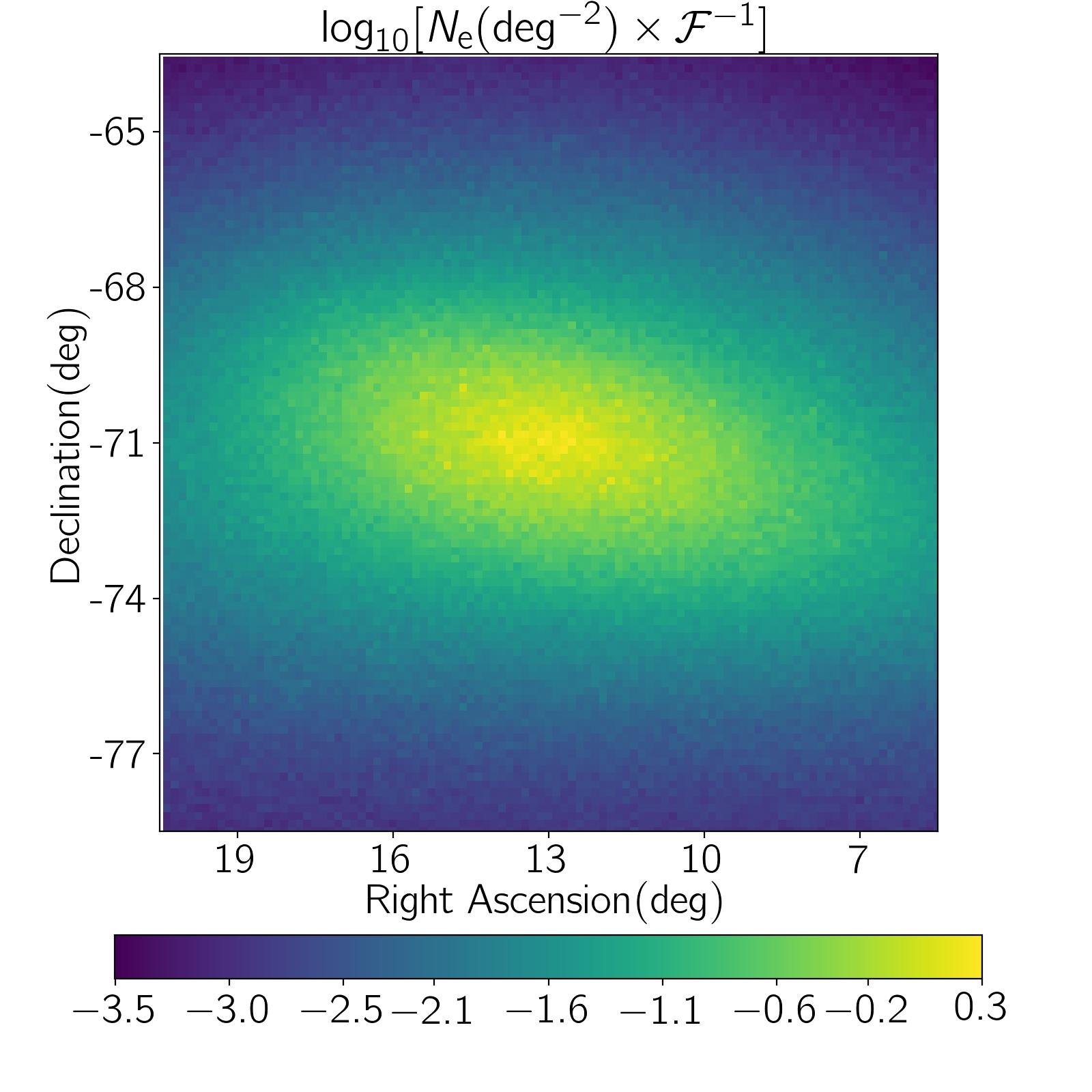}
    \caption{Same as Figure \ref{map1}, but they are resulted from simulations towards SMC.}\label{map2}
\end{figure*}

In Figures \ref{map1} and \ref{map2}, we show maps of several parameters over the 2D space representing the sky plane (made from right ascension $\alpha$ and declination $\delta$) derived from simulations of photometrically detectable microlensing events by applying the first mass function due to IBHs toward the LMC, and SMC, respectively. In the following, we explain their panels.

\noindent The first panels show the projected number density of stars in the unit of angular area over the sky plane at the distance of MCs $(D_{\rm{MC}})$, as given by:  
\begin{eqnarray}
N_{\star}\big(\alpha, \delta, D_{\rm{MC}}\big)=\int_{x=0}^{D_{\rm{MC}}} \sum_{i=1}\frac{\rho_{i}\big(x, \alpha, \delta\big)}{\overline{M_{i}}} x^{2} dx,
\label{nstar1}
\end{eqnarray} 
where, $x$ is a variable shows distances from the observer up to MCs, $\rho_{i}(x, \alpha, \delta)$ refers to the spatial density distribution in the given direction ($\alpha, \delta$) and distance $x$ due to the $i$th structure which could be the Galactic thin and thick disks, the Galactic bulge and stellar halo, the MC stellar halo, its disk or bulge (SMC done not have any bulge) which all of them are rewritten in Appendix \ref{app1}. $\overline{M_{i}}$ refers to the average mass of stars inside the $i$th structure. These maps are used to determine the number of blending stars at the position of any source star based on $N_{\rm{b}}=\Omega_{\rm{PSF}}\times N_{\star}\big(\alpha, \delta, D_{\star}\big)$, where $\Omega_{\rm{PSF}}=\pi\times\big(\rm{FWHM}\big/2\big)^{2}$, where the PSF FWHM values depend on the applied filter and weather conditions. In our simulations, we apply $\rm{FWHM}=1.22,~1.10,~0.99,~0.97,~0.95,~0.94$ arcsec related to the Rubin $ugrizy$ filters, respectively \citep{2025ApJChoi}.

\begin{deluxetable*}{c c c c c c c c c c c}[t]
\tablecolumns{11}
\centering
\tablewidth{0.95\textwidth}\tabletypesize\footnotesize
\tablecaption{Average lensing parameters of photometrically detectable microlensing events due to IBHs by considering four lens mass functions (offered by Equation \ref{mfs}) in the Rubin observations towards LMC and SMC.\label{tab1}}
\tablehead{\colhead{MF}&\colhead{$\log_{10}[\overline{M_{\rm{l}}}]$}&\colhead{$\overline{D_{\rm{l}}}$}&\colhead{$\overline{u_{0}}$}&\colhead{$\overline{f_{\rm{b}}}$}&\colhead{$\overline{m_{\rm{base}, r}}$}&\colhead{$\overline{\mu_{\rm{rel}}}$}&\colhead{$\overline{t_{\rm{E}}}$}&\colhead{$\log_{10}[\overline{\pi_{\rm{E}}}]$}&\colhead{$\log_{10}[\overline{\theta_{\rm{E}}}]$}&\colhead{$f_{\rm{Halo}}$}\\
&$(\rm{M_{\odot}})$ & $(\rm{kpc})$ &  & & $(\rm{mag})$&$(\rm{mas}/\rm{days})$& $(\rm{years})$& &$(\rm{mas})$&$[\%]$}
\startdata    
\multicolumn{11}{c}{Simulations towards LMC}\\ 
$\rm{MF}_{1}$ & $3.30\pm2.29$ & $16.8\pm9.2$ & $0.5\pm0.1$ & $0.90\pm0.09$ & $27.2\pm2.4$ & $28.4\pm11.8$ & $9.0\pm1.4$ & $-1.90\pm-2.42$ & $2.09\pm1.58$ & $68.6\pm19.1$\\
$\rm{MF}_{2}$ & $2.73\pm1.80$ & $17.0\pm9.1$ & $0.8\pm0.1$ & $0.91\pm0.08$ & $26.8\pm2.2$ & $21.5\pm7.0$ & $4.4\pm0.5$ & $-1.47\pm-2.02$ & $1.73\pm1.22$ & $68.1\pm18.9$\\
$\rm{MF}_{3}$ & $2.34\pm1.51$ & $17.1\pm9.2$ & $1.0\pm0.1$ & $0.91\pm0.08$ & $26.6\pm2.2$ & $17.8\pm6.0$ & $2.7\pm0.3$ & $-1.21\pm-1.75$ & $1.47\pm0.97$ & $67.8\pm19.1$\\
$\rm{MF}_{4}$ & $1.25\pm0.55$ & $17.3\pm9.3$ & $1.1\pm0.1$ & $0.91\pm0.07$ & $26.4\pm2.1$ & $14.0\pm4.9$ & $1.1\pm0.1$ & $-0.93\pm-1.45$ & $0.93\pm0.40$ & $67.2\pm19.4$\\
\tableline
\multicolumn{11}{c}{Simulations towards SMC}\\
$\rm{MF}_{1}$ & $3.33\pm1.99$ & $6.4\pm6.0$ & $0.4\pm0.1$ & $0.99\pm0.03$ & $23.2\pm0.5$ & $41.7\pm11.0$ & $8.8\pm1.2$ & $-1.81\pm-2.57$ & $2.25\pm1.42$ & $92.9\pm10.1$\\
$\rm{MF}_{2}$ & $2.82\pm1.71$ & $7.4\pm7.0$ & $0.6\pm0.1$ & $0.99\pm0.03$ & $22.7\pm0.6$ & $31.9\pm8.7$ & $4.8\pm0.7$ & $-1.39\pm-2.11$ & $1.92\pm1.16$ & $91.3\pm11.8$\\
$\rm{MF}_{3}$ & $2.47\pm1.52$ & $8.1\pm7.8$ & $0.6\pm0.1$ & $0.99\pm0.03$ & $22.4\pm0.6$ & $27.1\pm7.7$ & $3.1\pm0.4$ & $-1.12\pm-1.83$ & $1.68\pm0.96$ & $90.1\pm13.2$\\
$\rm{MF}_{4}$ & $1.32\pm0.66$ & $8.9\pm8.7$ & $0.7\pm0.1$ & $0.99\pm0.02$ & $22.0\pm0.6$ & $21.5\pm6.6$ & $1.2\pm0.2$ & $-0.79\pm-1.48$ & $1.10\pm0.42$ & $88.5\pm14.9$\\
\enddata
\tablecomments{$f_{\rm{Halo}}$ displays the fraction of detectable halo-lensing events from all detectable events.}
\end{deluxetable*}

\begin{deluxetable*}{c c c c c c c c c }
\tablecolumns{9}
\centering
\tablewidth{0.93\textwidth}\tabletypesize\footnotesize
\tablecaption{Average statistical parameters of photometrically detectable microlensing events due to IBHs by considering four lens mass functions in the Rubin observations towards LMC and SMC. \label{tab2}}
\tablehead{\colhead{MF}&\colhead{$\overline{\mu_{\star, \rm{n}}}$}&\colhead{$\log_{10}[\overline{\Delta \theta_{0, \rm{n}}}]$}&\colhead{$\epsilon_{\rm{D}}$}&\colhead{$\epsilon_{\rm{L}}$}&\colhead{$\overline{\tau_{\rm{IBH}}}$}&\colhead{$\log_{10}[\overline{\Gamma_{\rm{obs}}}\times \mathcal{F}^{-1}]$}&\colhead{$\log_{10}[\overline{N_{\star, \rm{Rubin}}}]$}&\colhead{$\log_{10}[\overline{N_{\rm{e}}}\times \mathcal{F}^{-1}]$}\\
& & &$[\%]$ & $[\%]$ & $\times 10^{6}\times\mathcal{F}^{-1}$ & $(\rm{star}^{-1}. \rm{yr}^{-1})$ & $(\rm{deg}^{-2})$& $(\rm{deg}^{-2})$}
\startdata  
\multicolumn{9}{c}{Simulations towards LMC}\\ 
$\rm{MF}_{1}$ & $4.42\pm1.29$ & $0.80\pm0.31$ & $1.89\pm0.97$ & $25.11\pm9.29$ & $0.50\pm0.33$ & $-7.84\pm-8.06$ & $6.50\pm6.59$ & $-0.15\pm0.16$\\
$\rm{MF}_{2}$ & $2.67\pm0.58$ & $0.45\pm-0.17$ & $1.88\pm0.96$ & $40.84\pm11.27$ & $0.50\pm0.33$ & $-7.48\pm-7.86$ & $6.50\pm6.59$ & $0.15\pm0.42$\\
$\rm{MF}_{3}$ & $1.97\pm0.37$ & $0.19\pm-0.41$ & $1.89\pm0.97$ & $51.05\pm12.87$ & $0.50\pm0.33$ & $-7.28\pm-7.70$ & $6.50\pm6.59$ & $0.34\pm0.59$\\
$\rm{MF}_{4}$ & $1.31\pm0.13$ & $-0.32\pm-0.98$ & $1.88\pm0.98$ & $58.87\pm14.98$ & $0.50\pm0.33$ & $-7.05\pm-7.52$ & $6.50\pm6.59$ & $0.55\pm0.79$\\
 \tableline
 \multicolumn{9}{c}{Simulations towards SMC}\\
$\rm{MF}_{1}$ & $12.54\pm2.82$ & $1.28\pm0.70$ & $6.85\pm1.37$ & $24.31\pm4.67$ & $0.28\pm0.14$ & $-8.25\pm-8.83$ & $6.08\pm6.32$ & $-1.05\pm-0.69$\\
$\rm{MF}_{2}$ & $7.70\pm1.73$ & $0.98\pm0.44$ & $6.80\pm1.37$ & $34.53\pm5.93$ & $0.28\pm0.14$ & $-7.91\pm-8.50$ & $6.07\pm6.32$ & $-0.71\pm-0.34$\\
$\rm{MF}_{3}$ & $5.41\pm1.24$ & $0.75\pm0.24$ & $6.74\pm1.35$ & $38.09\pm6.21$ & $0.28\pm0.14$ & $-7.72\pm-8.34$ & $6.07\pm6.31$ & $-0.52\pm-0.16$\\
$\rm{MF}_{4}$ & $2.77\pm0.72$ & $0.22\pm-0.26$ & $6.61\pm1.35$ & $37.77\pm5.85$ & $0.28\pm0.14$ & $-7.47\pm-8.13$ & $6.07\pm6.32$ & $-0.29\pm0.07$\\
\enddata
\tablecomments{Here, $\mu_{\star, \rm{n}}$ is the average angular source velocity during the Rubin observing window normalized to the angular source velocity at the baseline without lensing effect. $\Delta\theta_{0, \rm{n}}$ is the angular distance between two images at the time of the magnification peak normalized to the Rubin astrometric accuracy.}
\end{deluxetable*}

\noindent The second panels represent the lensing optical depth due to detectable stars (which pass the mentioned criteria i, ii). In these maps and for each given direction ($\alpha, \delta$), we average the optical depths related to distances of all detectable stars. The microlensing optical depth due to detectable stars at the magnification peak is usually labeled with DIA (Difference Image Analysis). The lensing optical depth due to IBHs is evaluated by:  
\begin{eqnarray}
\tau_{\rm{IBH}}(\alpha, \delta, D_{\star})=\frac{4 \pi G~u_{0, \rm{m}}^{2}}{c^{2}~D_{\star}}\times~~~~~~~~~~~~~~~~~~~~~~~&& \nonumber\\
\int_{0}^{D_{\star}} x\big(D_{\star}-x\big)\rho_{\rm{IBH}}(x, \alpha, \delta)~dx=\mathcal{F} \times \tau_{\rm{tot}}(D_{\star}, \alpha, \delta).~~~~~~~~~&&
\label{tau}
\end{eqnarray} 
While simulating microlensing events, we select $u_{0, \rm{m}}=3$ as the maximum lens impact parameter. We consider IBHs with the mass range $[3, 5000] M_{\odot}$ as the lens objects so in this integration we use the spatial density distribution of IBHs, i.e., $\rho_{\rm{IBH}}(x, \alpha, \delta)$. This spatial distribution has not been uniquely determined. Therefore, we assume the dependence of IBHs in this mass range spatial distribution on the observing direction and distance is similar to the known spatial stellar and dark-matter density, but it is smaller by the factor $\mathcal{F}$. Hence, in Equation \ref{tau} we apply $\rho_{\rm{IBH}}(x, \alpha, \delta)=\mathcal{F}\times\rho_{\rm{tot}}(x, \alpha, \delta)$, and accordingly $\tau_{\rm{IBH}}(D_{\star}, \alpha, \delta)=\mathcal{F}\times\tau_{\rm{tot}}(D_{\star}, \alpha, \delta)$, where $\tau_{\rm{tot}}$ is the lensing optical depth due to the total Galactic and MCs density distributions in a given direction and up to an indicated distance.
    
\noindent As a rough estimation on $\mathcal{F}$, we can assume $\mathcal{F}$ as the ratio of the total mass of BHs in our galaxy, i.e., $\sim10^{8}-10^{9}M_{\odot}$ \citep[see, e.g.][]{2025Bambi}, to the total dynamic mass of our galaxy ($M_{\rm{dy}}\sim10^{12}M_{\odot}$). In that case, $\mathcal{F}$ should be $\sim10^{-4}-10^{-3}$. 
For a more realistic estimation of $\mathcal{F}$ we can evaluate this fraction based on the initial stellar mass function in the Galaxy and dark-matter mass function, as follows: 
\begin{eqnarray}
\mathcal{F}=\frac{\int_{20M_{\odot}}^{\infty} M~\eta_{\rm{ST}}~dM+f\times\int_{3M_{\odot}}^{5000M_{\odot}}M~\zeta_{\rm{DM}}~dM}{\int_{0}^{\infty} M~\eta_{\rm{ST}}~dM+f\times\int_{0}^{\infty} M~\zeta_{\rm{DM}}~dM},
\label{mathf}
\end{eqnarray}    
where, $\eta_{\rm{ST}}=dN/dM$ is the initial stellar mass function which depends on the mass, $\zeta_{\rm{DM}}$ is the mass function in the Galactic DM halo, and $f=5\%$ is the fraction of the Galactic halo DM in the form of compact objects. Here by adopting the mass function in the Galactic disk and halo as reported in Appendix \ref{app3} (where the microlenses potentially locate while observing MCs), the resulting $\mathcal{F}$ is $\simeq0.04$. In Equation \ref{mathf}, the low limit of  $20M_{\odot}$ refers to the fact that stars with $M\geq20M_{\odot}$ will be converted to black holes. Since $\eta_{\rm{ST}}$ tends to zero rapidly for massive objects as $M^{-3}$, so the up limit of infinity in the first integral (instead of the most massive star at birth) does not change the value of $\mathcal{F}$ significantly. However, in the following and while estimating the number of detectable events we consider several values for $\mathcal{F}$.

The optical depth is higher for brighter parts of MCs, because detectable stars in these directions are mostly inside MCs (at farther distances) where the spatial density distribution maximizes. However, for brighter parts of MCs the blending effects are worse, as shown in the third panels. In these panels, for each given direction the reported blending parameter in $r$-band ($f_{\rm b, r-\rm{band}}$) is the average value of the blending parameters due to detectable stars. Higher blending effects for the lines of sight towards the MCs' center make the lower lensing detection efficiency $(\epsilon_{\rm L})$ as represented in the forth panels of Figures \ref{map1} and \ref{map2}. We note that detectable stars towards the tight central parts of MCs are brighter than ones towards the MCs' halo. For that reason, the lensing detection efficiency for tight central parts is somewhat higher than that for MCs' halo parts. 

Since the projected mass density in lines of sight towards MCs' center is higher, so the lens objects for potential microlensing events towards these directions are located inside rather MCs than the Galactic halo. This point can be found in fifth panels of Figure \ref{map1} and \ref{map2}. Here, $f_{\rm{Halo}}$ is the fraction of detectable halo-lensing events to all detectable microlensing events. Although almost all microlensing events far from the MCs' centers are halo-lensing events, towards the MCs' center only $20$-$30$ per cent of microlensing events are halo-lensing ones.   

\noindent Accordingly, for events toward the MCs' center $\pi_{\rm{rel}}$, $\theta_{\rm{E}}$ (see the sixth panels), and as a result $\pi_{\rm E}$ are less than those for events far from the MCs' centers. Generally, for self-lensing events the relative errors in the lensing and physical parameters are higher than those for halo-lensing events, because of their small $\pi_{\rm{E}}$ and $\theta_{\rm{E}}$ values. Three microlensing events shown in Figure \ref{fig1} confirm this point as well.

\begin{figure*}
    \centering
    \includegraphics[width=0.49\textwidth]{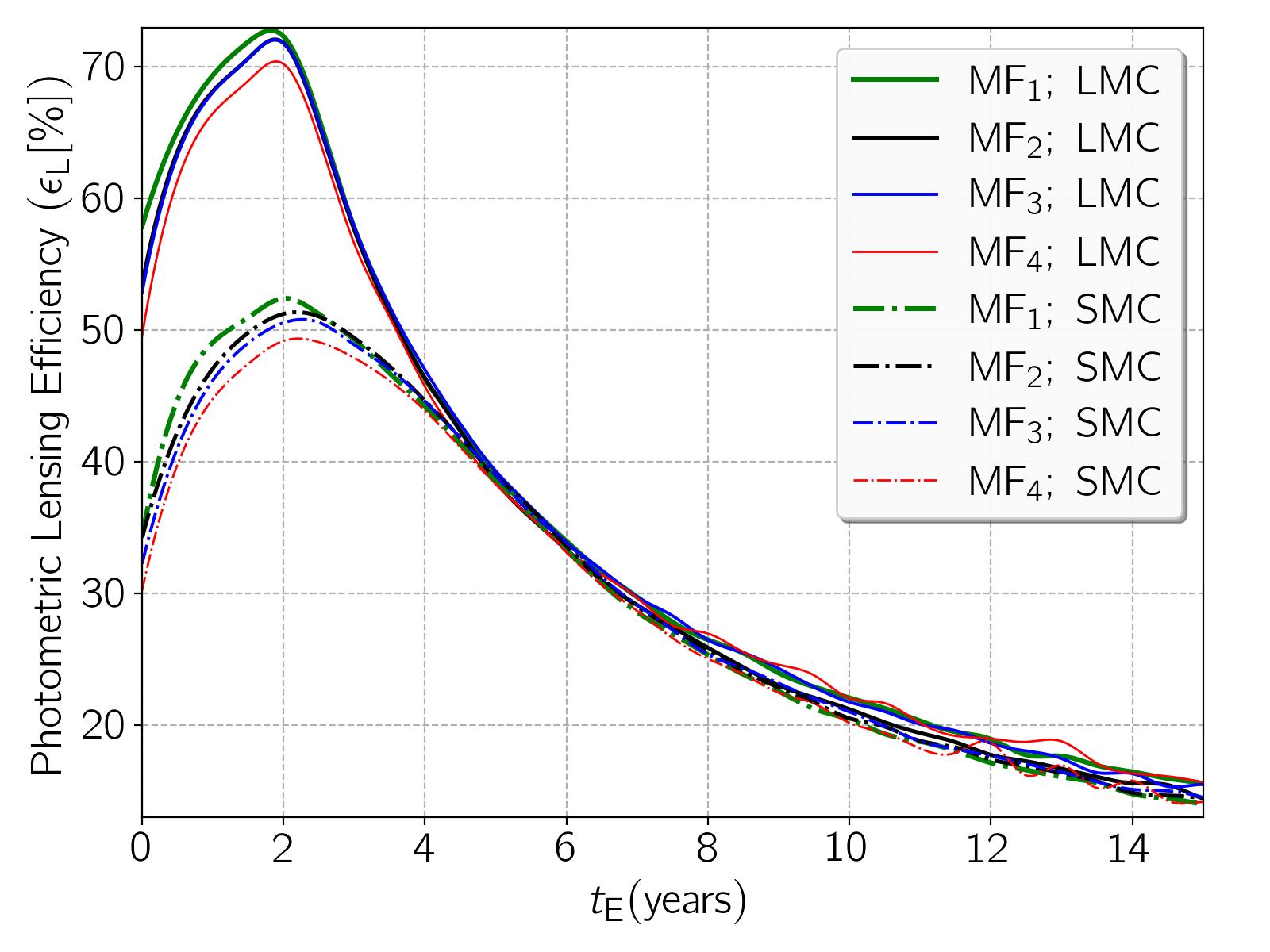}
    \includegraphics[width=0.49\textwidth]{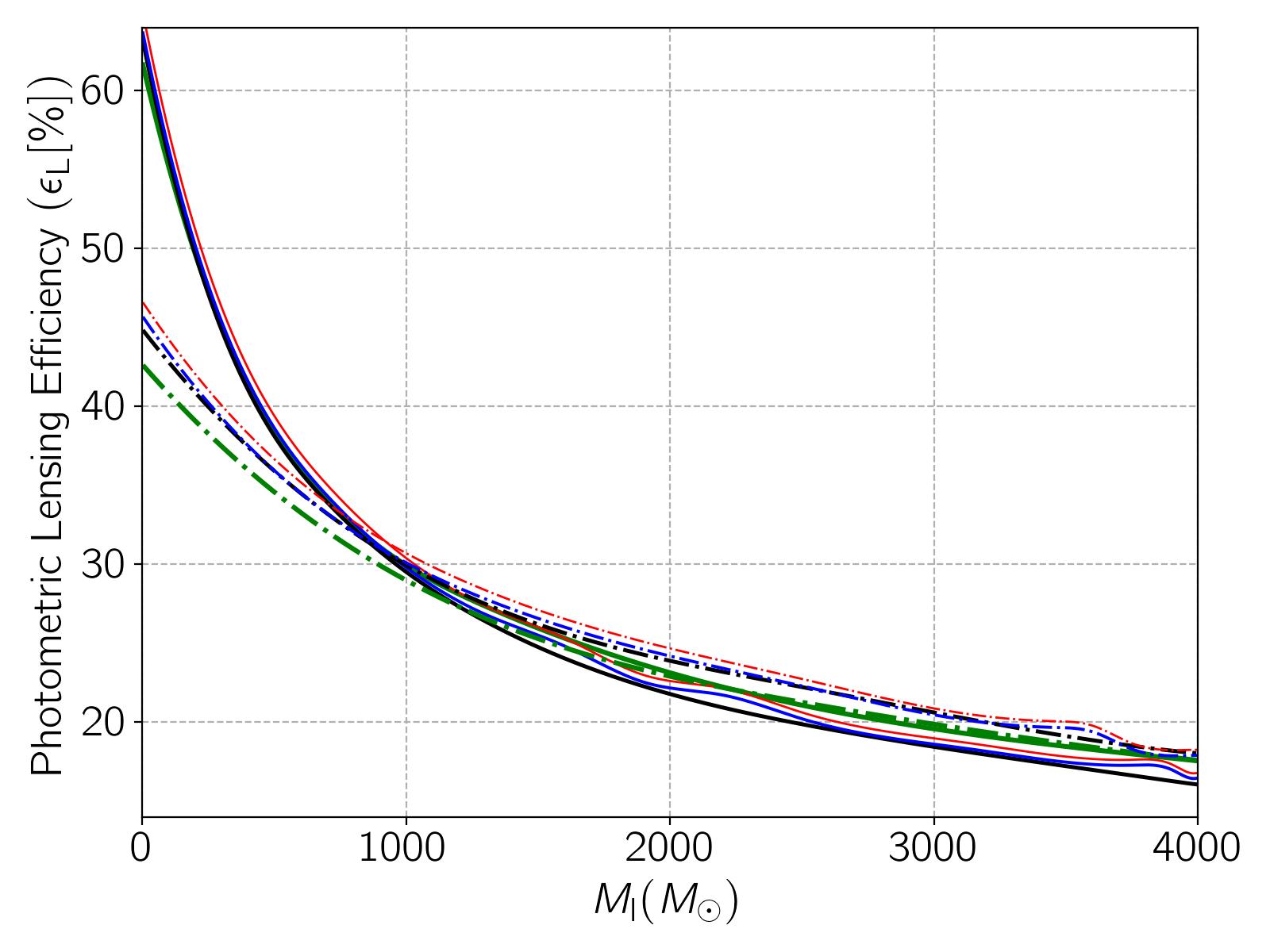}
    \includegraphics[width=0.49\textwidth]{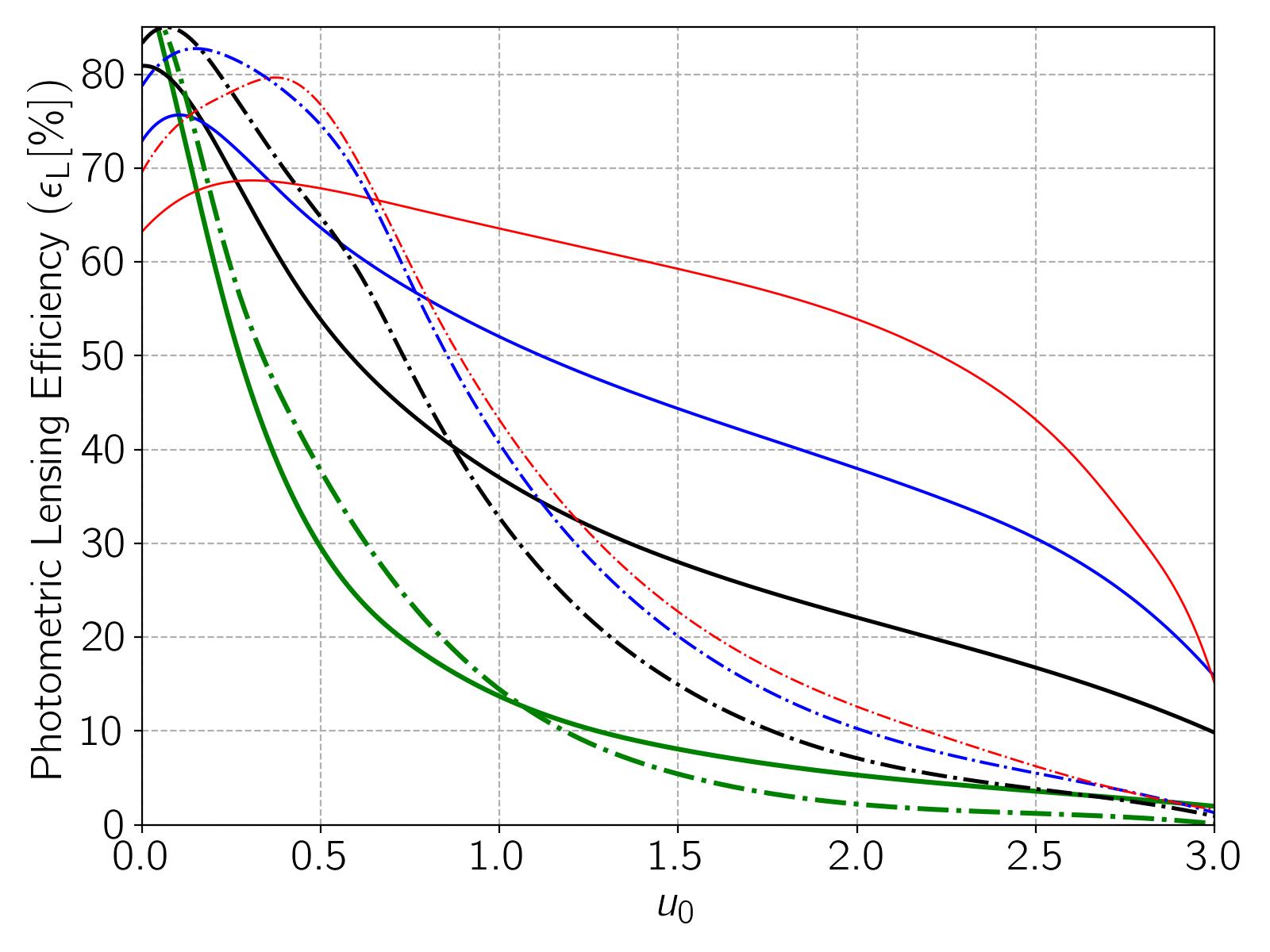}
    \includegraphics[width=0.49\textwidth]{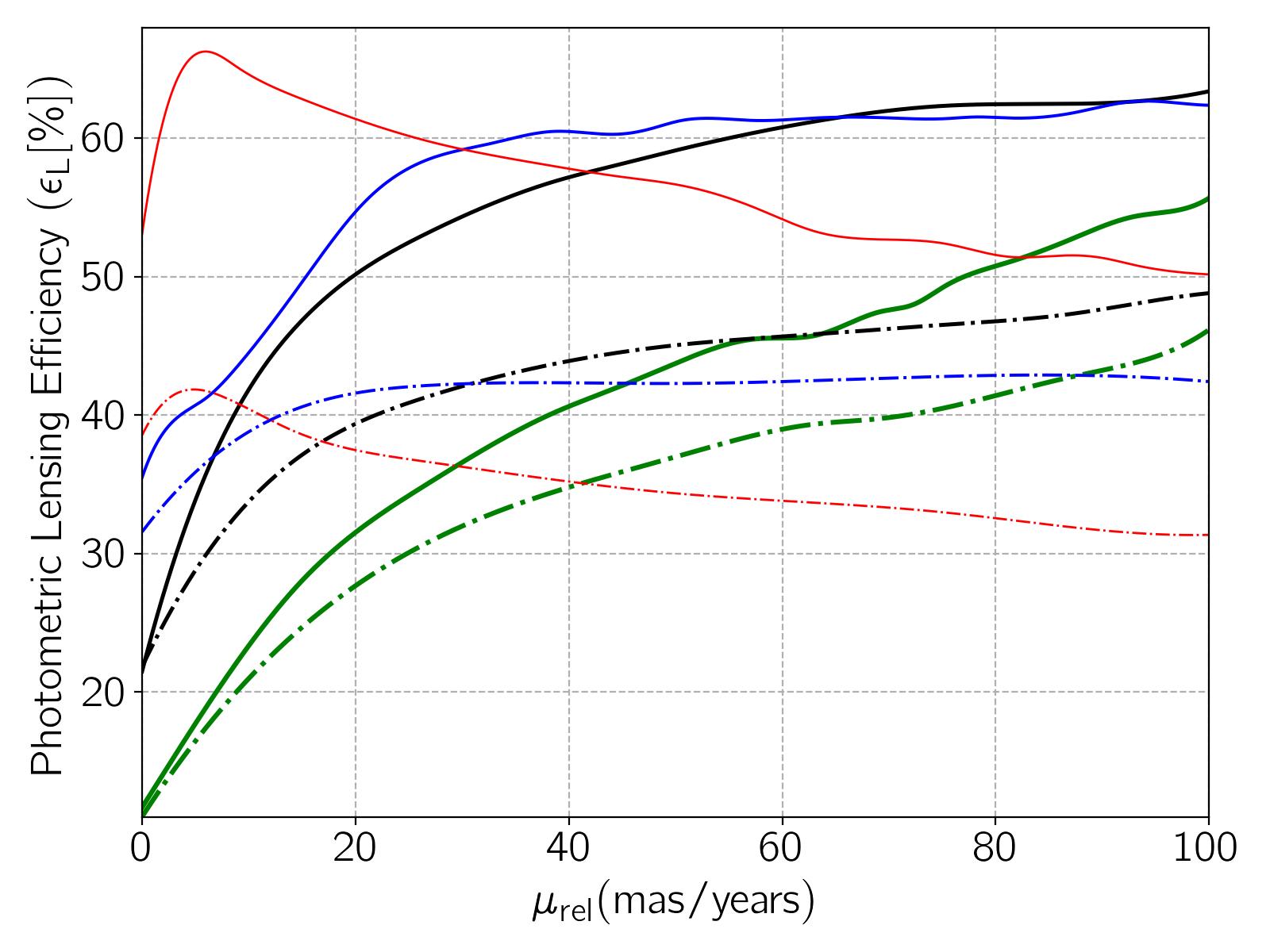}
    \caption{The Rubin efficiency for detecting microlensing events of detectable source stars due to IBHs versus four parameters. The different curves are due to different mass densities (MFs which are given by Equation \ref{mfs}), and for simulations towards LMC (solid curves) and SMC (dot-dashed curves).}\label{plotef}
\end{figure*}

We evaluate the microlensing event rate $\Gamma$ in a given direction $(\alpha, \delta)$ using:
\begin{eqnarray}
\Gamma(\alpha, \delta)= \frac{2}{\pi~u_{0, \rm{m}}}~\mathcal{F}~\tau_{\rm{tot}}(\alpha, \delta)~\overline{\epsilon_{\rm{L}}(t_{\rm E}, \alpha, \delta)\Big/t_{\rm E}}, 
\label{rate}
\end{eqnarray}
where, $\epsilon_{\rm{L}}(t_{\rm{E}}, \alpha, \delta)$ is the Rubin photometric efficiency for detecting a microlensing event with the time scale $t_{\rm{E}}$ and in a given line of sight ($\alpha, \delta$). This lensing efficiency depends on the time scale $t_{\rm{E}}$ as depicted in the first panel of Figure \ref{plotef}. In simulations we calculate this efficiency numerically and by making a sample of detectable microlensing events for any given line of sight. In Subsection \ref{effi}, we will study the dependence of this efficiency on four relevant parameters. In Equation \ref{rate} the over-line refers to the average value over $n$ uniform bins for $t_{\rm E}$ in the range $t_{\rm E} \in [0,~50~\rm{years}]$, so that $\overline{\epsilon_{\rm{L}}(t_{\rm{E}}, \alpha, \delta)/t_{\rm{E}}}=\frac{1}{n}\sum_{i=1}^{n} \epsilon_{\rm L}(t_{\rm{E}, i}, \alpha, \delta)/t_{\rm E, i}$, for $n=100$. The maps of $\Gamma$ from simulating microlensing events due to IBHs towards LMC and SMC (for $\mathcal{F}=1$) are depicted in the seventh panels of Figure \ref{map1} and \ref{map2}.

To estimate the number of microlensing events due to IBHs, we first evaluate the number of background stars towards MCs that the Rubin telescope will detect by: 
\begin{eqnarray}
&&N_{\star, \rm{Rubin}}\big(D_{\rm{MC}}, \alpha, \delta\big)=\nonumber\\\int_{x=0}^{D_{\rm{MC}}}&&\sum_{i=1}\frac{\rho_{i}\big(x, \alpha, \delta\big)}{\overline{M_{i}}} \epsilon_{\rm{D}}(x, \alpha, \delta) x^{2}dx,
\end{eqnarray}
\noindent which is similar to Equation \ref{nstar1}, but by adding the Rubin efficiency for resolving background stars $\epsilon_{\rm{D}}$ which is a function of observing direction and distance from the observer. The maps of $N_{\star, \rm{Rubin}}$ in the logarithmic scale towards LMC and SMC are represented in the eighth panels. By comparing two maps of $N_{\star}$ and $N_{\star, \rm{Rubin}}$ (the first and eighth panels), one can find that the Rubin detection efficiency is a few percent and its depends on the observing direction. Finally, we estimate the number of detectable events as $N_{\rm{e}}(\alpha, \delta)= N_{\star, \rm{Rubin}}(\alpha, \delta)~\Gamma(\alpha, \delta)~T_{\rm{obs}}$ which towards LMC and SMC (by assuming $\mathcal{F}=1$) are shown in the last panels of Figure \ref{map1}, and \ref{map2}, respectively. In the next subsection, we study the effects of different IBHs mass functions on the results of Monte-Carlo simulations.

\subsection{Effects of Different IBHs Mass Functions}\label{MFs}
We repeat simulations for four mass functions which are mentioned in Equation \ref{mfs}. To compare their results, in Tables \ref{tab1} and \ref{tab2} we report the average values of lensing and statistical parameters due to all detectable events in all directions, and discuss on them in following.    
 
{\bf Table \ref{tab1}:} According to its second column, by increasing the power index ($\beta$) in the IBHs mass function from zero to two (towards steeper slopes) the average mass of lenses in detectable events decreases from $\sim2000M_{\odot}$ to $\sim20M_{\odot}$. Hence, by increasing the power index in the IBHs mass function the detectable events have on average shorter time scales, smaller $\theta_{\rm E}$ and larger parallax amplitude. We note that very long-duration microlensing events with the time scale comparable with the observing time window (e.g., the events detectable by assuming the first mass function) may be mistakenly classified as long periodic variables  like miras.

Microlensing due to very massive lens objects are detectable often when their lens objects are closer to the observer with higher proper motions, so that they have time scales within (shorter than) the Rubin observing window. We note that the lensing time scale is proportional to $t_{\rm E}\propto \sqrt{M_{\rm l}}/\mu_{\rm{rel}}$. For that reason, by increasing the power index in the IBHs mass function the average lens distance increases somewhat as reported in the third column of Table \ref{tab1}. 

\noindent Based on the third detectability criterion (III), i.e., $\rm{FWHM}\leq T_{\rm{obs}}$, in a limited observing window detectable microlensing events with longer time scales (due to more massive lens objects) have on average smaller lens impact parameters. Accordingly, the average lens impact parameters increase by enhancing the power index in the IBHs mass function, which is brought in the forth column of Table \ref{tab1}.

\begin{figure}
    \centering
    \includegraphics[width=0.49\textwidth]{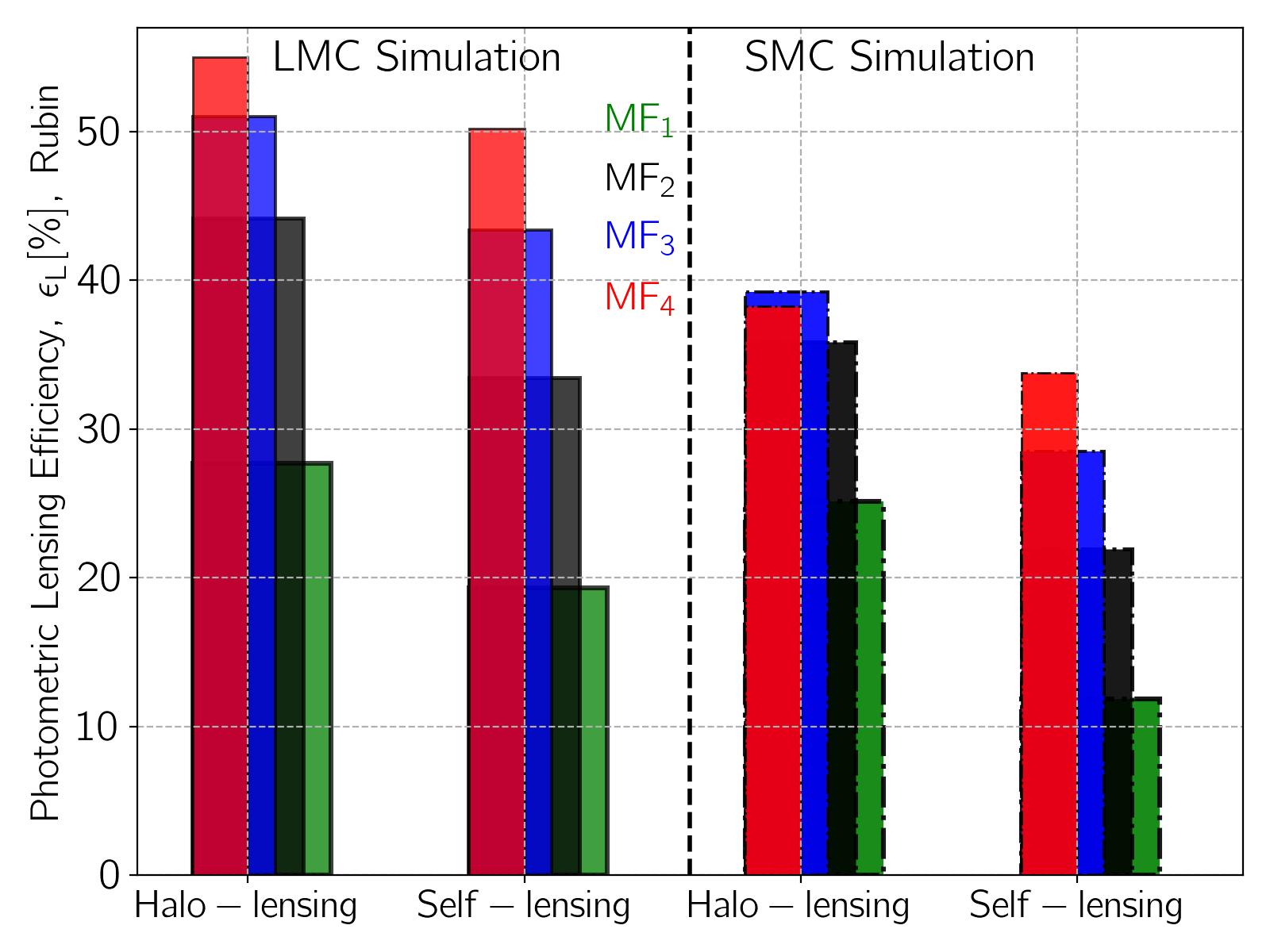}
    \caption{The photometric lensing efficiency of the Rubin telescope for detecting halo-lensing and self-lensing events from simulations towards LMC and SMC. The simulations are repeated for four lens mass densities as specified with four colors.}
    \label{selfhalo}
\end{figure}

For the last mass function (i.e., $\rm{MF}_{4}$ with $\beta=2$) on average detectable events have $\overline{t_{\rm{E}}}\simeq1.1$ years. During this time span, the Rubin telescope will record only $\sim50-100$ data points. These short-duration microlensing events with such a low number of data points should have bright source stars (with a good photometric accuracy) to pass the detection criteria I and II. The source stars within the central parts of MCs are brighter than sources inside halo parts. Hence, if the real IBHs mass function has a large power index (e.g., $\beta\sim2$ in which the majority of BHs have low masses) self-lensing events which most of them are towards the MCs' central parts have a higher detectability rate. Figure \ref{selfhalo} confirms this point. We note that self-lensing events have lower $\mu_{\rm{rel}}$ values than halo-lensing events, which is approved by the seventh column of Table \ref{tab1}.
\begin{deluxetable}{c c c c c c c}
\tablecolumns{7}
\centering  
\tablewidth{0.49\textwidth}\tabletypesize\footnotesize
\tablecaption{Each entrance reports $N_{\rm{e}, \rm{tot}}$, i.e., the expected number of detectable microlensing events due to IBHs toward MCs in the Rubin observations from two 2D spaces $\alpha \in[75, 90]$ and $\delta \in [-75, 60]$ degree for LMC, and $\alpha \in [5.7, 20.7]$ and $\delta \in [-78.7, -63.7]$ degree for SMC, by assuming six values for $\mathcal{F}$.\label{tab3}}
    \tablehead{\colhead{$\mathcal{F}:$}&\colhead{$5e$-$5$}&\colhead{$1e$-$4$}&\colhead{$5e$-$4$}& \colhead{$1e$-$3$}&\colhead{$5e$-$3$}&\colhead{$1$}}
    \startdata	
    \multicolumn{7}{c}{Simulations towards LMC}\\
    $\rm{MF}_{1}$ & $0.008$ & $0.016$ & $0.080$ & $0.160$ & $0.799$ & $159.867$\\
    $\rm{MF}_{2}$ & $0.016$ & $0.032$ & $0.159$ & $0.318$ & $1.590$ & $317.959$\\
    $\rm{MF}_{3}$ & $0.024$ & $0.049$ & $0.245$ & $0.490$ & $2.448$ & $489.599$\\
    $\rm{MF}_{4}$ & $0.040$ & $0.079$ & $0.396$ & $0.792$ & $3.960$ & $792.020$\\
    \tableline
    \multicolumn{7}{c}{Simulations towards SMC}\\
    $\rm{MF}_{1}$ & $0.001$ & $0.002$ & $0.010$ & $0.020$ & $0.101$ & $20.171$\\
    $\rm{MF}_{2}$ & $0.002$ & $0.004$ & $0.022$ & $0.044$ & $0.220$ & $43.998$\\
    $\rm{MF}_{3}$ & $0.003$ & $0.007$ & $0.034$ & $0.068$ & $0.340$ & $67.967$\\
    $\rm{MF}_{4}$ & $0.006$ & $0.012$ & $0.058$ & $0.116$ & $0.580$ & $116.003$\\
    \enddata
\end{deluxetable}

Finally, we compare the properties of detectable events towards LMC and SMC. Their source distances are on average $\sim50,~60$ kpc, respectively. In the SMC direction $\mu_{\rm{rel}}$ values are on average higher than those towards LMC. Because, projected source velocities decrease with distance from the observer. Considering the detectability criterion (III), detectable events towards SMC have closer lens objects, larger $\theta_{\rm E}$ values which results higher parallax amplitudes on average.

{\bf Table \ref{tab2}:} In this table, we report the statistical parameters derived from simulations. In its second column $\mu_{\star, \rm{n}}$ for each detectable microlensing event is estimated by the ratio of source star's angular velocity during observing window to its value at the baseline (without lensing). Here, the over-line means averaging over all photometrically detectable microlensing events. Since most of microlensing events due to IBHs have long durations and extremely large $\theta_{\rm{E}}$ values, hence lensing-induced deflections are rather straight lines (during the Rubin $10$-year mission), and their lensing-induced deviations from straight lines (with circle/ellipse shapes see, e.g., the last panel of Figure \ref{fig1}) are mostly small. $\overline{\mu_{\star, \rm{n}}}$ shows the average variation in the source angular velocity due to the lensing effect with respect to the baseline. Accordingly, the angular source velocity enlarges. This variation is higher for events towards SMC because of their larger $D_{\star}$ values and as a result larger $\theta_{\rm{E}}$ values.

The third column of Table \ref{tab2} reports $\Delta\theta_{0, \rm{n}}=\Delta\theta_{0}/\sigma_{\rm{a}}$ which is the angular separation of two images at the time of maximum magnification normalized to the Rubin astrometric accuracy. However, the Rubin astrometric accuracy for resolving two close stars is worse than $\sigma_{\rm{a}}$ by the factor larger than $5$ which depends on (1) the signal to noise ratio (SNR), and (2) the difference between magnitudes of the two close stars which will be explained more in Subsection \ref{astrop}. On average, detectable events have $\Delta \theta_{0, \rm n}\gtrsim1.5$, which means that for some of them their two images at the maximum magnification are resolvable. However, these two images should additionally be bright enough to be realized (i.e., their apparent magnitudes should be less than the Rubin detection thresholds). We study this point in Subsection \ref{astrop}.

Towards MCs, source stars are detectable in the Rubin observations by the efficiency $\epsilon_{\rm D}\sim 2-7\%$, as mentioned in the forth column of Table \ref{tab2}. This efficiency weakly depends on the applied lens mass density, because of the third criterion (iii). In that criterion, we label simulated microlensing events with the number of data points less than three as ones with undetectable source stars. For detectable source stars, the Rubin efficiency to realize their lensing signals is $\epsilon_{\rm L}\sim20-50\%$. Since towards SMC detectable source stars are brighter (see the sixth column of Table \ref{tab1}), the Rubin efficiency to detect their lensing signals are higher. We note that the stellar number density towards the LMC is higher than that towards the SMC by one order of magnitude. This implies higher blending effects and lower detection efficiencies towards the LMC than those towards the SMC.

The average optical depths and the number of detectable stars in the Rubin observations $N_{\star, \rm{Rubin}}$ (mentioned in the sixth and eighth columns of Table \ref{tab2}) do not depend on applied lens mass densities. The second one for observations towards SMC is lower than that in the Rubin observations towards LMC, which results  the number of microlensing events from LMC observations is somewhat higher than the number of events from observations towards SMC.  

In Table \ref{tab2}, and for reported optical depth values, lensing event rate and number of detectable events, we put aside the fraction of IBHs's mass in the total Galactic mass which was discerned by $\mathcal{F}$. In Table \ref{tab3}, we consider several values for $\mathcal{F}$ and estimate the number of detectable events due to IBHs in the Rubin observations from the projected $15\times15~\rm{deg}^{2}$ angular areas over LMC and SMC. For $\mathcal{F}\sim 5\times 10^{-3}$ the number of detectable microlensing events $N_{\rm{e}, \rm{tot}}$ reaches to one.

\subsection{Photometric Rubin Efficiency for Detecting Lensing}\label{effi}
Here, we study dependence of the lensing detection efficiency of Rubin $\epsilon_{\rm L}$ on the lensing parameters. In Figure \ref{plotef}, the photometric lensing efficiencies in the Rubin observations are depicted versus four parameters $t_{\rm E}$, $M_{\rm l}$, $u_{0}$, and $\mu_{\rm{rel}}$, respectively. In each panel, eight curves are shown with four different colors and widths (related to simulations by considering different lens mass densities which are specified in the legend of the first panel) and two line styles including solid (related to simulations towards LMC) and dot-dashed (due to simulations towards SMC). For four mentioned parameters, we consider $100$ bins in their corresponding ranges. For $i$th bin related to parameter $p$ with the interval $[p_{i},~p_{i+1}]$, we calculate the number of detectable microlensing events in which $p\in [p_{i},~p_{i+1}]$ to total number of simulated events with detectable source stars and $p$ values in the given interval as the lensing detection efficiency. We note that curves had small fluctuations and we smoothed them without losing their trends using the python package \texttt{scipy.interpolate.splrep} \citep{2020SciPyNMeth}. 
\begin{figure*}
\centering
\includegraphics[width=0.32\textwidth]{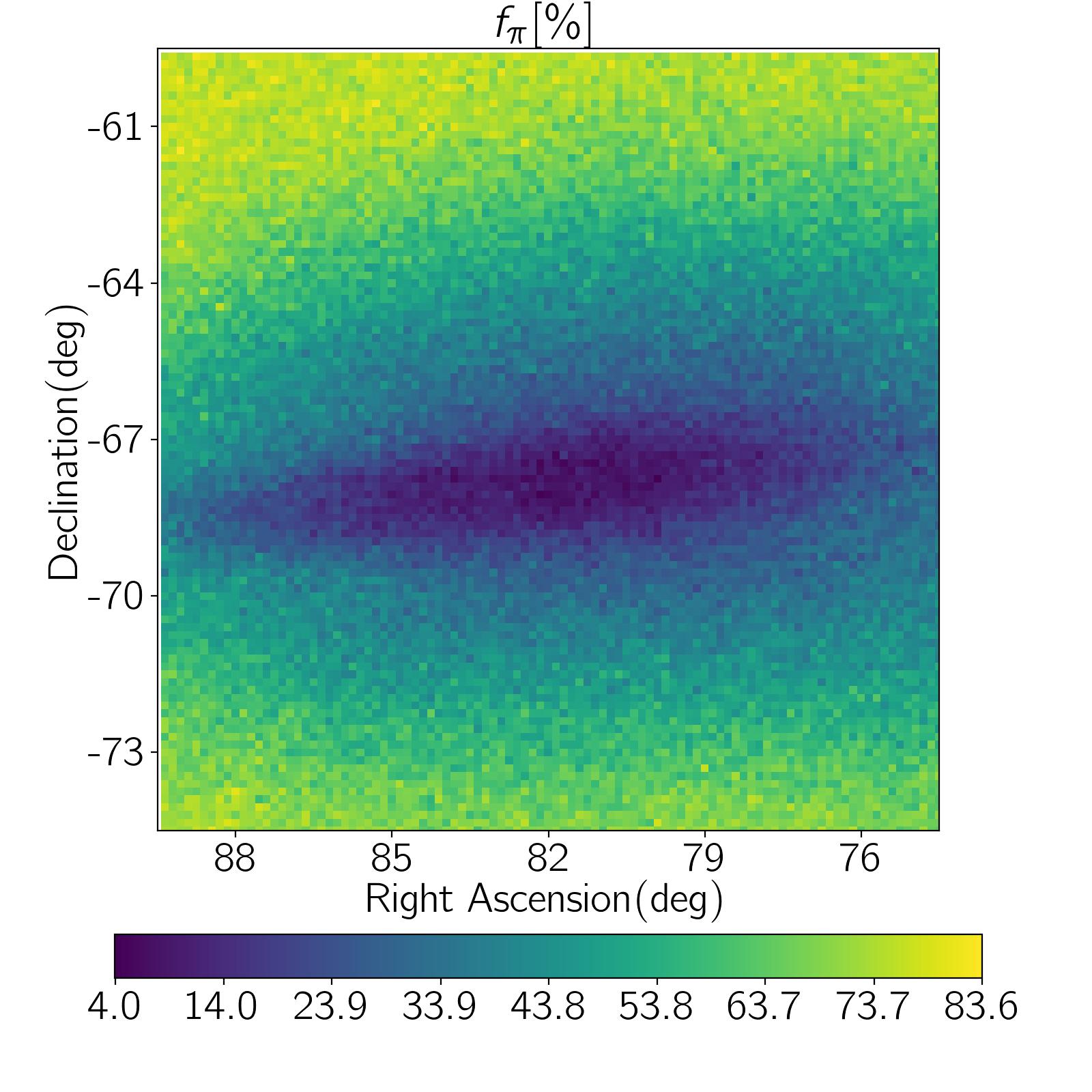}
\includegraphics[width=0.32\textwidth]{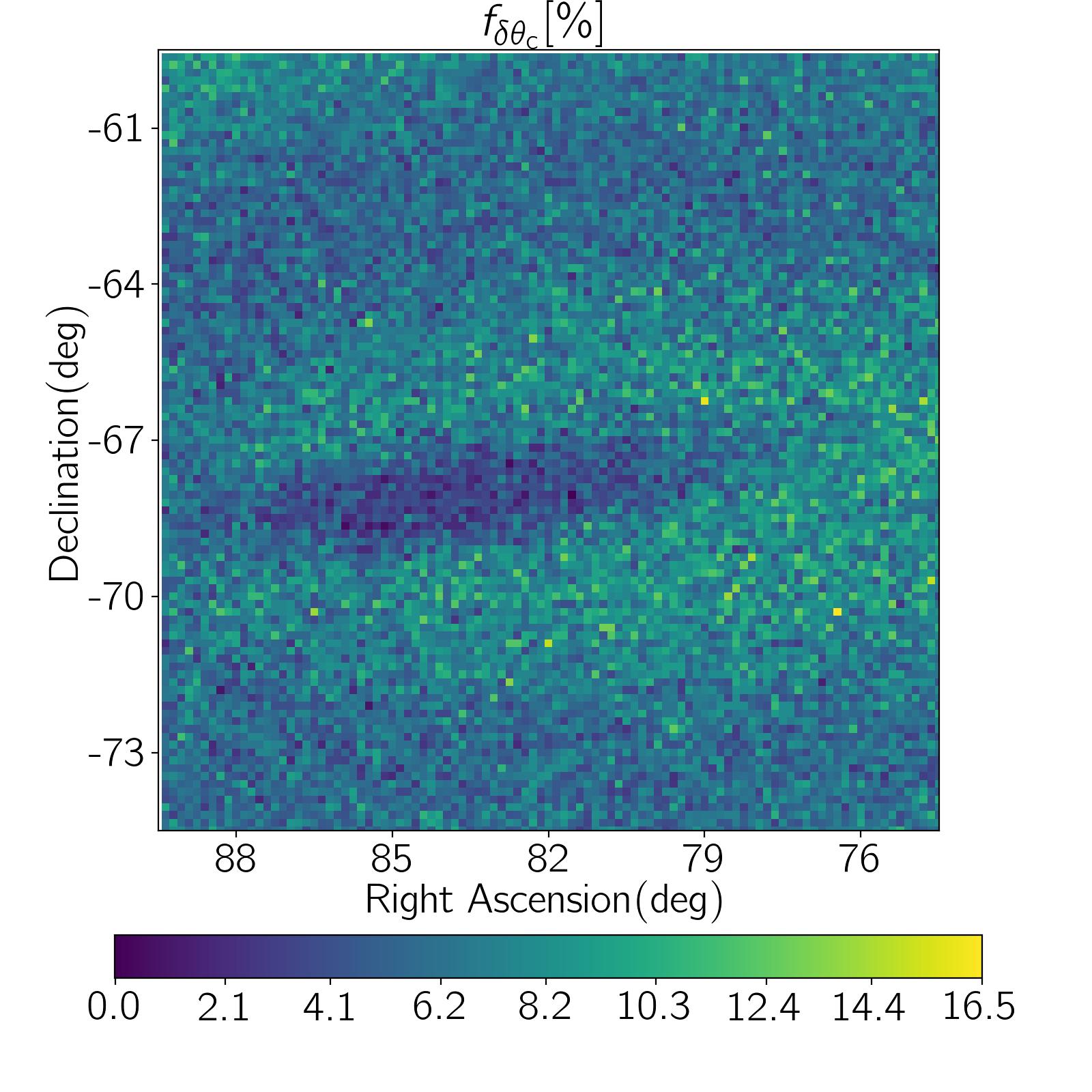}
\includegraphics[width=0.32\textwidth]{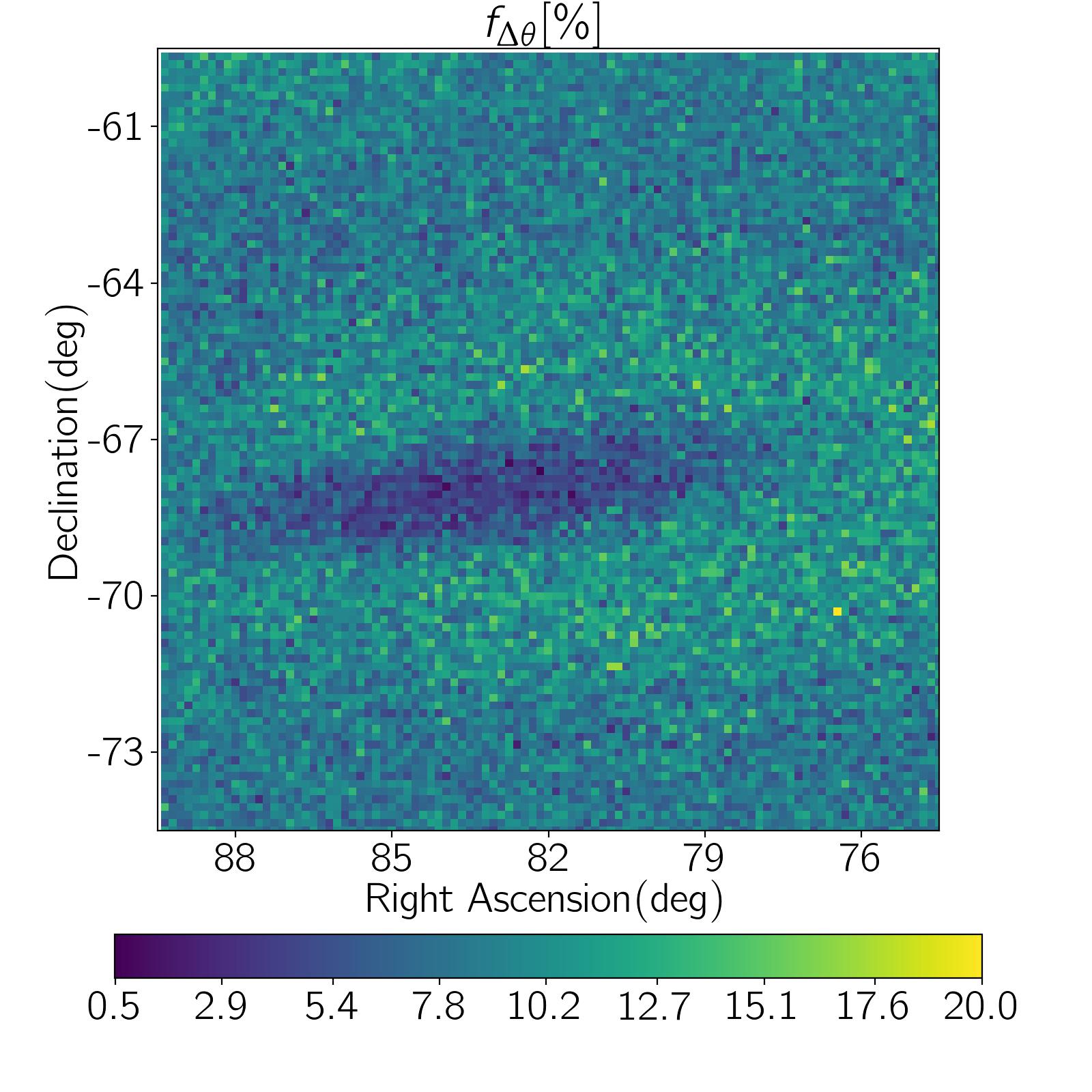}
\includegraphics[width=0.32\textwidth]{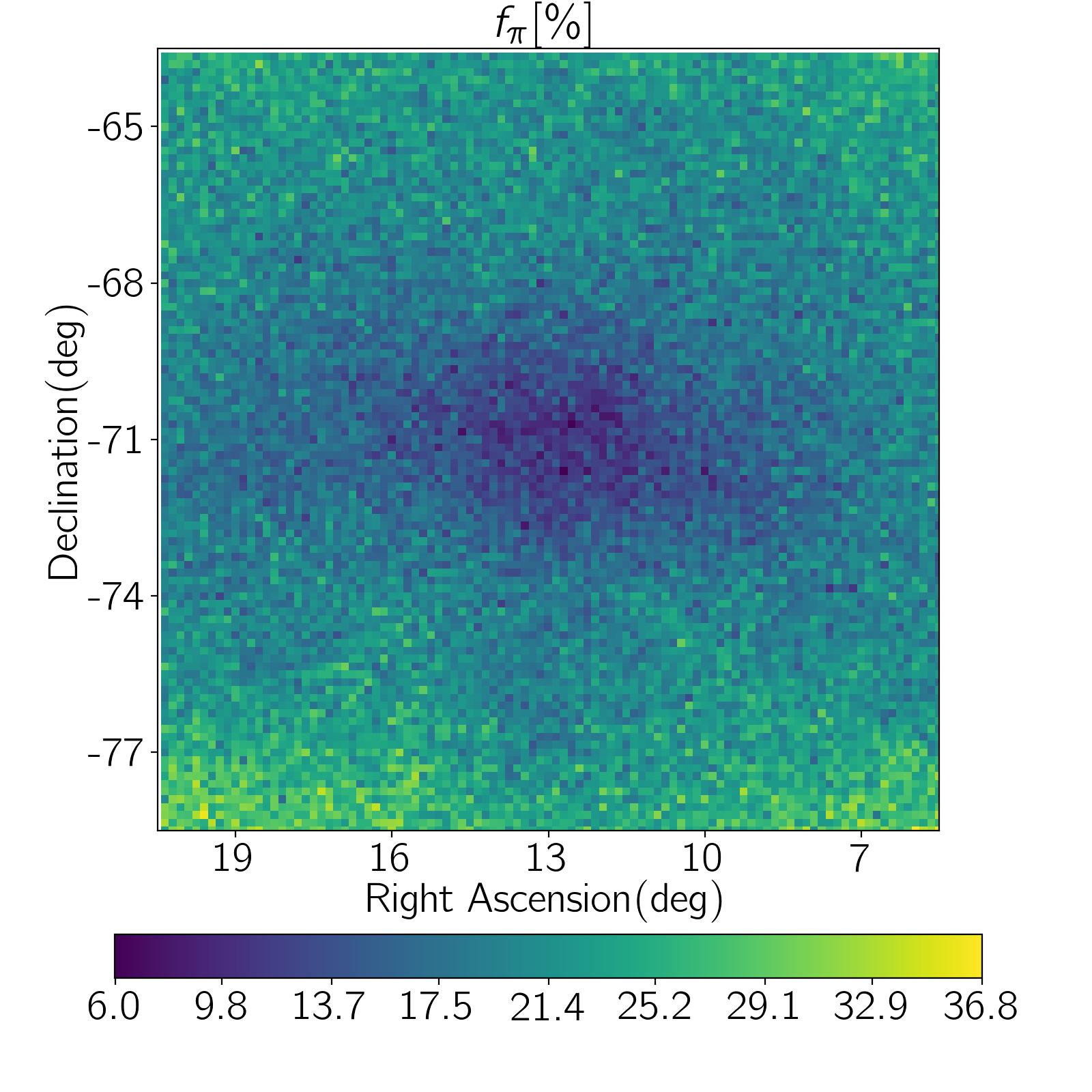}
\includegraphics[width=0.32\textwidth]{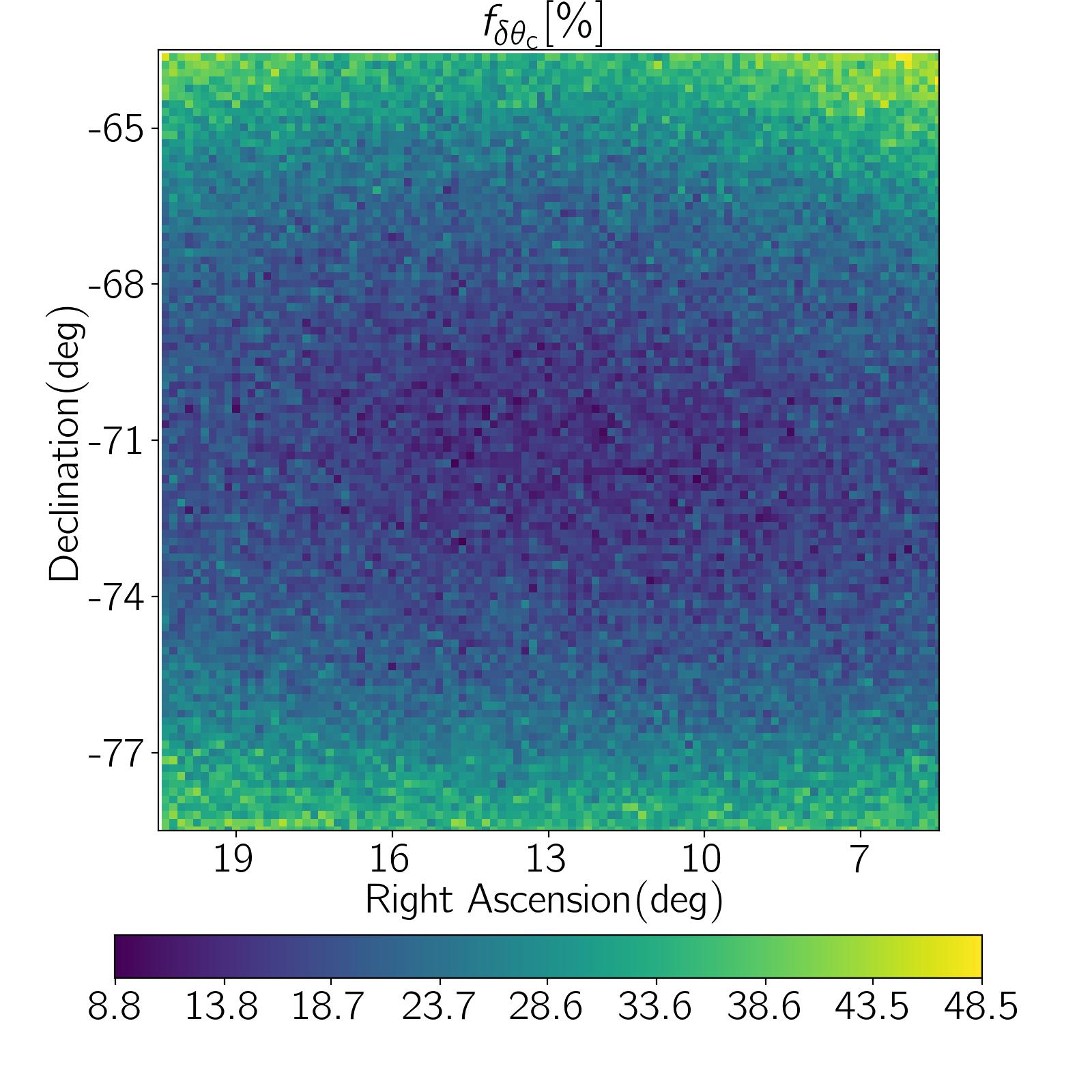}
\includegraphics[width=0.32\textwidth]{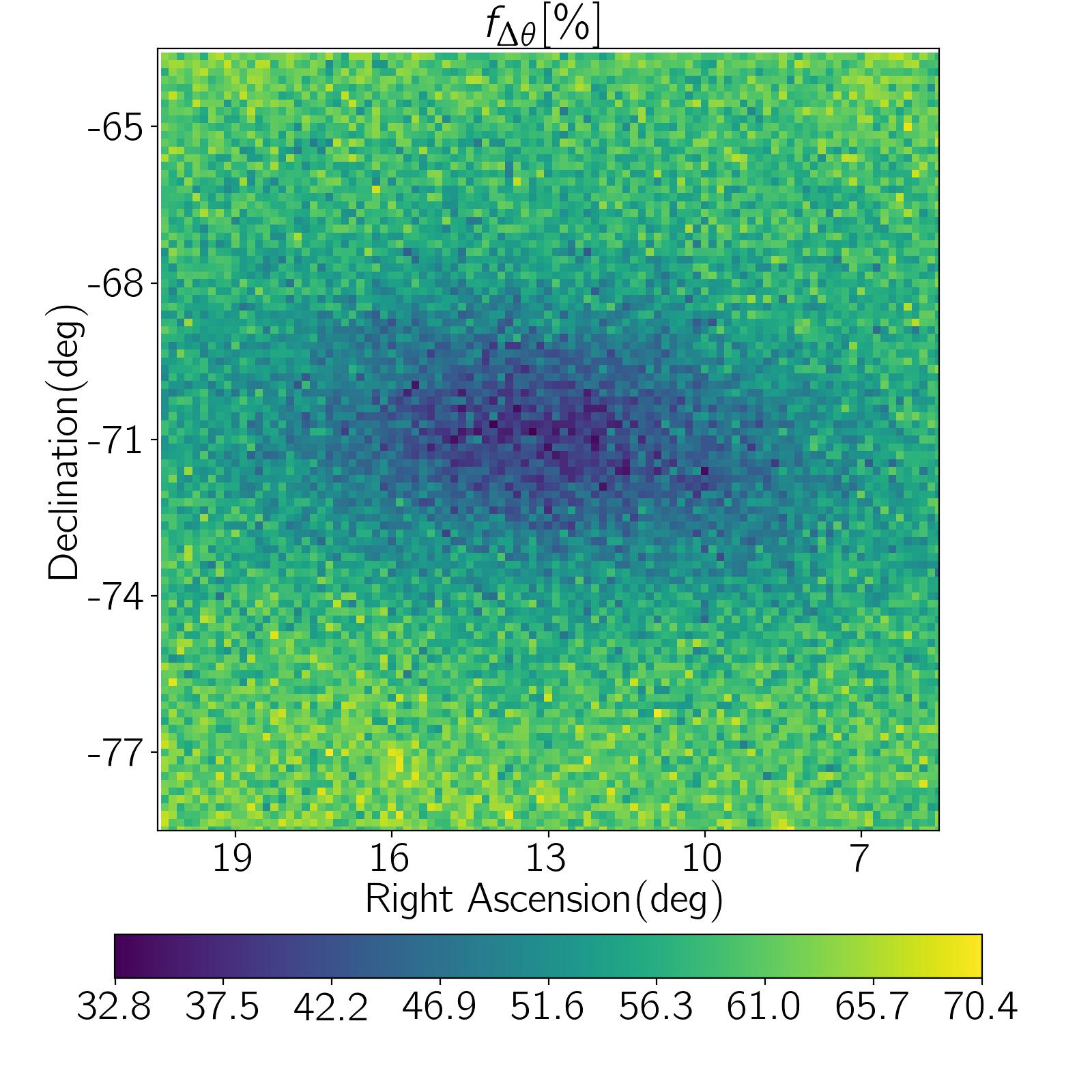}
\caption{From left to right, the maps represent $f_{\rm{\pi}}[\%]$ the fraction of photometrically detectable microlensing events in which parallax effects are realizable in their light curves, $f_{\delta\theta_{\rm{c}}}[\%]$ the fraction of detectable events with realizable lensing-induced astrometric deflections in their source trajectories, and $f_{\Delta\theta}[\%]$ the fraction of detectable microlensing events in which two images are resolvable in at least three data points ($N_{\Delta\theta}\ge3$). Top and bottom maps are based on simulations towards LMC and SMC, respectively.}\label{last}
\end{figure*}

According to Figure \ref{plotef}, the lensing detection efficiency maximizes for $t_{\rm E}\sim1-4$ years and low-mass BHs (i.e., $M_{\rm l}\lesssim100 M_{\odot}$) as lens objects. The lensing detection efficiency maximizes for $u_{0}\lesssim 0.5$. If IBHs' mass function has a slower slope with a lower power index, resulting microlensing events have longer time scales. To pass the detectability criterion $\rm{FWHM}\leq T_{\rm{obs}}$, these long-duration microlensing events should have smaller lens impact parameters. For that reason, the lensing detection efficiency maximizes for the smaller lens impact parameter for the IBHs mass density $\rm{MF}_{1}$ (green curves) in comparison with the mass density $\rm{MF}_{4}$ (red curves).

The lens-source relative velocities for self-lensing events are $\sim0.1-1\rm{mas}/\rm{years}$ while for halo-lensing events are $\sim1-100\rm{mas}/\rm{years}$. According to the last panel of Figure \ref{plotef}, the Rubin detection efficiency for self-lensing events is similar to that for halo-lensing events if IBHs mass density is $\rm{MF}_{4}$ (with power index $\beta=2$). For other mass functions the detection efficiency for halo-lensing events is noticeably higher than that for self-lensing ones. Its reason was explained beforehand as well. By taking the lens mass from the mass density $\rm{MF}_{4}$, simulated events have on average shorter time scales. In these events the source star should be bright with low photometric uncertainties to pass the first and second detectability criteria (I, II). The source stars towards the central parts of MCs are brighter than stars from MCs' halo parts. 

\noindent We also show this point in Figure \ref{selfhalo}. Here, the lensing detection efficiencies for self-lensing and halo-lensing events toward LMC (left part) and SMC (right part) by applying four lens mass densities are represented with the aid of different colors. Accordingly, detection efficiencies for halo- and self-lensing towards SMC are similar for $\rm{MF}_{4}$, while they are significantly different for other mass densities. 

Another parameter which significantly impacts on the lensing detection efficiency is the blending parameter $f_{\rm b}$, so that by changing the blending parameter from zero to one the detection efficiency improved from zero to $\sim 60\%$. In the next subsection, we study the Rubin efficiencies for detecting parallax effects, astrometric deflections in source trajectories, and two resolved images for photometrically detectable microlensing events.

\subsection{Rubin Efficiency for Discerning Parallax and Astrometry}\label{astrop}
As mentioned in Section \ref{sec3}, our applied detectability criteria (I, II, III) were based on the Rubin photometric observations. Meanwhile, we assumed that best-fitted models for astrometric source trajectories were the true ones, and only evaluated the errors in relevant parameters (e.g., $\theta_{\rm E}$) by evaluating Fisher and Covariance matrices. Here, we determine what fractions of these photometrically detected microlensing events (which pass those three detectability criteria) have (a) discernible parallax effects, (b) recognizable lensing-induced astrometric deflections in their source trajectories, and (c) their lensing-induced images are resolvable.

Towards MCs and for self-lensing events discerning the parallax effect in light curves is barely possible (see the third light curve in Figure \ref{fig1}), since these events have $\overline{\pi_{\rm{E}}}\sim10^{-3}$. The parallax amplitudes in halo-lensing events are on average $\overline{\pi_{\rm{E}}}\gtrsim10^{-2}$. Hence, the parallax effect can make detectable footprints in halo-lensing light curves. We evaluate the possibility of discerning the parallax effect in detectable microlensing light curves based on $\Delta\chi^{2}_{\pi}$, which is the difference between $\chi^{2}$ values from fitting true microlensing light curves to photometric data with and without considering the parallax effect. If $\Delta\chi^{2}_{\pi}\geq2N_{\rm{data}}$, the parallax effect is recognizable.
\begin{deluxetable}{c c c c c c c}
    \tablecolumns{7}
    \centering  
    \tablewidth{0.48\textwidth}\tabletypesize\footnotesize
    \tablecaption{Average some statistical parameters of photometrically detectable microlensing events due to IBHs by considering four lens mass functions in the Rubin observations towards LMC and SMC.\label{tabn}}
    \tablehead{\colhead{MF}&\colhead{$\overline{\Delta\chi^{2}_{\pi}\big/N_{\rm{data}}}$}&\colhead{$\overline{\Delta\chi^{2}_{\delta\theta_{\rm{c}}}\big/N_{\rm{data}}}$}&\colhead{$\overline{N_{\Delta\theta}}$}&\colhead{$f_{\pi}$}&\colhead{$f_{\delta\theta_{\rm{c}}}$}&\colhead{$f_{\Delta\theta}$}\\
        &&&&$[\%]$&$[\%]$&$[\%]$}
    \startdata	
    \multicolumn{7}{c}{Simulations towards LMC}\\
    $\rm{MF}_{1}$ & $196.2\pm224.2$ & $1.27\pm1.65$ & $6.1\pm3.4$ & $48.3$ & $6.5$ & $8.9$\\
    $\rm{MF}_{2}$ & $236.6\pm230.7$ & $0.35\pm0.33$ & $1.6\pm1.0$ & $42.2$ & $2.4$ & $2.8$\\
    $\rm{MF}_{3}$ & $299.3\pm280.0$ & $0.13\pm0.16$ & $0.7\pm0.5$ & $41.2$ & $1.0$ & $1.4$\\
    $\rm{MF}_{4}$ & $397.2\pm355.0$ & $0.01\pm0.01$ & $0.1\pm0.2$ & $41.9$ & $0.0$ & $0.5$\\
    \tableline
    \multicolumn{7}{c}{Simulations towards SMC}\\
    $\rm{MF}_{1}$ & $28.4\pm28.4$ & $6.08\pm3.62$ & $49.5\pm13.7$ & $20.1$ & $22.6$ & $55.6$\\
    $\rm{MF}_{2}$ & $32.3\pm27.5$ & $1.85\pm1.17$ & $20.5\pm5.9$ & $21.2$ & $11.0$ & $30.6$\\
    $\rm{MF}_{3}$ & $40.8\pm30.2$ & $0.79\pm0.59$ & $9.4\pm2.9$ & $24.8$ & $5.2$ & $16.3$\\
    $\rm{MF}_{4}$ & $58.1\pm38.8$ & $0.04\pm0.05$ & $0.5\pm0.5$ & $33.2$ & $0.2$ & $1.6$\\
    \enddata
    \tablecomments{$N_{\Delta\theta}$ represents the number of data points in which both images have the apparent magnitudes less than the Rubin detection threshold, and (2) their angular separation is comparable with the Rubin astrometric accuracy to resolve two close stars.}
\end{deluxetable}

\noindent For realizing astrometric deflections in source trajectories, we calculate $\Delta\chi^{2}_{\delta\theta_{\rm{c}}}$ which is the difference between $\chi^{2}$ values from fitting true source trajectories with and without considering the lensing-induced deflections (i.e., $\boldsymbol{\theta}_{\star}$ and $\boldsymbol{\theta}_{\star}-\boldsymbol{\delta\theta}_{\rm c}$) to the synthetic astrometric data. We assume that the events with $\Delta \chi^{2}_{\delta\theta_{\rm{c}}}\geq2N_{\rm{data}}$ have recognizable astrometric deflections in their source trajectories.

\noindent To evaluate the possibility of detecting two separated lensing-induced images for each event, we calculate the number of data points ($N_{\Delta\theta}$) in which (1) two images are detectable, i.e., their apparent magnitudes (Equation \ref{appm}) are less than the Rubin detection threshold, and (2) their angular separation is comparable with the Rubin astrometric accuracy to resolve two close stars as $\Delta\theta(=\theta_{\rm E}\sqrt{u^{2}+4})\geq1\sigma_{\rm{r}}$. Here, $\sigma_{\rm{r}}=\mathcal{D}\sigma_{\rm{a}}$ is the Rubin astrometric accuracy to resolve two close stars and the factor $\mathcal{D}\in[5,~100]$ for SNR$\in [5,~1000]$. At the faint detection limit with SNR$=5$, this factor is $\sim5$ while it reaches to $\sim20$ for SNR$=100$, if two stars have similar magnitudes. $\mathcal{D}$ will be even higher by $40\%$ and three times for two close stars by different magnitudes with differences $\in[2,~3],~\rm{and} \ge3$ mag, respectively (\v{Z}.~Ivezi\'{c}, personal communication, June, 2026). We assume resolving two lensing-induced images is possible, if at least three recorded data points meet the mentioned conditions, i.e., $N_{\Delta\theta}\ge3$.

In Figure \ref{last} we show the maps of $f_{\pi}[\%]$ which represents the fraction of photometrically detected events with recognizable parallax effects based on $\Delta\chi^{2}_{\pi}\geq2N_{\rm{data}}$, $f_{\delta\theta_{\rm{c}}}[\%]$ which shows the fraction of detectable microlensing events with $\Delta\chi^{2}_{\delta\theta_{\rm{c}}}\geq2N_{\rm{data}}$, and $f_{\Delta\theta}[\%]$ the fraction of events with resolvable images in at least three data points ($N_{\Delta\theta}\ge3$), respectively. In this figure, top and bottom maps are based on simulations by adopting the IBHs mass function MF$_{1}$ ($\beta=0$) towards LMC and SMC, respectively.

According to these maps, the probability of discerning parallax effects in halo-lensing events is high and usually $\gtrsim20\%$, while this probability for self-lensing events is low and mostly $\lesssim10\%$. Based on the fifth panels of Figure \ref{map1} and \ref{map2} most of microlensing events towards the MCs' central parts are self-lensing events, and almost all of microlensing events away from the MCs' central parts are halo-lensing ones. On average, discerning astrometric deflections in projected source trajectories for $\gtrsim 25\%$ of halo-lensing and $\lesssim10\%$ of self-lensing events (towards the MCs' central parts) is possible. We note that towards SMC, $\theta_{\rm{E}}$ values are on average somewhat higher (see Table \ref{tab1}), so towards SMC $f_{\delta\theta_{\rm{c}}}$ values are higher that those in the LMC microlensing events.

\begin{figure*}
    \centering
    \includegraphics[width=0.49\textwidth]{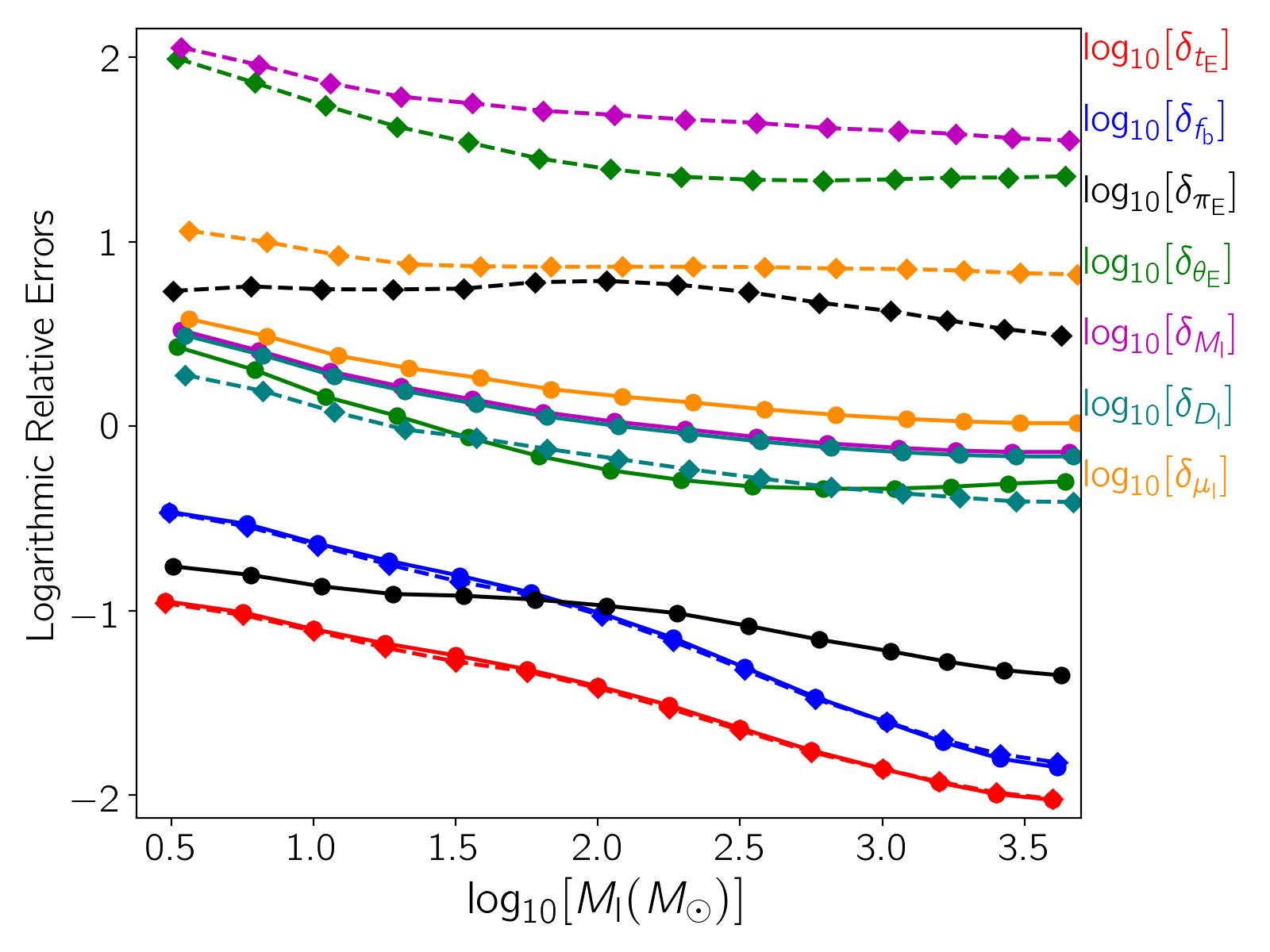}
    \includegraphics[width=0.49\textwidth]{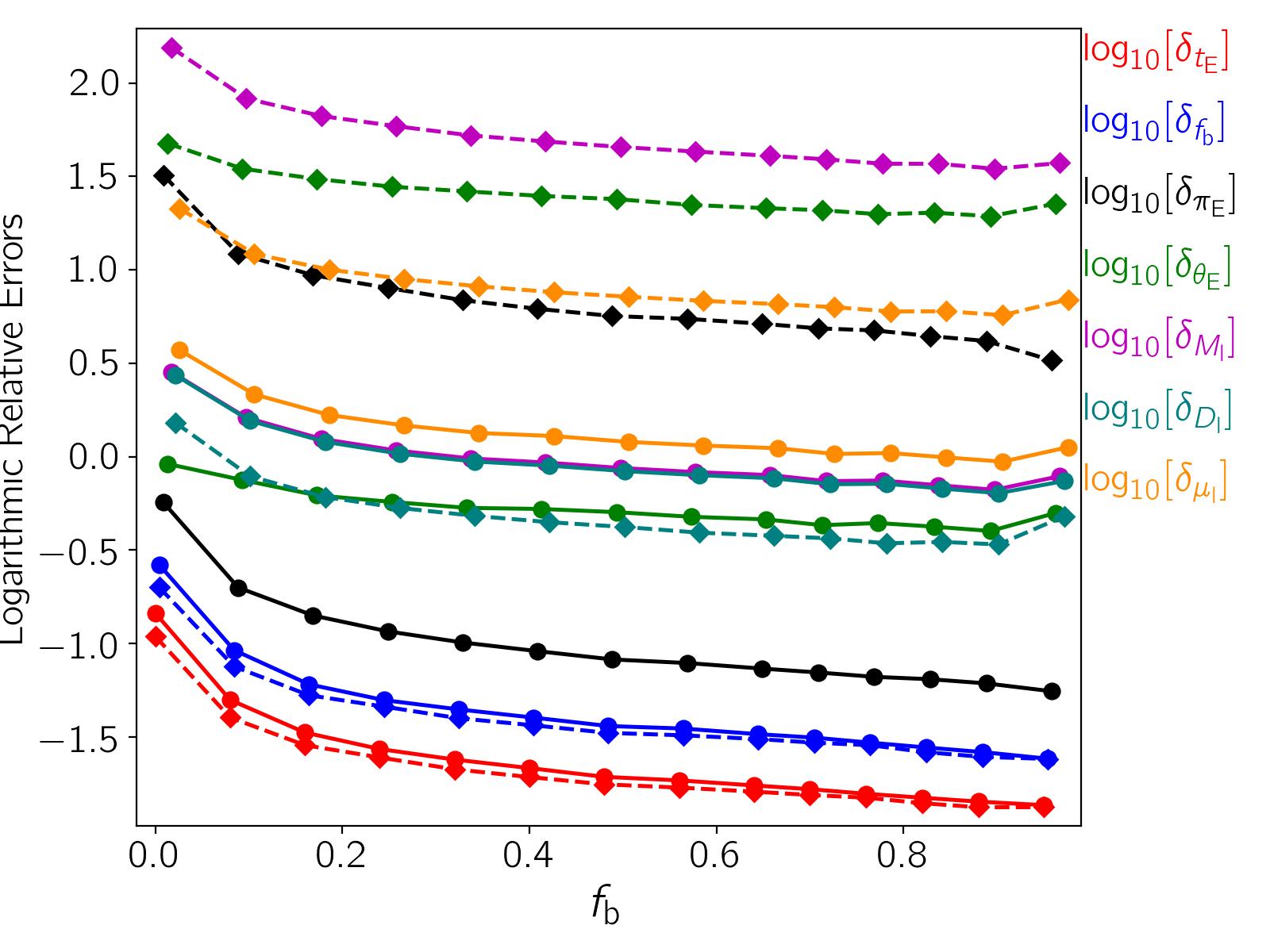}
    \includegraphics[width=0.49\textwidth]{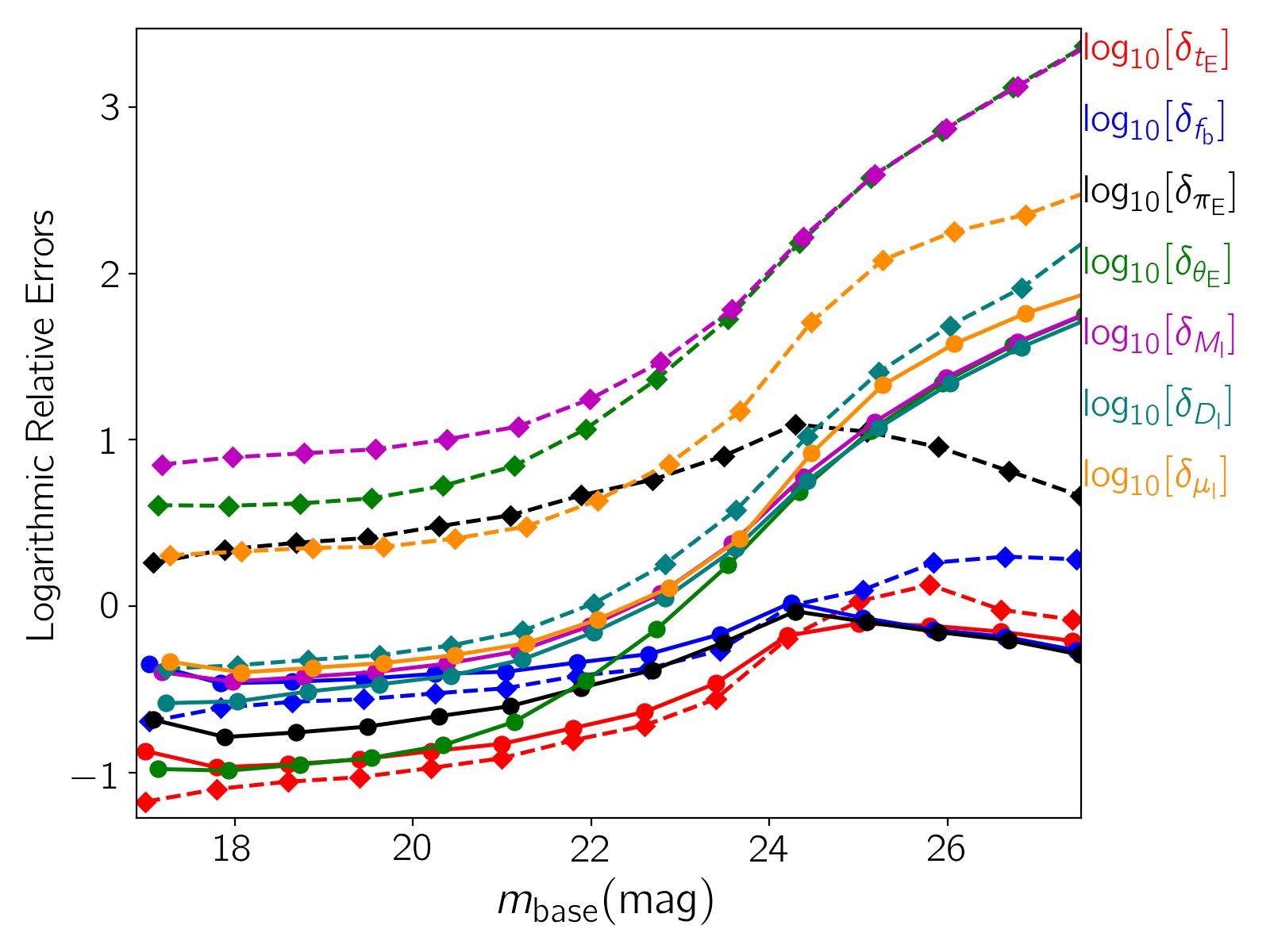}
    \includegraphics[width=0.49\textwidth]{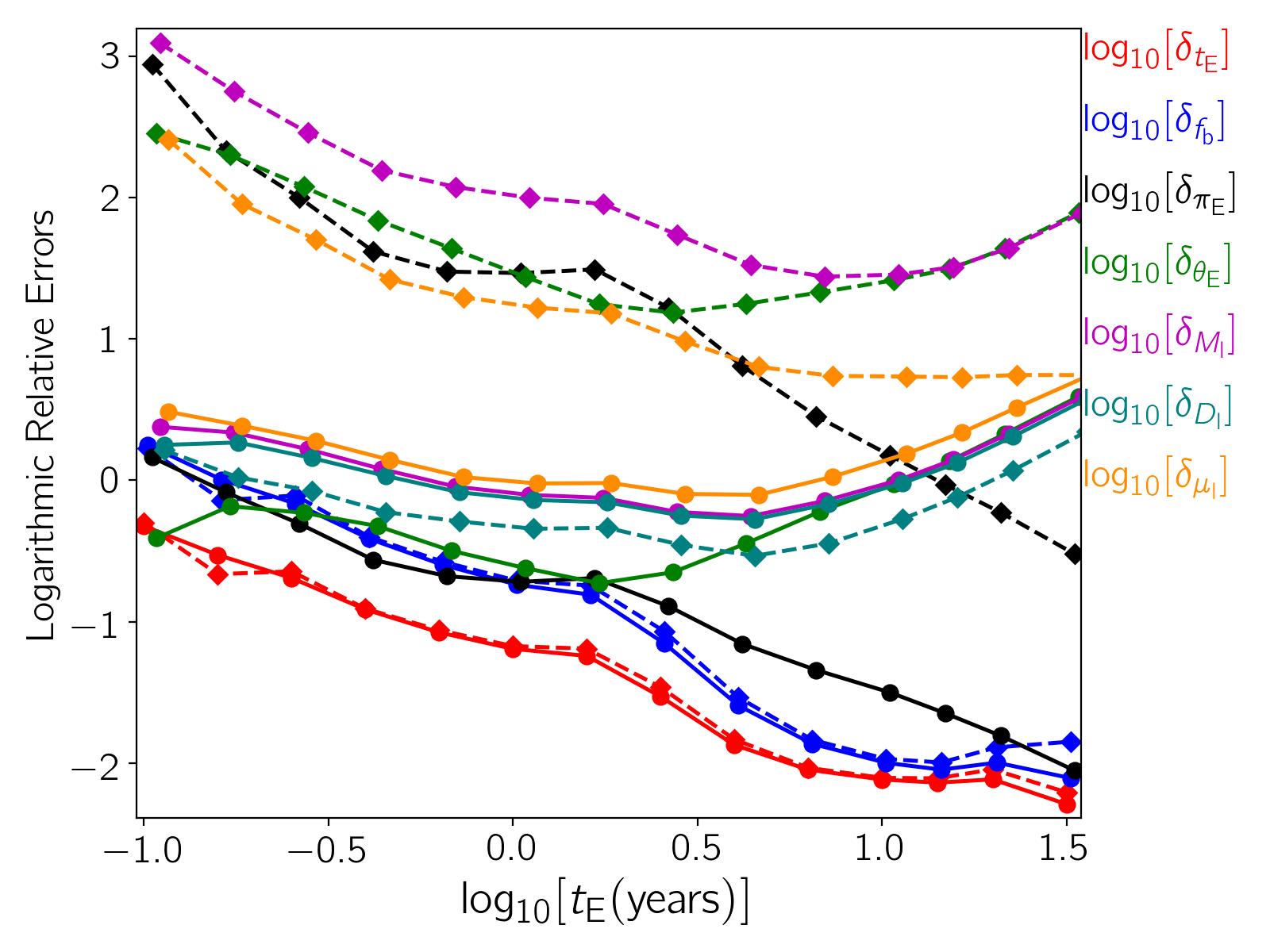}
    \caption{The average relative errors in seven lensing and physical parameters (in the logarithmic scale) which are specified with different colors over detectable microlensing events versus four parameters including $\log_{10}[M_{\rm{l}}(M_{\odot})]$, $f_{\rm{b}}$, $m_{\rm{base}}(\rm{mag})$, and $\log_{10}[t_{\rm{E}}(\rm{years})]$. The relative errors due to detectable halo-lensing and self-lensing events are shown by solid curves with circle symbols and dashed curves with diamond symbols, respectively.}\label{Error1}
\end{figure*}

\noindent Comparing the second and third (up and down) panels of Figure \ref{last} reveals that the probability of resolving two images (in at least three data points) in photometrically detectable microlensing events due to IBHs \textit{towards the SMC} $f_{\Delta\theta}$ is more than the probability of measuring the lensing-induced astrometric deflections for such events $f_{\delta\theta_{\rm{c}}}$. Because, discerning the astrometric deflections depends strongly on the number of data points (its criterion is $\Delta\chi^{2}_{\delta\theta_{\rm{c}}}\geq2N_{\rm{data}}$), while for resolving two images we demand that two lensing-induced images be resolvable in at least only three data points. The number of visits towards SMC is lower than the number of visits towards  the LMC by $\sim 60$, so that discerning astrometric deflection based on that criterion is harder than resolving two images. Both resolving images and discerning the astrometric deflections are helpful for measuring $\theta_{\rm E}$.

\begin{figure*}
    \centering
    \includegraphics[width=0.49\textwidth]{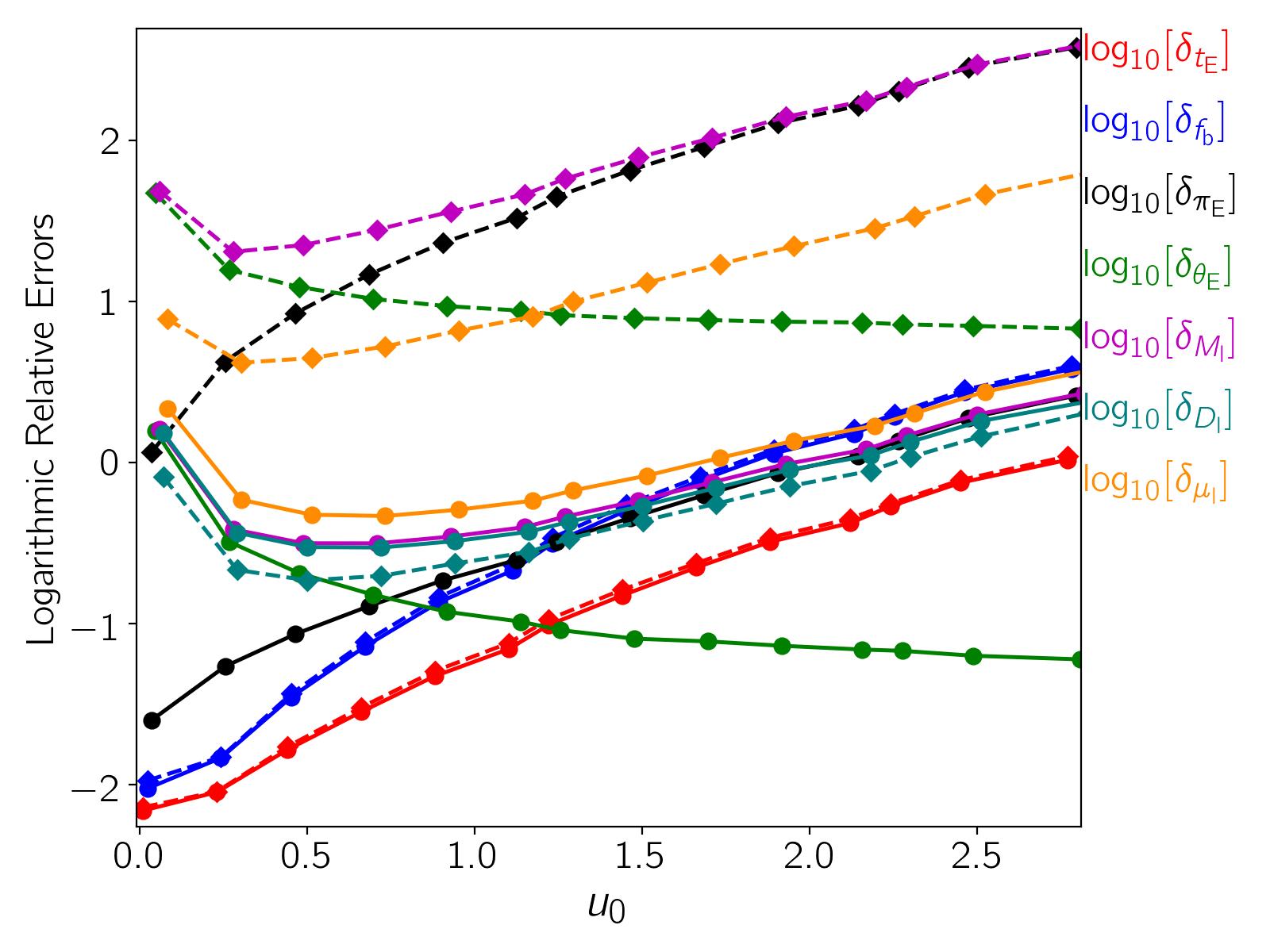}
    \includegraphics[width=0.49\textwidth]{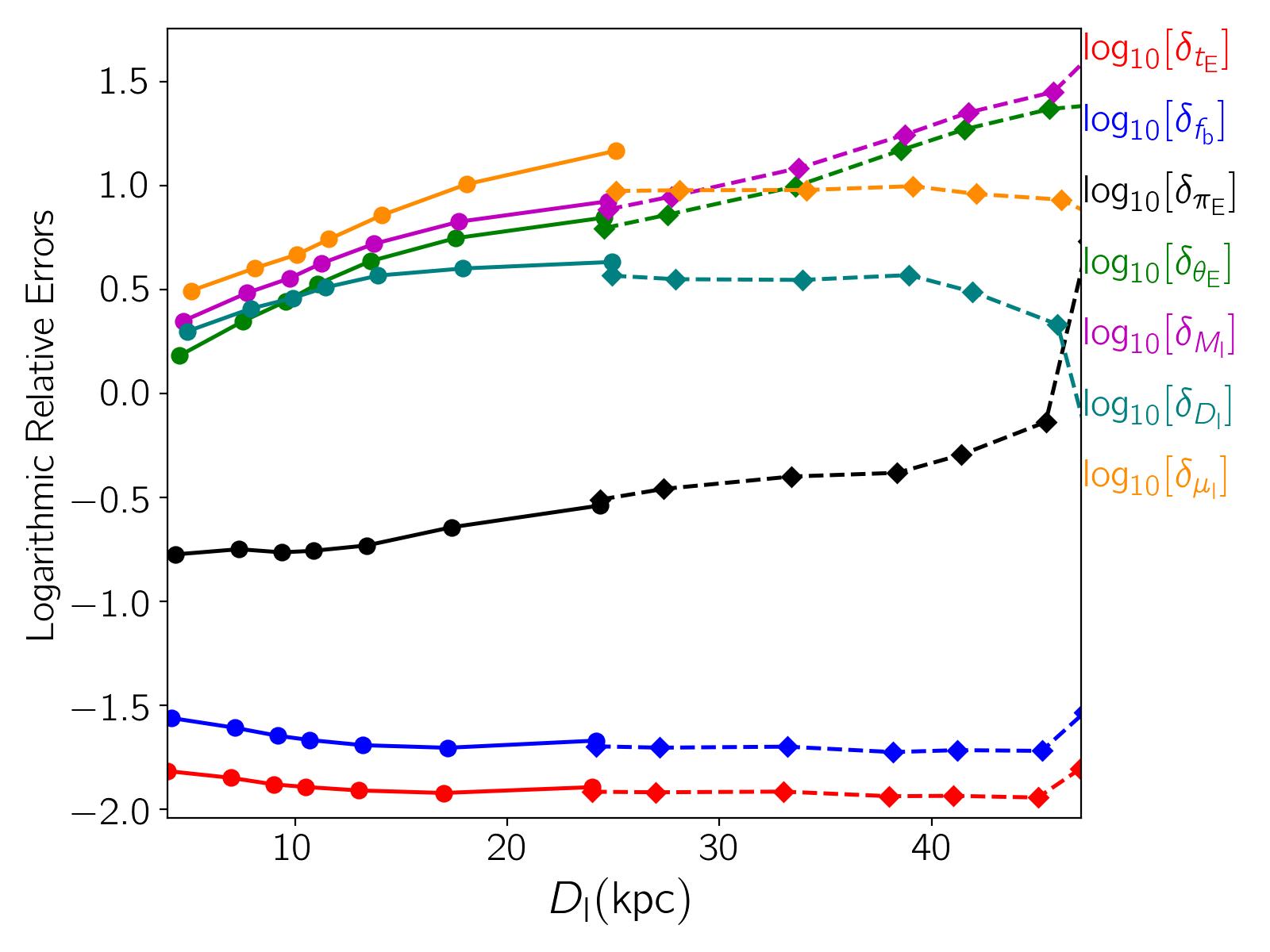}
    \includegraphics[width=0.49\textwidth]{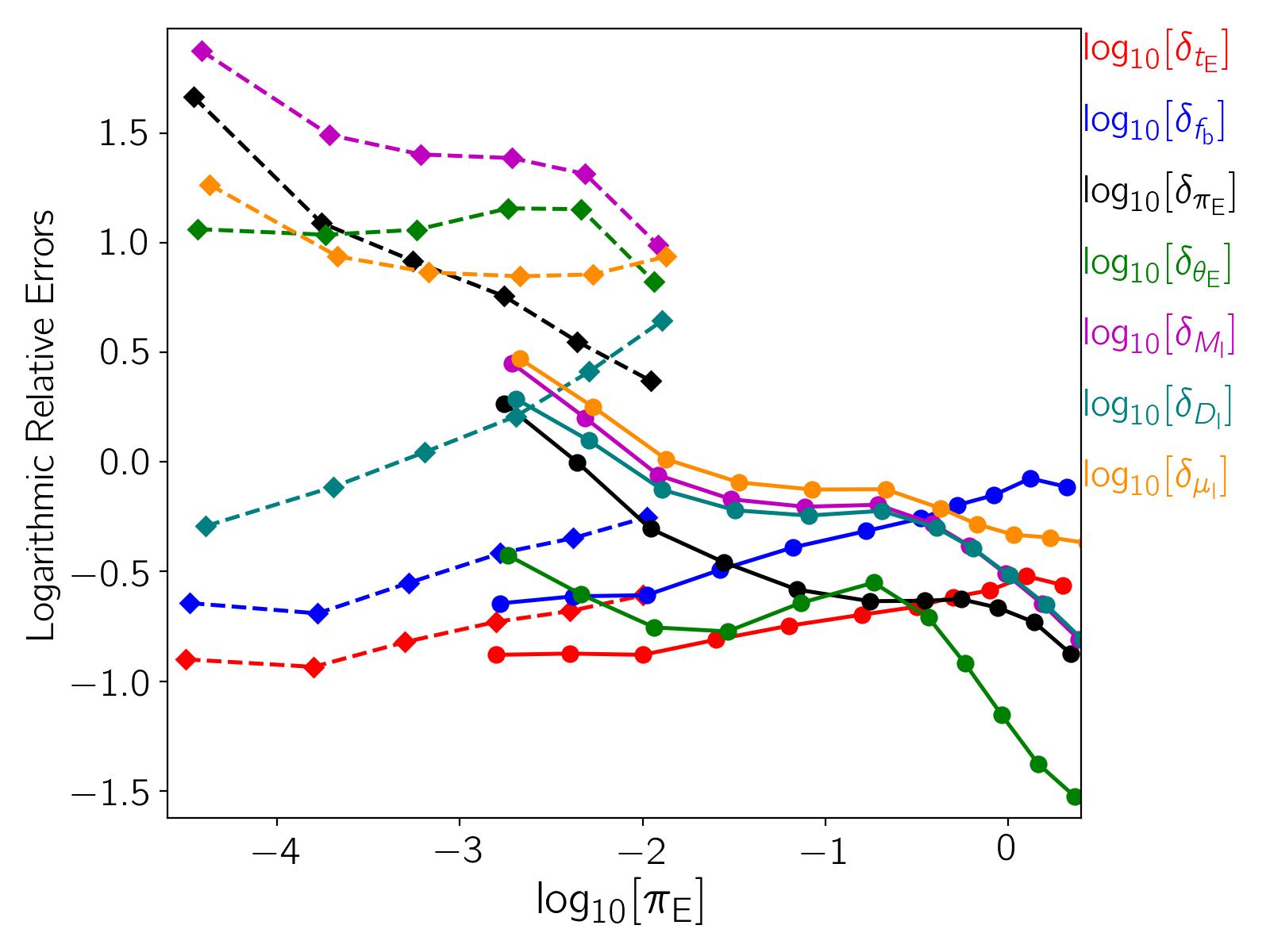}
    \includegraphics[width=0.49\textwidth]{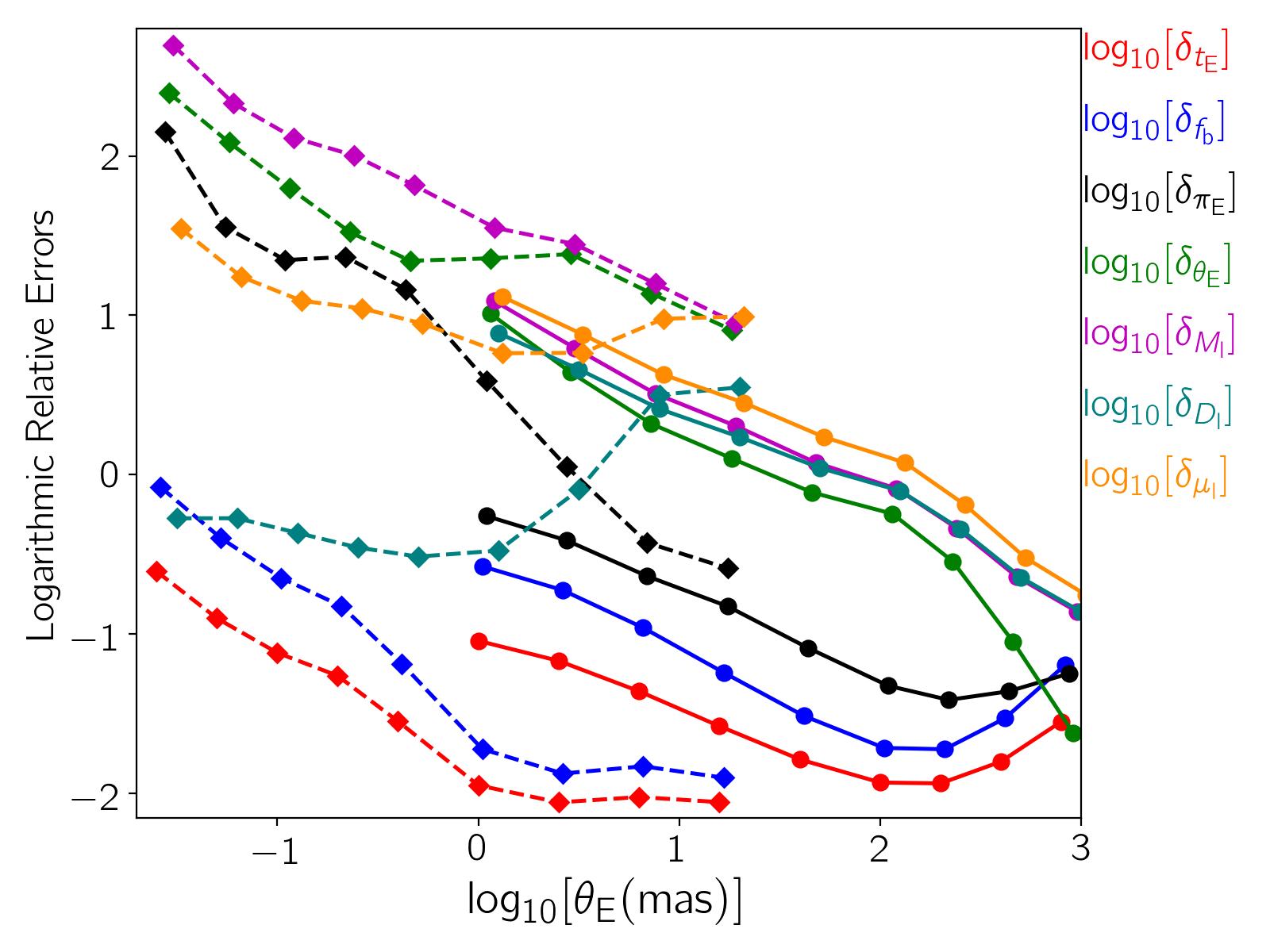}
    \caption{Same as Figure \ref{Error1} but versus four other parameters.}\label{Error2}
\end{figure*}

In Table \ref{tabn}, we report the average values of $\Delta\chi^{2}_{\pi}\big/N_{\rm{data}}$, $\Delta\chi^{2}_{\delta\theta_{\rm{c}}}\big/N_{\rm{data}}$, $N_{\Delta\theta}$, $f_{\pi}[\%]$, $f_{\delta\theta_{\rm{c}}}[\%]$, and $f_{\Delta\theta}[\%]$ over the 2D maps of detectable lensing events towards LMC and SMC resulting from simulations by applying different lens mass functions. Accordingly, by increasing the power index of the lens mass function ($\beta$) the probability of measuring the parallax effect increases while both probabilities of realizing astrometric deflections and resolving two images considerably decrease. In fact, the parallax amplitude decreases by the lens mass while $\theta_{\rm E}$ enhances by the lens mass. By increasing the power index $\beta$ the average value of the lens mass decreases.

In real observations, discerning lensing-induced astrometric deflections through the Rubin observations will be likely challenging, because of the blending effects and motions of blending stars. In simulations, we did not consider astrometric fluctuations due to blending stars's motions while generating source trajectories ($\boldsymbol{\theta}_{\star}$ given by Equation \ref{tets}). However, by applying the second criterion for selecting detectable source stars (ii), high-blended source stars are removed from our sample and labeled as undetectable source stars. This point can be found from the third panels of Figure \ref{map1} and \ref{map2}. In these maps the blending parameters of detectable stars are on average higher than $\sim0.6$.

\subsection{Relative Errors Evaluation}\label{error}
For all photometrically detected microlensing events we estimate the errors in their parameters by calculating Fisher and Covariance matrices numerically. There are two Fisher matrices related to the photometric $\boldsymbol{\mathcal{A}}$ and astrometric $\boldsymbol{\mathcal{B}}$ observing data. The microlensing light curves depend on 7 parameters including $p_{k}\in u_{0},~t_{0},~t_{\rm E},~f_{\rm b},~\pi_{\rm{E}}$, $\xi$,~$m_{\rm{base}}$; for $k=\{1, 2, ..., 7\}$ (we ignore the finite-source size in all simulations). We put aside two un-relevant parameters $t_{0}$ and $m_{\rm{base}}$ to speed up the calculations. These two parameters do not change the derived lens physical parameters. The astrometric source trajectory is a function of four parameters containing $p_{k}\in \theta_{\rm{E}}$, $\mu_{\star, 1}$, $\mu_{\star, 2}$, $\pi_{\rm{E}}$; for $k=\{1, 2, 3, 4\}$. The parallax effect changes both light curves and astrometric trajectories. We determine the errors in the parallax amplitude from both photometric and astrometric observations, and take the smaller error. The $ij$th element of these two Fisher matrices are given by:
\begin{eqnarray}
\boldsymbol{\mathcal{A}}_{ij}&=& \sum_{k=1}^{N_{\rm{data}}}\frac{1}{\sigma_{\rm p, k}^{2}}\frac{\partial^{2}A(t_{\rm k})}{\partial p_{i}\partial p_{j} }, \nonumber \\
\boldsymbol{\mathcal{B}}_{ij}&=& \sum_{k=1}^{N_{\rm{data}}}\frac{1}{2\sigma_{\rm a, k}^{2}}\Big(\frac{\partial^{2}\theta_{\star, 1}(t_{\rm k})}{\partial p_{i}\partial p_{j} } + \frac{\partial^{2}\theta_{\star, 2}(t_{\rm k})}{\partial p_{i}\partial p_{j} }\Big), 
\label{fisher}
\end{eqnarray}
where, $N_{\rm{data}}$ is the number of data points, $A(t_{\rm{k}})$ is the magnification factor at the time $t_{\rm{k}}$, $\theta_{\star, 1}$ and $\theta_{\star, 2}$ are two components of the angular source position (given by Equation \ref{tets}) at the time $t_{\rm k}$. We numerically calculate the inverses of Fisher matrices $\boldsymbol{\mathcal{A}}^{-1}$, and $\boldsymbol{\mathcal{B}}^{-1}$ which are the so-called Covariance matrices, then one can determine the errors as the square root of diagonal elements of covariance matrices, i.e., $\sqrt{\boldsymbol{\mathcal{A}}^{-1}_{ii}}$ and $\sqrt{\boldsymbol{\mathcal{B}}^{-1}_{ii}}$ \citep[see, e.g.,][]{2023AJSajadianSahu,2015AJsajadian}. We report the relative errors which are the estimated error in a given parameter normalized to its true value. For instance, the relative error in the lensing time scale $t_{\rm E}$ is defined as $\delta_{t_{\rm{E}}}=\sigma_{t_{\rm{E}}}/t_{\rm{E}}$, where $\sigma_{t_{\rm{E}}}$ is the error while extracting $t_{\rm{E}}$ derived from the diagonal elements of Covariance matrices. 

In Figure \ref{Error1} and \ref{Error2}, we depict the average values of the relative errors in seven lensing and physical parameters in the logarithmic scale versus eight relevant parameters in different panels. The relative errors contain $\delta_{t_{\rm E}}$ (dark red color), $\delta_{f_{\rm{b}}}$ (dark blue color), $\delta_{\pi_{\rm E}}$ (black color), $\delta_{\theta_{\rm E}}$ (dark green color), $\delta_{M_{\rm l}}$ (magenta color), $\delta_{D_{\rm l}}$ (teal color), and $\delta_{\mu_{\rm l}}$ (orange color), and are shown by solid curves with filled circles (related to halo-lensing events) and dashed curves with filled diamonds (related to self-lensing events).

The first panel of Figure \ref{Error1} represents the average values of relative errors versus the lens mass. Accordingly, for halo-lensing events enhancing the lens mass decreases all relative errors. Halo-lensing events due to more massive lens objects have longer durations, closer lens objects with larger parallax amplitudes and larger angular Einstein radii.  All of these changes cause lower relative errors. For self-lensing events, the parallax amplitudes become smaller for more massive lens objects which makes higher relative errors in parallax, lens mass and its velocity. The relative errors in $\theta_{\rm E}$ for self-lensing events is higher than those for halo-lensing events by $1-2$ orders of magnitude owing to the large difference of their $\pi_{\rm{rel}}$ values.

The second panel of Figure \ref{Error1} shows the relative errors versus the blending parameter. All relative errors decrease with the blending parameter, because the events with higher blending parameters generally have brighter source stars with lower photometric and astrometric uncertainties. The relative errors depend on the baseline magnitudes as represented in the third panel of Figure \ref{Error1}. Accordingly, the relative errors enhance for fainter source stars at the baseline as one expects.  

Th last panel of Figure \ref{Error1} displays the relative errors versus the Einstein crossing time. All relative errors (except $\delta_{\theta_{\rm E}}$ and as a result the lens mass, its distance and its proper velocity) decrease in longer events, because the number of data points will be more in longer events. However, for very long-duration microlensing events the astrometric deflections in the source trajectories are rather straight lines (see, e.g., the astrometric deflection due to first microlensing event shown in Figure \ref{fig1}) which makes hard discerning their amplitudes ($\theta_{\rm{E}}$). Also, very long duration microlensing events happens mostly when $D_{\rm l}\sim D_{\rm s}$ which have very small lens-source relative velocities. These types of microlensing events have very small $\theta_{\rm E}$ values with large relative errors.

In the first panel of Figure \ref{Error2}, the relative errors are depicted versus the lens impact parameter. Accordingly all errors (except $\delta_{\theta_{\rm E}}$) increase with the lens impact parameter (or lower magnification factor). Unlikely, the relative errors in the angular Einstein radius decrease with the lens impact parameter, because the astrometric deflection changes from a straight line (a part of an elongated ellipse) to a circle by increasing the lens impact parameter. 

The next panel of Figure \ref{Error2} shows the relative errors versus the lens distance from the observer (resulting from simulations towards LMC). For this parameter halo-lensing and self-lensing events are separated. Generally, self-lensing events have larger relative errors than halo-lensing events. Self-lensing events have smaller $\pi_{\rm E}$, and $\theta_{\rm E}$ values than halo-lensing events, although they have similar time scales.   By increasing the lens distance almost all relative errors increase except the Einstein crossing time and the blending parameter.


Two next panels of Figure \ref{Error2} show the relative errors versus the parallax amplitude and the angular Einstein radius. Both of these two parameters in self-lensing events have lower orders of magnitude than those in halo-lensing events. Accordingly, solid and dashed curves are separated from each other. Generally in self-lensing events the relative errors are higher than those in halo-lensing ones.
\begin{deluxetable}{c c c c c c c c c c}
\tablecolumns{10}
\centering
\tablewidth{0.48\textwidth}\tabletypesize\footnotesize
\tablecaption{The fractions of photometrically detectable microlensing events with the relative errors in given parameters (different columns) less than the given thresholds ($\mathcal{T}$) by considering four lens mass densities ($\rm{MF}$s) in simulations towards LMC and SMC (different rows). \label{tab4}}
\tablehead{\colhead{MF}&\colhead{$\mathcal{T}$}&\colhead{$\delta_{t_{\rm E}}$}&\colhead{$\delta_{f_{\rm{b}}}$}&\colhead{$\delta_{\pi_{\rm{E}}}$}&\colhead{$\delta_{\theta_{\rm{E}}}$}&\colhead{$\delta_{M_{\rm{l}}}$}&\colhead{$\delta_{D_{\rm{l}}}$}&\colhead{$\delta_{\mu_{\rm{l}}}$} &\colhead{$\delta_{\rm{lens}}$}\\
& $[\%]$ & $[\%]$ &$[\%]$ &$[\%]$ &$[\%]$ &$[\%]$ &$[\%]$ &$[\%]$ & $[\%]$}
\startdata  
\multicolumn{10}{c}{Simulations towards LMC}\\ 
$\rm{MF}_{1}$ & $3$ & $76.4$ & $64.9$ & $22.8$ & $6.9$ & $0.7$ & $1.4$ & $0.4$ & $0.4$\\
$\rm{MF}_{2}$ & $3$ & $57.9$ & $44.1$ & $20.4$ & $4.3$ & $0.3$ & $0.5$ & $0.1$ & $0.1$\\
$\rm{MF}_{3}$ & $3$ & $45.9$ & $31.9$ & $18.0$ & $2.5$ & $0.1$ & $0.2$ & $0.1$ & $0.1$\\
$\rm{MF}_{4}$ & $3$ & $30.8$ & $17.7$ & $15.0$ & $0.4$ & $0.0$ & $0.0$ & $0.0$ & $0.0$\\
\tableline
$\rm{MF}_{1}$ & $6$ & $84.4$ & $75.4$ & $37.0$ & $12.5$ & $4.0$ & $7.6$ & $2.3$ & $2.2$\\
$\rm{MF}_{2}$ & $6$ & $68.4$ & $55.1$ & $32.5$ & $8.8$ & $1.8$ & $3.1$ & $1.0$ & $0.9$\\
$\rm{MF}_{3}$ & $6$ & $57.5$ & $42.5$ & $28.4$ & $5.6$ & $0.8$ & $1.5$ & $0.5$ & $0.4$\\
$\rm{MF}_{4}$ & $6$ & $43.4$ & $27.3$ & $24.0$ & $1.5$ & $0.1$ & $0.2$ & $0.1$ & $0.0$\\
\tableline
$\rm{MF}_{1}$ & $10$ & $88.5$ & $80.7$ & $46.0$ & $17.7$ & $8.9$ & $16.4$ & $5.6$ & $5.5$\\
$\rm{MF}_{2}$ & $10$ & $75.0$ & $62.0$ & $41.0$ & $13.3$ & $4.7$ & $7.9$ & $2.8$ & $2.7$\\
$\rm{MF}_{3}$ & $10$ & $65.3$ & $49.8$ & $36.3$ & $9.1$ & $2.5$ & $4.3$ & $1.5$ & $1.4$\\
$\rm{MF}_{4}$ & $10$ & $52.5$ & $34.7$ & $31.6$ & $3.3$ & $0.4$ & $0.7$ & $0.3$ & $0.2$\\
\tableline
$\rm{MF}_{1}$ & $20$ & $92.9$ & $86.3$ & $55.4$ & $25.3$ & $17.8$ & $31.2$ & $13.0$ & $12.8$\\
$\rm{MF}_{2}$ & $20$ & $82.9$ & $70.2$ & $51.2$ & $20.6$ & $11.4$ & $18.5$ & $7.9$ & $7.7$\\
$\rm{MF}_{3}$ & $20$ & $75.4$ & $59.1$ & $46.6$ & $15.6$ & $7.3$ & $12.1$ & $5.0$ & $4.7$\\
$\rm{MF}_{4}$ & $20$ & $65.0$ & $45.0$ & $42.8$ & $7.5$ & $2.2$ & $3.5$ & $1.6$ & $1.3$\\
\tableline
\multicolumn{10}{c}{Simulations towards SMC}\\
$\rm{MF}_{1}$ & $3$ & $25.3$ & $19.2$ & $8.3$ & $27.2$ & $1.5$ & $2.2$ & $1.1$ & $1.0$\\
$\rm{MF}_{2}$ & $3$ & $18.6$ & $12.9$ & $7.5$ & $20.7$ & $0.9$ & $1.3$ & $0.7$ & $0.5$\\
$\rm{MF}_{3}$ & $3$ & $14.9$ & $9.7$ & $7.3$ & $14.0$ & $0.5$ & $0.8$ & $0.4$ & $0.3$\\
$\rm{MF}_{4}$ & $3$ & $9.8$ & $5.4$ & $6.7$ & $3.5$ & $0.1$ & $0.1$ & $0.1$ & $0.0$\\
\tableline
$\rm{MF}_{1}$ & $6$ & $38.3$ & $30.3$ & $15.0$ & $40.8$ & $6.1$ & $7.7$ & $4.7$ & $4.4$\\
$\rm{MF}_{2}$ & $6$ & $30.0$ & $21.6$ & $13.8$ & $33.0$ & $4.2$ & $5.5$ & $3.3$ & $2.9$\\
$\rm{MF}_{3}$ & $6$ & $25.2$ & $16.9$ & $13.4$ & $24.6$ & $2.8$ & $3.8$ & $2.2$ & $1.9$\\
$\rm{MF}_{4}$ & $6$ & $18.4$ & $10.6$ & $12.7$ & $9.8$ & $0.7$ & $1.1$ & $0.7$ & $0.4$\\
\tableline
$\rm{MF}_{1}$ & $10$ & $49.1$ & $39.7$ & $21.4$ & $51.1$ & $12.1$ & $14.2$ & $9.9$ & $9.3$\\
$\rm{MF}_{2}$ & $10$ & $40.3$ & $29.5$ & $20.0$ & $42.8$ & $9.2$ & $11.2$ & $7.6$ & $6.9$\\
$\rm{MF}_{3}$ & $10$ & $35.0$ & $23.8$ & $19.7$ & $33.8$ & $6.8$ & $8.6$ & $5.6$ & $4.9$\\
$\rm{MF}_{4}$ & $10$ & $27.4$ & $16.1$ & $19.0$ & $17.2$ & $2.6$ & $3.7$ & $2.4$ & $1.6$\\
\tableline
$\rm{MF}_{1}$ & $20$ & $64.1$ & $53.3$ & $32.1$ & $64.0$ & $23.1$ & $26.1$ & $20.0$ & $19.2$\\
$\rm{MF}_{2}$ & $20$ & $56.1$ & $42.0$ & $30.8$ & $56.0$ & $19.5$ & $22.4$ & $16.8$ & $15.9$\\
$\rm{MF}_{3}$ & $20$ & $51.1$ & $35.4$ & $30.9$ & $46.9$ & $16.4$ & $19.3$ & $14.1$ & $13.0$\\
$\rm{MF}_{4}$ & $20$ & $43.5$ & $26.2$ & $30.7$ & $29.6$ & $9.9$ & $12.6$ & $8.5$ & $7.1$\\
\enddata
\tablecomments{$\delta_{\rm{lens}}$ means all three relative errors in the lens mass, distance and its angular velocity are less than the given threshold value $\mathcal{T}$.}
\end{deluxetable}

In Table \ref{tab4}, we offer the results of relative error calculations by another method. In this table, each entrance represents the fraction (in per cent) of detectable microlensing events with the relative errors in the given parameter (its column) less than the given threshold value ($\mathcal{T}$) by considering a lens mass density $\rm{MF}$ (its row) in simulations towards LMC and SMC. The last column $\delta_{\rm{lens}}$ means considering three relative errors in the lens mass, distance and its angular velocity together. Accordingly, in $\lesssim1\%$ of detectable events, the lens physical parameters (including mass, distance and angular velocity) can be specified with the relative errors less than $3\%$. The fraction of detectable events with small relative errors by increasing the power index $\beta$ from zero to two in the lens mass density decreases. Its main reason is that by increasing the power index the average lens mass decreases which generates events with shorter durations. The number of data points is lower for shorter events. When the number of data points is higher the relative errors are lower. The fractions of detectable events with low relative errors (less than $3\%$) in two parameters $\theta_{\rm E}$ and $t_{\rm E}$ are higher than $0.3\%$ and $10\%$, respectively. These two parameters are proportional to $\propto\sqrt{M_{\rm{l}}}$ and have large values with small errors.   
\begin{deluxetable}{c  c  c  c c c c c c}
\tablecolumns{9}
\centering
\tablewidth{0.49\textwidth}\tabletypesize\footnotesize
\tablecaption{Each entrance reports the ruled-out upper limit on the mass of IBHs, $M_{\rm l}(M_{\sun})$, to entirely make DM halos at the $95\%$ C.L. (where $N_{\rm{e}, \rm{tot}}(M_{\rm l})=3$) resulting from simulations towards LMC (second row) and SMC (third row) by considering five values of $\mathcal{F}$. \label{tab5}}
\tablehead{$\mathcal{F}:$&$7e$-4 & $1e$-3 & $2e$-3 & $3e$-3 & $4e$-3 & $6e$-3 & $8e$-3 & $1e$-2}
\startdata
$\rm{LMC}$& $3.7$ &   $9.1$ &  $49.5$  &  $119.9$ &  $207.5$  &  $435.9$ &  $713.6$  &  $1031.8$\\
\tableline
$\rm{SMC}$& $--$ &   $--$ &  $1.3$  &  $3.6$ &  $7.6$  &  $22.2$ &  $50.1$  &  $86.9$\\
\enddata
\end{deluxetable} 
\begin{figure}
    \centering
    \includegraphics[width=0.49\textwidth]{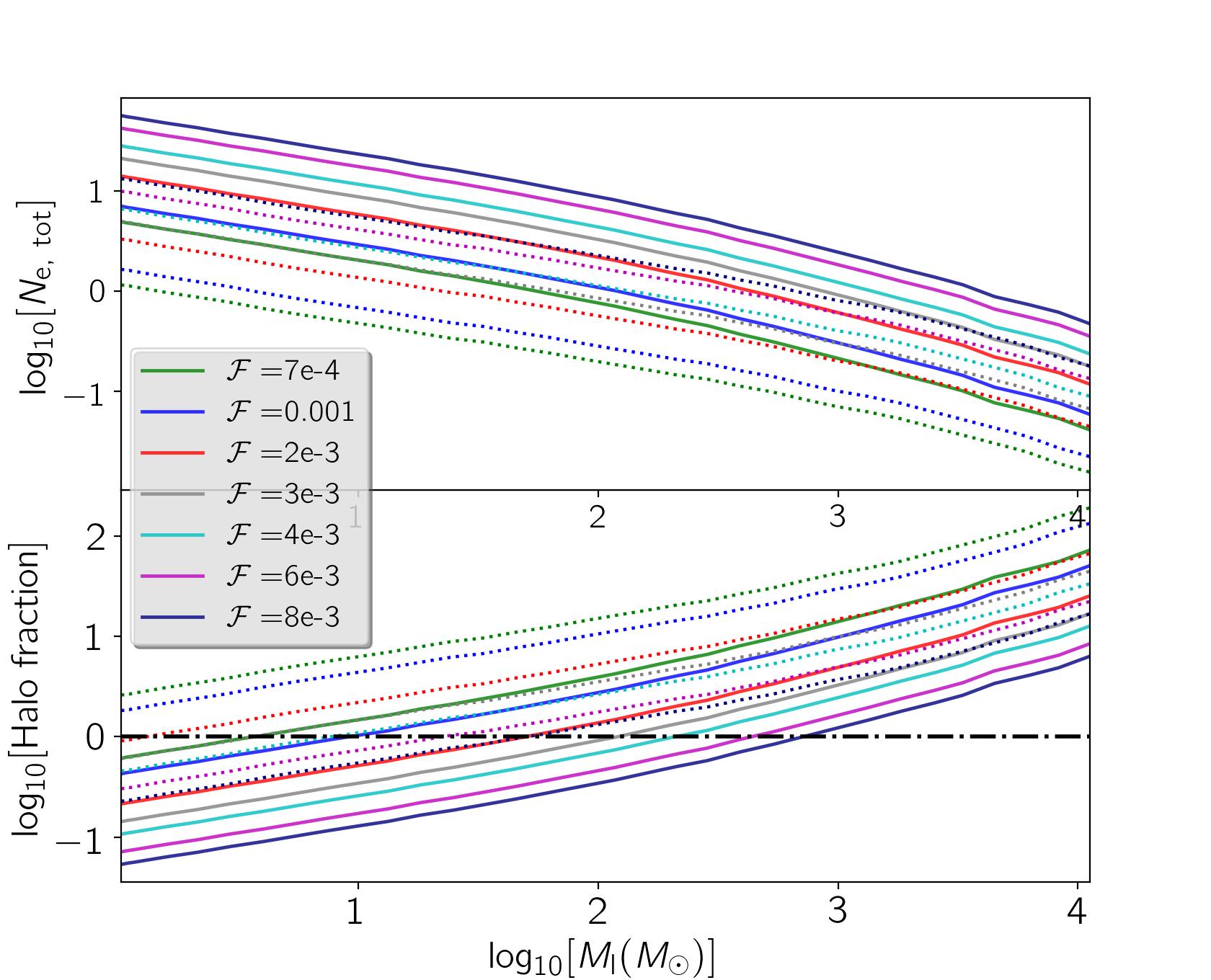}
    \caption{Top panel: the predicted number of detectable IBHs versus their mass by assuming $100\%$ contribution of IBHs in the halo dark matter for four values of $\mathcal{F}$ in the Rubin observations toward LMC (solid curves) and SMC (dot-dashed curves). Bottom panel: $95\%$ C.L. upper limit on the exclusion of IBHs to make entire DM halos versus their mass.}\label{nevent33}
\end{figure} 

\section{Expected Number of Detectable IBHs versus Mass}\label{sec5}
In this section, we calculate the expected number of detectable IBHs versus mass through microlensing observations during the Rubin $10$-year mission, by assuming that they make $100\%$ of dark matter (DM) halos (i.e., $f=1$). In these simulations, we consider discrete values for the lens mass in the logarithmic scale and in the range $\log_{10}[M_{\rm l}(M_{\odot})]\in[0.01,~4.05]$. For each lens mass, we calculate the overall $N_{\rm{e}}$ from two square-shape fields (with the area $15\times15~\rm{deg}^{2}$) towards LMC and SMC. In this regard, to derive the microlensing event rates and the lensing detection efficiency $\epsilon_{\rm L}(t_{\rm{E}}, \alpha, \delta)$ we use the efficiencies shown with blue (solid and dot-dashed) curves in the first panel of Figure \ref{plotef} which are due to the third lens mass function (log-uniform one). Also, we need $\mathcal{F}$ (the fraction of IBHs mass in the total mass of our galaxy) to estimate the number of events. We consider several values for this parameter which are reported in the first row of Table \ref{tab5}. In the top panel of Figure \ref{nevent33}, we plot the expected number of detectable events $N_{\rm{e}, \rm{tot}}$ versus IBHs mass in the logarithmic scale by considering seven values for $\mathcal{F}\in7e-4,~1e-3,~2e-3,~3e-3,~4e-3,~6e-3,~8e-3$ with green, blue, red, gray, cyan, magenta, and navy curves, respectively. Curves of $N_{\rm{e}, \rm{tot}}$ resulting from simulations toward LMC and SMC are depicted with solid and dotted curves, respectively.

If no microlensing event due to IBHs with a given mass is detected from the Rubin observations, the Poisson $95\%$ C.L. upper limit on the expected number of events from simulations and by assuming $100\%$ contribution of IBHs in the DM halo (i.e., $f=1$) will be three \citep{1996ApJAlcock,Griest2014,2021sajadian}. Accordingly, in the bottom panel of Figure \ref{nevent33} we shows $3/N_{\rm{e}}$ in the logarithmic scale as $95\%$ C.L. upper limit on the exclusion of IBHs to entirely make DM halos including Galactic and MCs' halos. The mass of IBHs with $N_{\rm{e}, \rm{tot}}(M_{\rm{l}})=3$ is the exclusion limit (on the mass of IBHs to make entire DM halos) which can be determined after the Rubin survey observations. We report these mass values for some given values of $\mathcal{F}$ and for simulations toward LMC (second row) and SMC (third row) in Table \ref{tab5}.  

\noindent Accordingly, the Rubin telescope potentially can exclude the full contribution of IBHs with the mass lower than $208$ and $8$ in halos while observing LMC and SMC, respectively by $95\%$ C.L., if $\mathcal{F}=4e-3$.


\section{Conclusions}\label{sec6}
In this work, we numerically studied one important output from the upcoming Rubin observations during its mission towards the Magellanic Clouds (MCs) which is detection and characterization of isolated black holes (IBHs). IBHs are detectable uniquely through survey microlensing observations during a long observing window, such as Vera C. Rubin observatory which is predicted to detect all sky during $\sim3$ days and its mission will last $10$ years. Its observing strategy is suitable to capture long-duration microlensing events due to IBHs specially ones located in the Galactic halo by observing MCs.   

We simulated microlensing events towards MCs due to IBHs with masses within a wide range $[3,~5000]M_{\odot}$ (including stellar-mass and massive BHs or primordial BHs) by considering four different mass densities as $dN/dM_{\rm{l}}\propto M_{\rm l}^{-\beta}$ where $\beta=0,~0.5,~1,~2$. We first extracted detectable source stars in at least two Rubin filters in the magnification peak and by considering the $r$-band blending parameter as a weight function. Accordingly, $\sim2\%$ and $\sim7\%$ of all stars in LMC and SMC were detectable by Rubin. For detectable stars we simulated microlensing events and extracted recognizable lensing events. Our detectability criteria for photometric lensing signals were (I) $\Delta \chi^{2}\ge2N_{\rm{data}}$ ($N_{\rm{data}}$ is the number of taken data points), (II) at least three data points should be above the baseline by more than three photometric accuracy, and (III) the light curves' width should be smaller than the Rubin observing window. Around $\sim25-60\%$ simulated events due to detectable source stars passed these three criteria. $f_{\rm{Halo}}\simeq70-90\%$ of these detectable events were halo-lensing (in which their lens objects were inside our galaxy). In lines of sight toward the central parts of MCs the fractions of halo-lensing events drop to $10-40\%$ among all detectable events. Self-lensing events, with lens objects within MCs, have $\theta_{\rm{E}}$ and $\pi_{\rm{E}}$ considerably smaller than those due to halo-lensing events by more than one order of magnitude which causes higher errors while extracting the physical parameters of their lens objects.    
 
We estimated the numbers of detectable stars towards LMC and SMC by Rubin as $\log_{10}[\overline{N_{\star, \rm{Rubin}}}(\rm{deg}^{-2})]\simeq 6.50,~6.07$ on average. We then estimated the microlensing events rates towards LMC and SMC due to IBHs as $\log_{10}[\overline{\Gamma_{\rm{obs}}}(\rm{star}^{-1}.\rm{yrs}^{-1} \times \mathcal{F}^{-1})]\simeq [-8.3,~-7.1]$, where $\mathcal{F}$ is the fraction of IBHs' mass in the total Galactic mass. We assumed that IBHs spatial distribution is similar to known stellar spatial distribution in our galaxy due to all objects, although it is smaller by the factor $\mathcal{F}$. Finally, we estimated the expected number of microlensing events in the Rubin observations from $15\times15~\rm{deg}^{2}$ fields projected on the sky plane while covering SMC (LMC) as $N_{\rm{e}, \rm{tot}} \times \mathcal{F}^{-1}\simeq 20 (160),~44 (318),~68 (490),~116 (792)$ by applying four lens mass functions with $\beta=0, 0.5, 1, 2$, respectively (see Table \ref{tab3}).

The astrometric deflections in source trajectories in these long-duration events are mostly straight lines with large amplitudes (large $\theta_{\rm{E}}$ values) specially for halo-lensing events (see, e.g., the first microlensing event was displayed in Figure \ref{fig1}).  For detectable halo-lensing events the efficiency for realizing astrometric deflections in source trajectories (the fraction of photometrically detectable events in which astrometric deflections are realizable) is $f_{\delta\theta_{\rm{c}}}\gtrsim10\%$, while for self-lensing (the events toward the central parts of MCs) this efficiency is lower than 10 per cent. Resolving two lensing-induced images (detecting separated images in at least three astrometric data points) for self-lensing events towards the SMC is doable with the efficiency $f_{\Delta\theta}\sim30\%$ which is higher than $f_{\delta\theta_{\rm{c}}}$. Its reason is that the number of visits towards the SMC is $\sim200-300$ which for most events this number of data points is not enough for passing the criterion $\Delta \chi^{2}_{{\delta\theta}_{\rm{c}}}\ge 2 N_{\rm{data}}$ and detecting their lensing-induced astrometric deflections. To resolve two lensing-induced images we demanded their angular separation be larger than the Rubin astrometric accuracy for resolving close stars ($\sigma_{\rm{r}}$) in at least only three data points. Detection of parallax effects in light curves of self-lensing events is a challenge, and only $\lesssim6\%$ of them have realizable parallax effects.

We evaluated the number of expected microlensing events (i) for discrete values of the IBHs' mass in the logarithmic scale and from the range $\log_{10}[M_{\rm{l}}(M_{\odot})]\in [0.01 4.05]$, and (ii) by assuming that IBHs make $f=100\%$ of DM halos. The resulting expected number of events decreases from $100$ to $0.1$ by IBHs' mass for $\mathcal{F}\sim 10^{-3}-10^{-2}$ (see Figure \ref{nevent33}). We concluded that the Rubin telescope would determine $95\%$ C.L. upper limits (for exclusion of IBHs to entirely make DM halos) on IBHs' mass $M_{\rm l}(M_{\odot})\lesssim 208,~8$ if $\mathcal{F}=4e$-$3$ from its survey observations during $10$ years towards LMC and SMC, respectively. 

\small{All simulations that have been done for this paper are available at: \url{https://github.com/SSajadian54/}}\\

\small{We are grateful from \v{Z}eljko~Ivezi\'{c} for his helpful comments and explanations about the Rubin astrometric accuracy. We especially thank the referee for the helpful comments and good suggestions. The work by S. Sajadian was supported by the Carlos Chagas Filho Foundation for Research Support of the State of Rio de Janeiro (FAPERJ). S. Sajadian thanks Centro Brasileiro de Pesquisas F\'{i}sicas-CBPF/MCTI (CBPF) for hospitality.}


\normalsize
\appendix
\section{Applied Spatial Distributions in the Milky Way and MCs}\label{app1}
Here, we report the spatial density distributions in our galaxy and MCs that we used to make lens and source populations in the Monte Carlo simulations.

{\bf Galactic density distributions:} There are two coordinate systems in our formalism: the observer coordinate $(D,~\theta,~\phi)$ which its center is over the observer, and the Galactic Cartesian coordinate $(x,~y,~z)$. Here, $D$ is the radial distance from the observer, $\theta=360-l$, $\phi=b$, where $(l,~b)$ are the Galactic longitude and latitude, respectively. The center of the Galactic Cartesian coordinate is the Galactic center, $x$ represents the Galactic center-Sun direction, and $z$ is normal to the Galactic plane. The relations converting these two coordinates are simply given by:
\begin{eqnarray}
x&=&D\cos\phi\cos\theta-D_{\odot},\nonumber\\
y&=&D\cos\phi\sin\theta,\nonumber\\
z&=&D\sin\phi,
\end{eqnarray} 
where, $D_{\odot}=8$ kpc is the Sun distance from the Galactic center in the Galactic plane. We take the galactic spatial distributions from the Besan\c{c}on model \citep{Robin2003,Robin2012}.

\noindent Accordingly, the spatial density distribution for the Galactic thin disk is given by:
\begin{eqnarray}  
\rho_{\rm{d}}^{\rm{Thin}}(x, y, z)=\sum_{i=1}^{7}\rho_{i}\Big(\exp\big[-\sqrt{0.25+\frac{r_{i}^{2}}{h_{1}^{2}}}\big]-\exp\big[-\sqrt{0.25+\frac{r_{i}^{2}}{h_{2}^{2}}}\big] \Big)+\rho_{0}\big(\exp\big[-\frac{r_{0}^{2}}{25\rm{kpc}^{2}}\big]-\exp\big[-\frac{r_{0}^{2}}{9\rm{kpc}^{2}}\big]\big),
\end{eqnarray}
where the summation is done over stellar populations with different ages, $r_{i}=\sqrt{R^{2}+z^{2}/\varepsilon_{i}^{2}}$, $R^{2}=x^{2}+y^{2}$, and  $\varepsilon_{i}=0.014,~0.0268,~0.0375,~0.0551,~0.0696,~0.0785,~0.0791,~0.0791$ kpc (the axis ratio). Also, $\rho_{i}=0.06,~0.768,~0.60,~0.384,~0.564,~0.48,~0.636,~0.384 M_{\odot}/\rm{pc}^{3}$ for $i\in [0,~7]$, $h_{1}=2.17$ kpc, and $h_{2}=1.33$ kpc.  Accordingly, the total mass of the Galactic thin disk will be $4.25\times10^{10} M_{\odot}$. 

In this model, the density distribution of the Galactic thick disk is given by:
\begin{eqnarray}
\rho_{\rm d}^{\rm{Thick}}(x, y, z)=\begin{cases} 
\rho_{0, \rm{Th}}\exp\big[-\frac{R-D_{\odot}}{2.5~\rm{kpc}}\big]\big(1-\frac{z^{2}}{0.8 \rm{kpc}^{2}}\big)~&|z|\leq0.4~\rm{kpc}\nonumber\\
1.32\times\rho_{0, \rm{Th}}\exp \big[-\frac{R-D_{\odot}}{2.5~\rm{kpc}}\big]\exp\big[-\frac{|z|}{0.8~\rm{kpc}}\big]~&\rm{elsewhere}
\end{cases}  
\end{eqnarray}    
where, $\rho_{0,~\rm{Th}}=0.0043M_{\odot}/\rm{pc}^{3}$, which offers the total mass of the Galactic thick disk being $0.8\times 10^{10}M_{\odot}$. 

\noindent For the Galactic stellar halo (it is also called the spheroid), the Besan\c{c}on model suggests the following density distribution: 
\begin{eqnarray}
\rho_{\rm{h}, \rm{ST}}(x, y, z)=\begin{cases}9.32\times 10^{-6} M_{\odot}/\rm{pc}^{3}&R_{\rm h}\leq0.5\rm{kpc}\nonumber\\
 \rho_{0, \rm h}\Big(\frac{r_{\rm{h}}}{D_{\odot}}\Big)^{-2.44} &\rm{elsewhere},\end{cases}
\end{eqnarray}
where, $r_{\rm h}^{2}=R^{2}+\big(\frac{z}{0.76 \rm{kpc}}\big)^{2}$, and $\rho_{0, \rm h}=5.677 \times10^{-5}M_{\odot}/\rm{pc}^{3}$.  According to this mass density, the total mass of the Galactic stellar halo is $1.2\times 10^{9}M_{\odot}$. 

\noindent In the Besan\c{c}on model, the following density distribution is introduced for the Galactic dark-matter halo: 
\begin{eqnarray}
\rho_{\rm{h}, \rm{DM}}(x, y, z)=\frac{\rho_{0, \rm{DM}}}{1+r^{2}/R_{\rm{c}}^{2}},
\label{dmro}
\end{eqnarray}
where, $R_{\rm{c}}=2.697$ kpc, $r^{2}=R^{2}+z^{2}$ is the radial distance from the Galactic center, and $\rho_{0,{DM}}=9.7 \times10^{-2}M_{\odot}/\rm{pc}^{3}$.  The characteristic density is indicated so that it offers the dark-matter density at the Sun position of $2.475\times10^{-3}M_{\sun}/\rm{pc}^{3}$, which offers the total mass of the Galactic dark-matter halo as  $1.8\times 10^{11}M_{\odot}$. 

\noindent The spatial density distribution of the Galactic bar that we used in our simulations was first offered by \citet{Robin2012}. For the Galactic bar, we first need to rotate $(x,~y)$ plane by $\alpha=12.89^{\circ}$ which is the bar inclination angle with respect to the Galactic center-Sun direction. Therefore, we introduce the Galactic bar coordinate system $(X, Y, Z)$ so that $X=x\cos \alpha+y\sin\alpha$, $Y=-x\sin\alpha+y\cos\alpha$, and $Z=z$. The Galactic bar density distribution has two parts $\rho_{\rm{S}}$ and $\rho_{\rm{E}}$ as following:  
\begin{eqnarray}
\rho_{\rm{S}}(X, Y, Z)=\begin{cases}\rho_{\rm{bar}}\cosh^{-2}(-R_{\rm{s}})~& R\leq R_{\rm{c}}\nonumber\\
\rho_{\rm{bar}}\cosh^{-2}(-R_{\rm{s}})\exp\big[-\frac{(R-R_{\rm{c}})^2}{0.25 \rm{kpc}^{2}}\big]&  \rm{elsewhere},
\end{cases}  
\end{eqnarray}  
where, $R=\sqrt{X^2+ Y^2}$,~and $R_{\rm{s}}^{c_{\rm{p}}}=\Big[\big|\frac{X}{X_{0}}\big|^{c_{\rm{n}}}+\big|\frac{Y}{Y_{0}}\big|^{c_{\rm n}}\Big]^\frac{c_{\rm{p}}}{c_{\rm{n}}}+\Big|\frac{Z}{Z_{0}}\Big|^{c_{\rm{p}}}$ ($R_{\rm{s}}$ is a non-dimensional parameter). The fixed parameters for this part of the bar density distribution are $X_{0}=1.46$ kpc, $Y_{0}=0.49$ kpc, $Z_{0}=0.39$ kpc, $R_{\rm c}=3.43$ kpc, $c_{\rm{p}}=3.007$, $c_{\rm{n}}=3.329$, and $\rho_{\rm{bar}}=4.145M_{\odot}/\rm{pc}^{3}$.

\noindent The second part of the Galactic bar density is given by:  
\begin{eqnarray}
\rho_{\rm{E}}(X, Y, Z)=\begin{cases} \rho_{\rm{bar}}\exp[-R_{\rm{s}}]~&R\leq R_{\rm{c}}\nonumber\\
\rho_{\rm{bar}}\exp[-R_{\rm{s}}]\exp\big[-\frac{(R-R_{\rm{c}})^2}{0.25~\rm{kpc}^2}\big]&\rm{elsewhere},\end{cases}  
\end{eqnarray}        
where, $X_{0}=4.44$ kpc, $Y_{0}=1.31$ kpc, $Z_{0}=0.80$ kpc, $R_{\rm c}=6.83$ kpc, $c_{\rm{p}}=2.786$, $c_{\rm{n}}=3.917$, and $\rho_{\rm{bar}}=0.0117M_{\odot}/\rm{pc}^{3}$. Hence, the Galactic bar density distribution is derived by $\rho_{\rm b}(X, Y, Z)= \rho_{\rm S}+\rho_{\rm E}$ which offers a bulge with the total mass of $1.7\times10^{10}$. 

\textbf{MCs density distributions}: For simulating MCs structures, we first convert the Galactic longitude and latitude $(l, b)$ to the right ascension $\alpha$ and declination $\delta$ (related to the equatorial coordinate system) using the Python package \texttt{astropy.coordinates} \citep{2013AAAstropy}. We then define two coordinate systems which are the observer and MCs coordinate systems. The Cartesian observer system, $(x_{\rm{o}},~y_{\rm{o}},~z_{\rm{o}})$, is specified by its $z_{\rm{o}}$-axis towards the observer, $x_{\rm{o}}$-axis anti-parallel to the right ascension axis and $y_{\rm{o}}$-axis parallel with the declination axis. The MCs coordinate system, $(x',~y',~z')$ is determined by $x'$ and $y'$ axes which are parallel with the semi-major and semi-minor axes of the MCs' disk, respectively. The centers of both coordinate systems are on the MCs center. We first determine $(x_{\rm{o}}, y_{\rm{o}}, z_{\rm{o}})$ from $(D, \alpha, \delta)$ using the converting relations in the spherical trigonometry \citep[e.g.,][]{2001ApJweinbergnikolaev,2012ApJsubramanian}:  
\begin{eqnarray}
x_{\rm{o}}&=&-D\cos\delta \sin(\alpha-\alpha_{\rm{MC}}), \nonumber\\
y_{\rm{o}}&=&D\sin\delta \cos\delta_{\rm{MC}}-D\cos\delta \sin\delta_{\rm{MC}}\cos(\alpha-\alpha_{\rm{MC}}), \nonumber\\
z_{\rm{o}}&=&D_{\rm{MC}}-D\cos\delta\cos \delta_{\rm{MC}}\cos(\alpha-\alpha_{\rm{MC}})-D \sin\delta \sin\delta_{\rm{MC}},
\end{eqnarray} 
where, $D_{\rm{MC}}=49.97~\rm{kpc},~\rm{and}~61.70$ kpc are the LMC and SMC distances from the observer. $\alpha_{\rm{MC}}=80.89,~13.19^{\circ}$ are the right ascension values of the LMC and SMC centers, $\delta_{\rm{MC}}= -68.24,~-71.17^{\circ}$ are the declination values of the LMC and SMC centers. To specify the MCs density distributions, we rotate this coordinate system by $\theta_{0}$ around $z_{\rm o}$, and then by $-i$ around the resulting $x'$-axis as following:
\begin{eqnarray}\label{convert}
x' &=&x_{\rm o}\cos\theta_{0}+y_{\rm{o}}\sin\theta_{0},\nonumber\\
y' &=&-x_{\rm{o}} \sin\theta_{0}\cos i+y_{\rm{o}}\cos\theta_{0} \cos i-z_{\rm{o}}\sin i,\nonumber\\
z' &=&-x_{\rm{o}} \sin\theta_{0}\sin i+y_{\rm{o}}\cos\theta_{0} \sin i + z_{\rm{o}}\cos i,
\end{eqnarray}
where, $\theta_{0}=80,~24$ and $i=34.7,~74.0$ are the projection angles for the LMC and SMC disks, respectively. For the LMC bar, these two projection angles are $\theta_{0}=20^{\circ}$ and $i=0^{\circ}$.

The LMC structures have been studied in many references \citep[see, e.g., ][]{VanderMarel2001a,VanderMarel2001b,VanDerMarel2002}. In this work, for the LMC disk we use the following spatial density distribution which was suggested in \citet{2000ApJGyuk}:
\begin{eqnarray}
\rho_{\rm{d}}^{\rm{LMC}}=\rho_{\rm d, 0}^{\rm{LMC}}\exp\big[-\frac{R'}{1.8 \rm{kpc}}\big]\exp\big[- \frac{|z'|}{0.3\rm{kpc}}\big], 
\end{eqnarray}
where, $\rho_{\rm d, 0}^{\rm  LMC}=0.6206M_{\odot}/\rm{pc}^{3}$, $R'=\sqrt{x'^{2}+y'^{2}}$.

\noindent The density distribution for the LMC bar that we use in our simulations is given by\citep{2000ApJGyuk}: 
\begin{eqnarray}
\rho_{\rm{b}}^{\rm{LMC}}= \rho_{\rm b, 0}^{\rm{LMC}}\exp\Big[-0.5 \Big((\frac{x'}{1.2\rm{kpc}})^{2}+\frac{y'^{2}+z'^{2}}{0.194\rm{kpc}^{2}}\Big)\Big],
\end{eqnarray}
where $\rho_{\rm{b}, 0}^{\rm{LMC}}=0.697 M_{\odot}/{\rm{pc}}^{3}$. 

\noindent The offered stellar halo distribution  for LMC and SMC is the spherical shape (spheroid) and is given by \citep{LMCstellarHalo}:   
\begin{eqnarray}
\rho_{\rm{h}, \rm{ST}}^{\rm{MC}}=\frac{\rho_{0, \rm{h}}}{\big(1+r'^{2} /R_{\rm{h}}^{2} \big)^{3/2}},
\label{halos}
\end{eqnarray}
where, $r'^{2}=x'^{2}+y'^{2}+z'^{2}$ is the radial distance from the LMC and SMC centers. Also, $R_{\rm{h}}=1.42,~1.5$ kpc, $\rho_{0, \rm{h}}=0.00195,~5e-05 M_{\odot}/{\rm{pc}}^{3}$ for LMC and SMC, respectively. We note that the total mass of the LMC spheroid resulting from this distribution is $\simeq 10^{9}M_{\odot}$.  

\noindent The dark-matter halo distribution for LMC and SMC is given by \citep{2000ApJGyuk,LMChalototalmass}:   
\begin{eqnarray}
\rho_{\rm{h}, \rm{DM}}^{\rm{MC}}=\frac{\rho_{0, \rm{h}}}{1+r'^{2} /R_{\rm{c}}^{2}},
\label{halos}
\end{eqnarray}
where, $R_{\rm{c}}=2.0,~1.5$ kpc, $\rho_{0, \rm{h}}=0.024,~0.000625 M_{\odot}/{\rm{pc}}^{3}$ for LMC and SMC, respectively. The total mass of the LMC spheroid resulting from this distribution is around $10^{11}M_{\odot}$, although less than $5\%$ of it is in the form of compact and massive objects which can act as microlenses.

SMC does not have any bulge structure, and only have one disk as well as its stellar halo. The density distribution of its disk as offered by \citep{2013MNRAS.435.1582C} has two parts due to two stellar populations which are given in the following:    
\begin{eqnarray}
\rho_{\rm {d}}^{\rm{SMC}}=\rho_{1}\exp\Big[\frac{-1}{2}\big((\frac{x'}{0.8 \rm{kpc}})^{2}+(\frac{y'}{3.5 \rm{kpc}})^{2}+(\frac{z'}{1.3 \rm{kpc}})^{2} \big)\Big]+ \rho_{2}\exp\Big[-\sqrt{(\frac{x'}{0.8 \rm{kpc}})^{2} +(\frac{y'}{1.2 \rm{kpc}})^{2}}\Big]\exp \Big[\frac{-z'^{2}}{8.8\rm{kpc}^{2}}\Big],
\end{eqnarray}
here, $\rho_{1}= 0.00425 M_{\odot}\rm{pc}^{-3}$, and $\rho_{2}= 0.0195 M_{\odot}\rm{pc}^{-3}$. We note that the total dynamic masses of the LMC and SMC are 0.15 and 0.005 times the Galactic dynamic mass, respect.  

\section{Velocity Distributions Applied in the Monte-Carlo Simulations}\label{app2}
In our galaxy, stars in the Galactic thin and thick disks have a global (or rotational) velocity which is a function of the radial distance from the Galactic center in the Galactic plane. \citet{2009AARahal} offered the following relation for this component of stellar velocity: 
\begin{eqnarray}
V_{\rm{T}}=V_{\rm{T}, \odot} \Big(1.00762 \big(\frac{R}{D_{\sun}}\big)^{0.0394}+0.00712\Big),
\end{eqnarray}
where, $V_{\rm{T}, \odot}=226\rm{km}/\rm{s}$ is the global velocity of the Sun around the Galactic center, and $R=\sqrt{D^{2}\cos^{2}\phi + D_{\odot}^{2}- 2D_{\odot} D \cos\phi\cos\theta}$ is the radial distance from the Galactic center projected on the Galactic plane.

\noindent In addition to the global velocity, all stars in our galaxy have dispersion velocities in the radial, transverse, and normal directions which are determined from Gaussian (or normal) distributions ($\mathcal{N}(\mu,~\sigma)$) with the zero mean ($\mu=0$) and the widths $\sigma_{\rm R}$, $\sigma_{\rm T}$, and $\sigma_{\rm Z}$. According to the Besan\c{c}on model, for stars inside the Galactic bulge, these widths (or velocity dispersions) are $113,~115,~100$ km$/$s, respectively. Also, for stars within the Galactic halo and thick disk, they are $131,~106,~85$ km/s, and $67,~51,~42$ km$/$s, respectively. For dark-matter halo, velocity dispersions in different directions are similar and $\simeq 120$km/s. These velocity dispersions for stars in the Galactic thin disk depend on stellar ages, as mentioned in Table (2) of \citet{Robin2003}, and we do not repeat them here. We note that the tangential velocity components of stars in the Galactic disks are given by $V_{\rm T}+\mathcal{N}(\mu=0,~\sigma=\sigma_{\rm T})$.

We determine three components of stellar velocities inside the LMC using $v_{\rm R}=-57+\mathcal{R}[-13, 13]$ km/s, $v_{\rm T}=-226+\mathcal{R}[-15, 15]$ km/s, and $v_{\rm Z}=221+\mathcal{R}[-19, 19]$ km/s which determine the velocity of the LMC's center of mass \citep{LMCcofmass,2020AAMoniez}. Here, $\mathcal{R}$ is a function that chooses a random number inside its given range uniformly. Additionally, for each component, we add a velocity dispersion using $\mathcal{N}(\mu=0,~\sigma=20.2\rm{km}/\rm{s})$. For the SMC the three components of its center of mass's velocity are $v_{\rm R}=19+\mathcal{R}[-18, 18]$ km/s, $v_{\rm T}=-153+\mathcal{R}[-21, 21]$ km/s, and $v_{\rm Z}=153+\mathcal{R}[-17, 17]$ km/s \citep{2013ApJKallivayalil}. In addition, we include the dispersion velocity' components by the aid of $\mathcal{N}(\mu=0,~\sigma=27.5 \rm{km}/\rm{s})$. For the dark-matter halo, the velocity dispersions are higher and $50$km/s.


\section{Initial mass Functions in the Galactic disk}\label{app3}
To evaluate $\mathcal{F}$ in Equation \ref{mathf} we take the initial mass functions which are explained here. In the Besan\c{c}on model and the Galactic disk the initial mass function is specified by $\eta(M)=dN\big/dM\propto M^{-\gamma}$, where $\gamma=1.6$ for $M\leq M_{\sun}$, and $\gamma=3$ for $M\geq M_{\sun}$. However, for brown dwarfs with masses in the range $13M_{\rm{J}}\leq M\leq0.08M_{\sun}$ we use their known mass function that is $\eta\propto M^{-0.7}$ \citep{2015ApJBHMASS}. For the mass function of free-floating planets, a log-uniform function is commonly used specially for microlensing observations and their interpretations \citep[see, e.g., ][]{2020AJJohnson,2025MNRASSbfree}. IBHs with the mass higher than 100 solar mass have mostly a dark-matter (DM) origin, so we use the DM mass function for these massive objects while estimating $\mathcal{F}$. The mass-function for dark-matter halos is known as $\zeta(M)\propto M^{-1.9}$ \citep{ShethMo20001,MASSFunctionHAlo}.

\bibliography{ref}{}
\bibliographystyle{aasjournal}
\end{document}